\documentclass[12pt,a4paper]{article}
\usepackage[utf8]{inputenc}
\usepackage{epsf}
\usepackage{latexsym,amssymb,euscript}
\usepackage[dvips]{graphicx}
\usepackage[numbers,sort&compress]{natbib}
\usepackage{amsmath}
\usepackage{nicefrac}
\usepackage{slashed}
\usepackage{booktabs}
\usepackage[linktocpage]{hyperref}
\usepackage{braket}
\usepackage{chngcntr}
\usepackage{bbold}
\usepackage{multicol}
\usepackage{graphics}
\usepackage{graphicx}
\usepackage{mciteplus}
\usepackage{caption}
\usepackage{subcaption}
\usepackage{pdfpages}
\usepackage[titletoc]{appendix}
\usepackage[normalem]{ulem}
\graphicspath{{./figures/}}
\hypersetup{
 linktocpage = false,
 urlcolor = urlblue,
 colorlinks = true,
 linkcolor = urlblue,
 anchorcolor = urlblue,
 citecolor = urlblue,
 pdfstartview = {XYZ null null 1.25} 
           }
\usepackage[left=2cm, right=2cm]{geometry}
\usepackage{pstricks}
\usepackage{color}
\usepackage{xcolor}
\definecolor{urlblue}{rgb}{0.2,0.4,0.7}
\definecolor{citegreen}{rgb}{0,0.4,0.2}
\definecolor{linkred}{rgb}{0.9,0.2,0.1}
\usepackage{float}
\usepackage{academicons}
\definecolor{orcidlogocol}{HTML}{A6CE39}
\usepackage{fancyhdr}
\newcommand{\drv}{{\rm d}}

\newcommand{\LQCD}{\Lambda_{\rm QCD}}
\newcommand{\MSb}{\overline{\rm MS}}

\newcommand{\LL}{{\rm LL/LO}}

\newcommand{\NLLp}{{\rm NLL/NLO^+}}

\newcommand{\HENLOp}{{\rm HE}\mbox{-}{\rm NLO^+}}

\newcommand{\ClLL}{{\cal C}_l^\LL}

\newcommand{\ClNLLp}{{\cal C}_l^\NLLp}

\newcommand{\ClHENLOp}{{\cal C}_l^{{\rm HE}\text{-}{\rm NLO}^+}}

\newcommand{\DY}{\Delta Y}

\newcommand{\vqTTa}{\langle {\vec q}_T^{\;2} \rangle}

\newcommand{\E}{{\cal E}}

\newcommand{\Jpsi}{J/\psi}

\newcommand{\BCs}{B_c(^1S_0)}
\newcommand{\Bss}{B_c(^3S_1)}

\newcommand{\XQq}{X_{Qq\bar{Q}\bar{q}}}
\newcommand{\Xcq}{X_{cq\bar{c}\bar{q}}}

\newcommand{\Xcu}{X_{cu\bar{c}\bar{u}}}
\newcommand{\Xcs}{X_{cs\bar{c}\bar{s}}}
\newcommand{\Xbu}{X_{bu\bar{b}\bar{u}}}
\newcommand{\Xbs}{X_{bs\bar{b}\bar{s}}}
\newcommand{\TQQ}{T_{4Q}}
\newcommand{\TQc}{T_{4c}}
\newcommand{\TQcZpp}{T_{4c}(0^{++})}
\newcommand{\TQcTpp}{T_{4c}(2^{++})}

\newcommand{{\HFNRevo}}{\tt HF-NRevo}

\newcommand{{\Jethad}}{\tt JETHAD}
\newcommand{{\symJethad}}{\tt symJETHAD}
\newcommand{{\Hell}}{\tt HELL}
\newcommand{{\RadISH}}{\tt RadISH}
\newcommand{{\Pegasus}}{\tt QCD-PEGASUS}
\newcommand{{\HOPPET}}{\tt HOPPET}
\newcommand{{\QCDNUM}}{\tt QCDNUM}
\newcommand{{\APFEL}}{\tt APFEL}
\newcommand{{\APFELpp}}{\tt APFEL++}
\newcommand{{\APFELppp}}{\tt APFEL(++)}
\newcommand{{\EKO}}{\tt EKO}

\newcommand{\eref}[1]{~\eqref{#1}}

\newcommand{\orcidFGC}{\href{https://orcid.org/0000-0003-3299-2203}{\includegraphics[scale=0.1]{logo-orcid.pdf}}}

\newcommand{\orcidAP}{\href{https://orcid.org/0000-0001-8984-3036}{\includegraphics[scale=0.1]{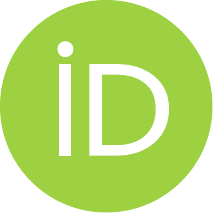}}}

\begin{document}

\begin{titlepage}

\begin{center}
  {\LARGE \bf Fully charmed tetraquarks from LHC to FCC: \vskip.025cm Natural stability from fragmentation}
\end{center}

\vskip 0.35cm

\centerline{
Francesco~Giovanni~Celiberto,$^{\;1\;\dagger}$ {\orcidFGC}
Gabriele~Gatto,$^{\;2,3\;\ddagger}$
and Alessandro~Papa$^{\;2,3\;\S}$ {\orcidAP}
}

\vskip .4cm

\centerline{${}^2$ {\sl Universidad de Alcal\'a (UAH), E-28805 Alcal\'a de Henares, Madrid, Spain} }
\vskip .18cm
\centerline{${}^2$ {\sl Dipartimento di Fisica, Universit\`a della Calabria,}}
\centerline{\sl I-87036 Arcavacata di Rende, Cosenza, Italy}
\vskip .18cm
\centerline{${}^3$ {\sl Istituto Nazionale di Fisica Nucleare, Gruppo collegato di Cosenza,}}
\centerline{\sl I-87036 Arcavacata di Rende, Cosenza, Italy}
\vskip 0.55cm

\begin{abstract}
\vspace{0.25cm}
\hrule \vspace{0.50cm}
We investigate the inclusive production of fully charmed tetraquarks, $T_{4c}(0^{++})$ or $T_{4c}(2^{++})$ radial excitations, in high-energy proton collisions.
We build our study upon the collinear fragmentation of a single parton in a variable-flavor number scheme, suited to describe the tetraquark formation mechanism from moderate to large transverse-momentum regimes.
To this extent, we derive a novel set of DGLAP-evolving collinear fragmentation functions, named {\tt TQ4Q1.0} determinations. 
They encode initial-scale inputs corresponding to both gluon and heavy-quark fragmentation channels, defined within the context of quark-potential and spin-physics inspired models, respectively.
We work within the NLL/NLO$^+$ hybrid factorization and make use of the {\tt JETHAD} numeric interface along with the {\tt symJETHAD} symbolic calculation plugin.
With these tools, we provide predictions for high-energy observables sensitive to $T_{4c}$ plus jet emissions at center-of-mass energies ranging from 14~TeV at the LHC to the 100~TeV nominal energy of the~FCC.
\vspace{0.50cm} \hrule
\vspace{0.75cm}
{
 \setlength{\parindent}{0pt}
 \textsc{Keywords}: \vspace{0.15cm} \\ 
 Exotic matter \\
 Tetraquarks \\
 Heavy flavor \\
 High-energy QCD \\
 Fragmentation \\
 Resummation \\ 
 Natural stability 
}
\end{abstract}

\vfill
$^{\dagger}${\it e-mail}:
\href{mailto:francesco.celiberto@uah.es}{francesco.celiberto@uah.es}

$^{\ddagger}${\it e-mail}:
\href{gabriele.gatto@unical.it}{gabriele.gatto@unical.it}

$^{\S}${\it e-mail}:
\href{alessandro.papa@fis.unical.it}{alessandro.papa@fis.unical.it}

\end{titlepage}

\tableofcontents
\clearpage

\section{Hors d'{\oe}uvre}
\label{sec:intro}

What is the true nature of exotic hadrons?
What information about their structure and dynamical formation mechanism(s) can we gather from hadronic collisions at present and new-generation colliders?
Although uncovering the fundamental dynamics behind exotic matter production is still challenging, recent advancements in all-order perturbative techniques and QCD factorization may unveil novel insights.

Within the realm of hadrons, quarkonium particles hold a special place.
They are mesons whose lowest Fock state consists of a heavy quark, $Q$, and its antiquark, $\bar Q$.
The exploration of quarkonium phenomena traces back to the ``First Quarkonium Revolution'' of November 1974.
In that period, a novel vector meson, with a mass around 3.1~GeV and possessing vector quantum numbers, was independently discovered by two distinct research groups. 

The newly observed meson was named $\Jpsi$, in homage to the institutions where its discovery took place: the Stanford Linear Accelerator Center (SLAC)~\cite{SLAC-SP-017:1974ind}, and the Brookhaven National Laboratory (BNL)~\cite{E598:1974sol}. 
In short order following these seminal announcements, the existence of the $\Jpsi$ was confirmed by ADONE in Frascati~\cite{Bacci:1974za}. 
This marked the onset of intensive investigations into quarkonium phenomena, yielding profound insights into the nature of the strong force and the underlying quark structure of hadrons.

While quarkonium mesons are considered ordinary hadrons, the QCD color neutrality allows for the existence of particles with more complex valence-parton configurations, giving rise to exotic hadrons. 
These exotic particles possess quantum numbers that defy explanation by the conventional three-quark or quark-antiquark configurations. 
Instead, they consist of more involved combinations of quarks, antiquarks, and gluons. 
Deciphering the internal structure of these exotic hadrons has been a major focus of the exotic spectroscopy.

Exotic hadrons can be broadly categorized into two groups: those containing active gluons, such as quark-gluon hybrids~\cite{Kou:2005gt,Braaten:2013boa,Berwein:2015vca} and glueballs~\cite{Minkowski:1998mf,Mathieu:2008me,D0:2020tig,Csorgo:2019ewn}, and those comprising multiple quarks, such as tetraquarks and pentaquarks~\cite{Gell-Mann:1964ewy,Jaffe:1976ig,Ader:1981db}.
The first exotic hadron, the $X(3872)$, was observed in 2003 by Belle at KEKB~\cite{Belle:2003nnu}.
This discovery heralded the dawn of what is known as the ``Exotic Matter Revolution'', or ``Second Quarkonium Revolution''. 
The $X(3872)$, believed to be composed of $c$ and $\bar{c}$ quarks~\cite{Chen:2016qju,Liu:2019zoy}, is a hidden-charm particle. 
In 2021, the LHCb experiment observed the first exotic state with open-charm flavor, the $X(2900)$ particle~\cite{LHCb:2020bls}.

Despite the $X(3872)$ having ordinary quantum numbers, its decay properties spoil isospin conservation, thus indicating a more complex inner structure beyond the traditional quarkonium states. 
Various alternative dynamical mechanisms have been proposed, surpassing the standard quarkonium scenario and going along the tetraquark direction.
They include: a compact diquark system~\cite{Maiani:2004vq,tHooft:2008rus,Maiani:2013nmn,Maiani:2014aja,Maiani:2017kyi,Mutuk:2021hmi,Wang:2013vex,Wang:2013exa,Grinstein:2024rcu}, a loosely bound meson molecule~\cite{Tornqvist:1993ng,Braaten:2003he,Guo:2013sya,Mutuk:2022ckn,Wang:2013daa,Wang:2014gwa,Esposito:2023mxw,Grinstein:2024rcu}, or a hadroquarkonium composed of quarkonium nucleus with an orbiting light meson~\cite{Dubynskiy:2008mq,Voloshin:2013dpa,Guo:2017jvc,Ferretti:2018ojb,Ferretti:2018tco,Ferretti:2020ewe}.

Highlights on the nature of the $X(3872)$ hadron may come out from studies on high-multiplicity proton collisions~\cite{Esposito:2020ywk}, as well as from applications of potential approaches to the hadronic thermal behavior~\cite{Armesto:2024zad}.
The first doubly charmed $T_{cc}^+$ tetraquark was observed at LHCb in 2021~\cite{LHCb:2021vvq,LHCb:2021auc}.
References~\cite{Fleming:2021wmk,Dai:2023mxm,Hodges:2024awq} describe this state as a nonrelativistic molecule of two $D$ mesons \emph{via} a well-suited nonrelativistic effective field theory, known as XEFT~\cite{Fleming:2007rp,Fleming:2008yn,Braaten:2010mg,Fleming:2011xa,Mehen:2015efa,Braaten:2020iye}.

Until recently, the $X(3872)$ stood as the sole exotic state observed in prompt proton collisions. However, the situation changed with the discovery of the aforementioned  $T_{cc}^+$~\cite{LHCb:2021vvq,LHCb:2021auc} and of a new resonance in the double $\Jpsi$ invariant-mass spectrum~\cite{LHCb:2020bwg}.
This resonance, known as $X(6900)$, is strongly believed to represent a favored candidate for the ground state ($0^{++}$) or, more likely, the radial excitation ($2^{++}$) of the fully charmed tetraquark, $\TQc$~\cite{Chen:2022asf}.

From a theory standpoint, $\TQQ$ tetraquarks perhaps represent the most straightforward exotic species to examine.
With the heavy-quark mass $m_Q$ well above the perturbative threshold, a fully heavy tetraquark can be conceptualized as a composite system comprising two nonrelativistic charms and two anticharms.
Its lowest Fock state, $|QQ\bar{Q}\bar{Q}\rangle$, remains uncontaminated by contributions from valence light quarks and dynamical gluons.
This bears a striking resemblance to quarkonia, whose leading state is just $|Q\bar{Q}\rangle$.
This strongly suggests that the theoretical methodologies employed to study quarkonia can also be applied to heavy tetraquarks.
Thus, in the same way that charmonia are wisely referred to as QCD ``hydrogen atoms''~\cite{Pineda:2011dg}, fully charmed tetraquarks can be seen either as QCD ``helium nuclei'' or ``hydrogen molecules'', depending on the embraced vision.

Despite extensive investigations into the mass spectra and decay properties of exotic hadrons since the discovery of the $X(3872)$, our understanding of their dynamical production mechanism(s) remains challenging. Only a few model-dependent studies, mostly based on color evaporation~\cite{Maciula:2020wri} and hadron-quark duality~\cite{Berezhnoy:2011xy,Karliner:2016zzc,Becchi:2020mjz}, have been proposed thus far.
The impact of multi-particle interactions in heavy-tetraquark production at hadron colliders was addressed in Refs.~\cite{Carvalho:2015nqf,Abreu:2023wwg}.
High-energy factorization studies on tetraquark structures were presented in Ref.~\cite{Cisek:2022uqx}.
Reference~\cite{Feng:2020qee} deals with the exclusive radiative emissions of $\TQc$ states at $B$ factories.
A quite recent analysis addresses $\TQc$ photoproduction at electron-ion colliding machines~\cite{Feng:2023ghc}.

The quite large $X(3872)$ cross sections at high transverse momenta, as recorded by LHC experiments~\cite{CMS:2013fpt,ATLAS:2016kwu,LHCb:2021ten}, yield significant implications for its underlying formation dynamics. 
These findings offer a unique chance for refining theoretical models and may lend support to production mechanisms natively connected to high-energy QCD, such as the leading-twist \emph{fragmentation} of a single parton into the observed hadron.

Building on modern quarkonium theory, a recent study provided the first calculation of the collinear fragmentation function (FF) governing the transition of an outgoing gluon to a $\TQc$ state within a potential-quark, nonrelativistic QCD (NRQCD) framework~\cite{Feng:2020riv}.
Few years later, evidence was provided that the $[g \to \TQc]$ fragmentation mechanism becomes dominant over direct short-distance production when the $\TQc$ is detected with a transverse momentum exceeding 20~GeV~\cite{Feng:2023agq}.
This result aligns with previous observations for quarkonia, where gluon fragmentation to $\Jpsi$ was found to dominate over direct production for transverse momenta approximately below 10~GeV~\cite{Doncheski:1993xm,Cacciari:1994dr}.

In the present study we will address the inclusive production, in high-energy proton scatterings, of two kinds of fully charmed tetraquarks: $T_{4c}(0^{++})$ bound states and their $T_{4c}(2^{++})$ radial excitations.
We will ground our analysis on the single-parton collinear fragmentation in a variable-flavor number scheme (VFNS), designed to describe the formation mechanism of heavy-flavored hadrons at moderate to large transverse momenta~\cite{Mele:1990cw,Cacciari:1993mq,Buza:1996wv}.

To this extent, we will derive a new set of collinear FFs, the {\tt TQ4Q1.0} functions, which incorporate initial-scale inputs from both gluon and charm fragmentation channels in a color-singlet configuration, respectively defined in the context of quark-potential NRQCD and spin-physics inspired models.
By making use of basic features of the newly developed heavy-flavor nonrelativistic evolution ({\HFNRevo}) scheme~\cite{Celiberto:2024mex,Celiberto:2024bxu}, we will perform a proper DGLAP evolution of these inputs, which consistently accounts for thresholds of the gluon and the charm quark.

As an application to phenomenology, we will work within the $\NLLp$ hybrid factorization scheme, where the resummation of next-to-leading (NLO) energy logarithms and beyond is consistently incorporated in the standard collinear picture.
We will make use of the {\Jethad} numeric interface along with the {\symJethad} symbolic calculation plugin~\cite{Celiberto:2020wpk,Celiberto:2022rfj,Celiberto:2023fzz,Celiberto:2024mrq,Celiberto:2024swu}, to provide predictions for high-energy observables sensitive to $T_{4c}$ plus jet emissions at center-of-mass energies ranging from 14~TeV~LHC to 100~TeV nominal energy of~FCC~\cite{FCC:2018byv,FCC:2018evy,FCC:2018vvp,FCC:2018bvk}.
The VFNS collinear fragmentation mechanism describing the production of a $\TQc$ state will shield the hybrid factorization from logarithmic instabilities.
The \emph{natural stability} resulting from $\TQc$ fragmentation will come as a key ingredient for our phenomenological analysis.

The structure of this work reads as follows. 
In Section~\ref{sec:FFs} we highlight the way the novel {\tt TQ4Q1.0} tetraquark FFs are built.
In Section~\ref{sec:hybrid_factorization} we introduce the $\NLLp$ hybrid factorization.  
Section~\ref{sec:results} is devoted to a phenomenological analysis of our high-energy observables. 
Finally, in Section~\ref{sec:conclusions} we draw our conclusions and provide some outlook.

\section{Tetraquark collinear fragmentation}
\label{sec:FFs}

In this Section we explain our strategy to build the two novel {\tt TQ4Q1.0} FF sets describing the production of $\TQcZpp$ and $\TQcTpp$ bound states.
In Section~\ref{ssec:HF_fragmentation} we briefly review basics of heavy-hadron fragmentation, proposing a short journey from heavy-light mesons to quarkonia and then tetraquarks.
Sections~\ref{ssec:FFs-g} and~\ref{ssec:FFs-Q} give insights on initial-scale inputs for the gluon and constituent heavy-quark to tetraquark fragmentation channel, respectively.
Section~\ref{ssec:FFs-TQ4Q10} contains a discussion on our DGLAP-evolved FFs and their properties.
All the symbolic calculations needed here were done by making use of {\symJethad}, the newly implemented \textsc{Mathematica} plugin of {\Jethad}, aimed at the symbolic computing of analytic expressions for high-energy QCD and the hadronic structure~\cite{Celiberto:2020wpk,Celiberto:2022rfj,Celiberto:2023fzz,Celiberto:2024mrq,Celiberto:2024swu}.

\subsection{Heavy-hadron fragmentation at a glance}
\label{ssec:HF_fragmentation}

The striking difference between the fragmentation production of light-flavored hadrons and heavy-flavored ones is directly connected to the fact that masses of heavy quarks (charm or/and bottom) entering the lowest Fock state of the latters are above the perturbative-QCD threshold.
Therefore, while light-hadrons' FFs are of a genuine nonperturbative nature, the initial-scale inputs of heavy-hadrons' FFs are thought to embody some perturbative components.

As for heavy-light hadrons, like $D$ mesons, $B$ mesons or $\Lambda_{c,b}$ baryons, 
one can imagine the initial-energy-scale fragmentation as a two-step process (see, \emph{e.g.}, Refs.~\cite{Cacciari:1996wr,Cacciari:1993mq,Kniehl:2005mk}). 
First, a parton $i$, produced in the hard scattering at large transverse momentum $|\vec q_T| \gg m_Q$, fragments into the constituent heavy quark $Q$ with mass $m_Q$: charm or anticharm for $D$ and $\Lambda_c$, bottom or antibottom for $B$ and $\Lambda_b$.
Since $\alpha_s(m_Q) < 1$, this step can be calculated within perturbative QCD at an initial scale of ${\cal O}(m_Q)$.
Being its time scale shorter than the hadronization, one usually refers to this part as the short-distance coefficient (SDC) of the $(i \to Q)$ fragmentation.
The first calculation of SDCs for singly heavy-flavored hadrons and the application to lepton-collider phenomenology was carried out by B.~Mele and P.~Nason~\cite{Mele:1990cw}.

Then, at larger times, the constituent heavy quark $Q$ hadronizes into the detected bound state.
This represents the fully nonperturbative component of the fragmentation.
Here, to extract the nonperturbative dynamics of hadronization at a fixed scale, one has to rely upon phenomenological or effective-theory inputs (for an incomplete list of the most popular input models, see Refs.~\cite{Kartvelishvili:1977pi,Bowler:1981sb,Peterson:1982ak,Andersson:1983jt,Collins:1984ms,Colangelo:1992kh}).
We also mention the nonrelativistic fragmentation approach of  Ref.~\cite{Braaten:1994bz}, based on the more general
heavy-quark effective theory (HQET)~\cite{Georgi:1990um,Eichten:1989zv,Isgur:1989vq,Shifman:1987rj,Grinstein:1992ss,Neubert:1993mb,Jaffe:1993ie}.
The HQET framework builds on flavor and spin symmetries due to the $m_Q \gg \LQCD$ hierarchy.
In the $(m_Q \to \infty)$ limit those symmetries are exact and radiative corrections can be organized into an expansion of powers of $\LQCD/m_Q$.

The last requirement to build a complete heavy-light hadron VFNS FF set is considering the energy evolution.
Starting from the initial-scale nonperturbative inputs described above and assuming that the latters contain no scaling-violation effects, one generally employs numeric strategies to solve the coupled DGLAP evolution equations at a given perturbative accuracy and obtain $\mu_F$-dependent FFs.

Let us now turn our attention to quarkonia, namely mesons whose leading Fock state is $|Q{\bar Q}\rangle$.\footnote{Some Authors consider also charmed $B$ mesons as (generalized) quarkonium states. 
Their lowest Fock level is $|c \bar b\rangle$ or $|\bar c b\rangle$.}
Here, the simultaneous presence of two heavy quarks makes the description of the quarkonium formation more intricate than the heavy-light hadron one.
Historically, the first mechanism proposed was the color-evaporation model (CEM)~\cite{Fritzsch:1977ay,Halzen:1977rs}, which assumes a complete color decorrelation between the $(Q \bar Q)$ produced in the hard scattering and the final-state quarkonium.
Main limitations of CEM come from the inability to provide information about polarization as well as to properly predict production rates of distinct quarkonia, as the $\Jpsi$ over $\chi_c$ one in photo- and hadroproduction~\cite{Beneke:1996yw,Beneke:1998re,Lansberg:2005aw,Lansberg:2006dh,Brambilla:2010cs}.

A second way to depict quarkonium formation builds on the color-singlet mechanism (CSM) \cite{Berger:1980ni,Baier:1981uk}.
In this framework, color and spin remain unaltered during the hadronization, and because the quarkonium must be color-neutral, the $(Q \bar Q)$ pair needs to be generated in a color-singlet state. Furthermore, since the quarkonium mass is slightly larger than $2 m_Q$, 
one can make a \emph{static approximation} by assuming the two constituent quarks are at rest in the meson frame. 
For $S$-wave states, the only nonperturbative contribution is a nonrelativistic Schr\"odinger wave function at the origin, $\Psi_{\cal Q}(0)$.
Since $\Psi_{\cal Q}(0) = 0$ for $P$-wave states, in that case one takes as input its first derivative, $\Psi_{\cal Q}^\prime(0)$.
The value of $\Psi_{\cal Q}(0)$ or $\Psi_{\cal Q}^\prime(0)$ is generally determined from measurements of quarkonium leptonic-decay widths.
The rise of infrared divergences at NLO in $P$-wave decay channels~\cite{Barbieri:1976fp,Bodwin:1992ye} represents, however, the deepest theoretical evidence for the incompleteness of the CSM.

In Ref.~\cite{Bodwin:1992ye} it was proven that those divergences are exactly canceled by a matching singularity embodied in radiative corrections to $P$-wave color-octet matrix elements.
Thus, it became clear that the inclusion of color-octet contributions is crucial to provide a consistent description of quarkonium formation.
This laid the groundwork for the development of an effective field theory, known as NRQCD, where both color-singlet and color-octet contribution are consistently accounted for~\cite{Caswell:1985ui,Thacker:1990bm,Bodwin:1994jh,Cho:1995vh,Cho:1995ce,Leibovich:1996pa,Bodwin:2005hm} (see Refs.~\cite{Grinstein:1998xb,Kramer:2001hh,QuarkoniumWorkingGroup:2004kpm,Pineda:2011dg} for pedagogical reviews).
The key idea of NRQCD is describing the physical quarkonium as a linear combination of all possible Fock states.
They are organized as a double expansion in the QCD running coupling $(\alpha_s)$ and in the relative velocity $v_{\cal Q}$ between $Q$ and $\bar Q$.

NRQCD also provides a systematic way to separate the short-distance dynamics and long-distance one encoded in the production mechanism.
Indeed, treating heavy-quark and antiquark field in the effective Lagrangian as nonrelativistic fields permits us to consistently formulate an established factorization between SDCs describing the perturbative production of the $(Q \bar Q)$ system and long-distance matrix elements (LDMEs) portraying the hadronization phase.
LDMEs are of a nonperturbative nature and need to be extracted from experimental data (generally from corresponding leptonic widths), calculated \emph{via} potential-model studies (see Ref.~\cite{Eichten:1994gt} and references therein), or gathered from lattice simulations~\cite{Lepage:1992tx,Davies:1994mp}.

NRQCD was specifically designed to operate under the assumption that the quarkonium primary production mechanism involves the \emph{short-distance} generation of a $(Q \bar Q)$ pair through hard scattering, which subsequently undergoes hadronization to form the physical state. 
The $Q$ and $\bar Q$ quarks are produced with a relative transverse separation of approximately $1/E$, with $E$ the characteristic energy scale of the process~\cite{Mangano:1995yd}.
Given that, in the large transverse-momentum regime one has $E \sim |\vec q_T|$, the short-distance mechanism is expected to rapidly fall off with $|\vec q_T|$.
This happens because at high $|\vec q_T|$ the time available to the particle pair to organize itself in the correct color state is smaller, roughly $1/|\vec q_T|$, or, equivalently, the available volume is quite confined, roughly $1/|\vec q_T|^3$, thus diminishing the probability amplitude to create the quarkonium state~\cite{Mangano:1995yd,Braaten:1996pv,Artoisenet:2009zwa}.

It is here that the \emph{fragmentation} mechanism comes into play.
Indeed, when the transverse momentum grows, a single parton produced in the hard scattering has enough energy to fragment into the observed quarkonium plus an inclusive hadronic radiation.
While the fragmentation mechanism typically involves higher perturbative orders compared to the short-distance one, it is enhanced by a $(|\vec q_T|/m_Q)^2$ power. 
Consequently, it becomes dominant at high energies~\cite{Braaten:1993rw,Kuhn:1981jy,Kuhn:1981jn,Cacciari:1994dr,Braaten:1994xb}.
By suitably adapting NRQCD studies to fragmentation, a leading-order (LO) calculation of the gluon and charm to $S$-wave color-singlet charmonium FF were performed in Refs.~\cite{Braaten:1993rw} and~\cite{Braaten:1993mp}, respectively.
The $P$-wave case soon followed~\cite{Braaten:1994kd,Ma:1995ci,Yuan:1994hn}.

Since a proper treatment of the fragmentation process must be grounded on \emph{collinear factorization}, it is necessary to establish a formally consistent link between nonrelativistic studies and a fragmentation-correlator viewpoint.
A modern application of heavy-flavor theory to quarkonium fragmentation builds upon recognizing NRQCD as a valuable and powerful framework for modeling FF initial-scale inputs~\cite{Kang:2011mg,Ma:2013yla,Ma:2014eja}.

We see two main advantages here.
First, NRQCD permits the factorization of these inputs as a convolution between perturbative SDCs and nonperturbative LDMEs in a manner analogous to heavy-flavored hadrons~\cite{Cacciari:1996wr,Cacciari:1993mq,Kniehl:2005mk}.
Furthermore, it provides a straightforward way for efficiently calculating SDCs and offers an intuitive physical interpretation of LDMEs.
It should be noted, however, that while NRQCD inputs for the short-distance case stand as a key ingredient for a production mechanism which, by its nature, starts a twist-3,\footnote{Since the hadronization in the short-distance production occurs after the emission of two heavy quarks from the hard scattering, that mechanism can also be interpreted as a low-$|\vec q_T|$, two-parton fragmentation, which goes clearly beyond the leading-twist approximation in the fragmentation correlator.} at large transverse momentum the corresponding SDCs act as \emph{power corrections} for the single-parton fragmentation correlator, which remains a genuine leading-twist quantity.

Finally, similarly to the heavy-light hadron case, one builds quarkonium VFNS FFs by switching standard DGLAP evolution on.
The {\tt ZCW19$^+$}~\cite{Celiberto:2022dyf,Celiberto:2023fzz} and {\tt ZCFW22}~\cite{Celiberto:2022keu,Celiberto:2024omj} sets represent a first determination of VFNS FFs for vector quarkonia and charmed $B$ mesons on top of gluon and heavy-quark initial-scale inputs from higher-order NRQCD~\cite{Braaten:1993rw,Chang:1992bb,Braaten:1993jn,Ma:1994zt,Zheng:2019gnb,Zheng:2021sdo,Feng:2021qjm,Feng:2018ulg}.

As recently pointed out, NRQCD factorization can also be employed to explore the true nature of di-$\Jpsi$ resonances~\cite{LHCb:2020bwg,ATLAS:2023bft,CMS:2023owd}, interpreting them as fully charmed-tetraquark states~\cite{Zhang:2020hoh,Zhu:2020xni}.
Indeed, to produce a $\TQc$ state, two charms and two anticharms need first be emitted at a short distance, approximately $\sim 1/m_c$. 
Then, asymptotic freedom permits to look at the fragmentation process as a two-step convolution between a short-distance phase and long-distance one.

\begin{figure*}[!t]
\centering
\includegraphics[width=0.425\textwidth]{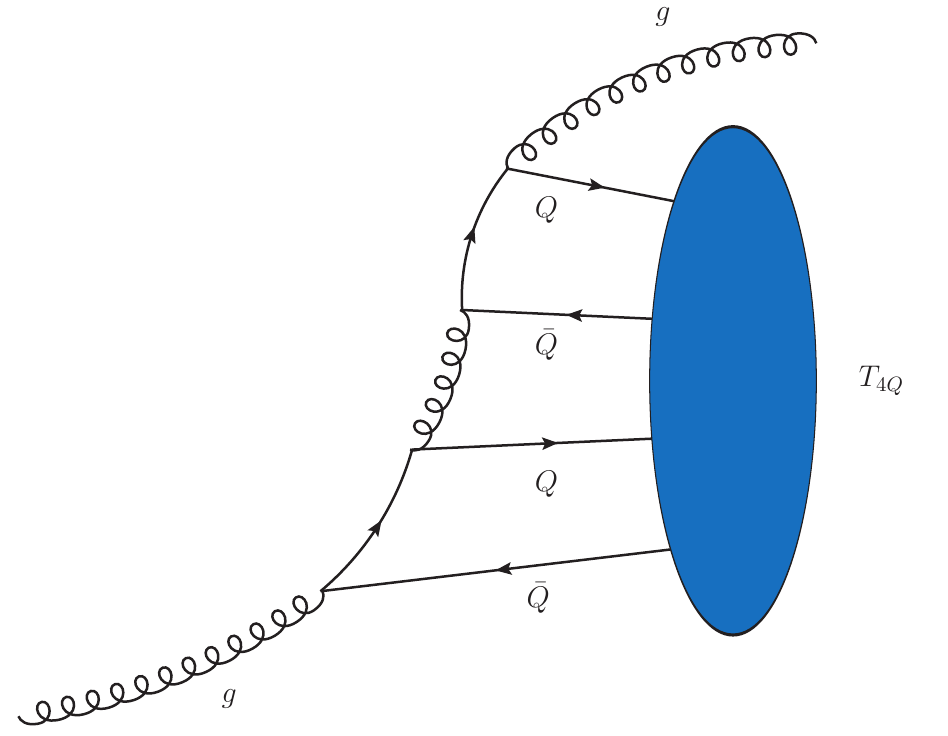}
\hspace{0.50cm}
\includegraphics[width=0.425\textwidth]{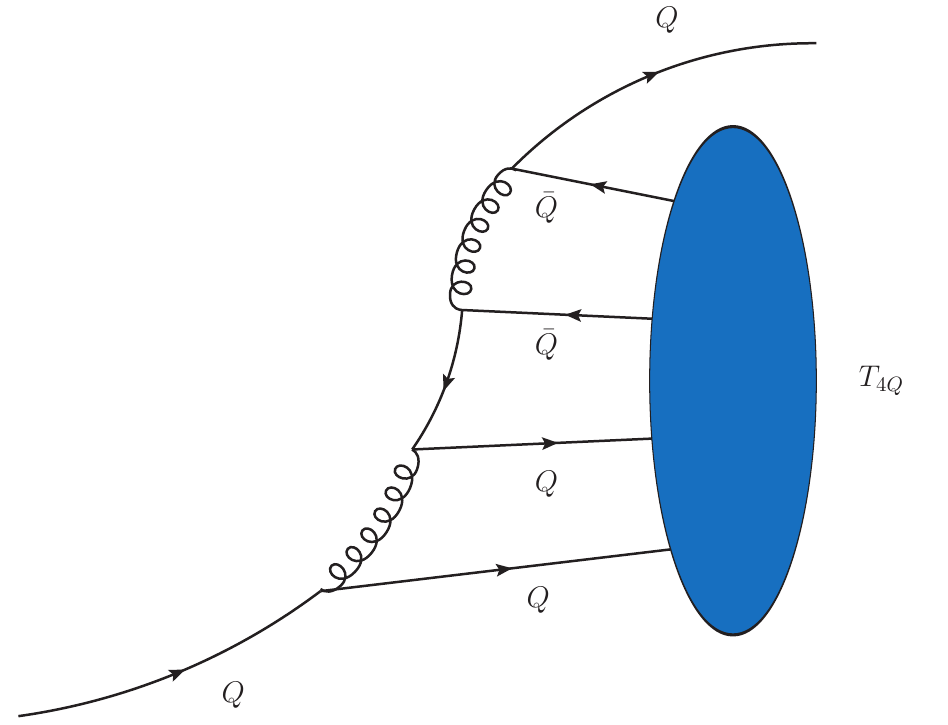}

\caption{LO representative diagrams for gluon (left) and constituent heavy-quark (right) to $\TQQ$ fragmentation channel in the color-singlet case.
Following a pure NRQCD vision, parton interactions in the left-hand side of each diagrams represent the short-distance coefficients (SDCs), while blue orange blobs are for the nonperturbative long-distance matrix elements (LDMEs).}
\label{fig:FF_diagram}
\end{figure*}

Working within this framework, a first calculation of the NRQCD initial-scale input for the $[g \to \TQc]$ color-singlet $S$-wave fragmentation channel (Fig.~\ref{fig:FF_diagram}, left diagram) was obtained in Ref.~\cite{Feng:2020riv} and then compared with the short-distance production at LHC kinematic ranges~\cite{Feng:2023agq}.
Conversely, the initial-scale input for the $[c \to \TQc]$ channel (Fig.~\ref{fig:FF_diagram}, right diagram) can be modeled by suitably adapting the Suzuki-model-like calculation~\cite{Suzuki:1977km,Suzuki:1985up,Amiri:1986zv} recently employed to address the fragmentation of heavy-light $\XQq$ states~\cite{Nejad:2021mmp}.
Starting from the latter, a first determination of VFNS FFs for those heavy-light tetraquarks, named {\tt TQHL1.0} functions~\cite{Celiberto:2023rzw}, was derived.

In the next subsections we will briefly describe initial-scale inputs for $\TQc$ fragmentation channels of Fig.~\ref{fig:FF_diagram}.
Then we will illustrate the strategy adopted to implement the DGLAP evolution with a proper choice of heavy-flavor thresholds.

\subsection{Gluon fragmentation channel}
\label{ssec:FFs-g}

As for the initial-scale input of the gluon fragmentation to a color-singlet $S$-wave $\TQc$ (Fig.~\ref{fig:FF_diagram}, left panel), we consider a recent calculation based on NRQCD factorization~\cite{Feng:2020riv}.
We consider fully charmed tetraquarks with total angular momentum, parity, and charge $J^{PC} = 0^{++}$ or~$2^{++}$.
Accounting for just the lowest-order contributions in the constituent-quarks' relative velocity, $v_{\cal Q}$, we write
\begin{equation}
 \label{Dg_FF_initial-scale}
 D^{\TQc(J^{PC})}_g(z,\mu_{F,0}) \, = \,
 \frac{1}{m_c^9}
 \sum_{[n]}
 {\cal D}^{(J^{PC})}_g(z,[n]) \, \langle {\cal O}^{\TQc(J^{PC})}([n]) \rangle
 \;,
\end{equation}
with $m_c = 1.5$~GeV being the charm-quark mass, ${\cal D}^{(J^{PC})}_g(z,[n])$ the SDCs for the $[g \to (cc\bar{c}\bar{c})]$ perturbative transition and $\langle {\cal O}^{\TQc(J^{PC})}([n]) \rangle$ the \emph{color-composite} LDMEs depicting the $\TQc(J^{PC})$ nonperturbative hadronization.
The composite quantum number $[n]$ runs over the combinations $[3,3]$,  $[6,6]$, $[3,6]$ and $[6,3]$.
Here we made use of the color diquark-antidiquark basis to decompose a color-singlet tetraquark state either as a $\bar{3} \otimes 3$ or a $6 \otimes \bar{6}$ configuration.
According to the Fermi--Dirac statistics, and taking in the $S$-wave state both the diquark and the antidiquark systems as well as the diquark-antidiquark cluster, the $\bar{3} \otimes 3$ system can have spin 0, 1 and 2, while the 
$6 \otimes \bar{6}$ one can only have spin 0.
We also note that ${\cal D}^{(J^{PC})}_g(z,[3,6]) \equiv {\cal D}^{(J^{PC})}_g(z,[6,3])$ and that $\langle {\cal O}^{\TQc(J^{PC})}([3,6]) \rangle \equiv \langle {\cal O}^{\TQc(J^{PC})}([6,3]) \rangle^*$. 

The SDCs describing the perturbative part of the $[g \to \TQcZpp]$ initial-scale FF read
\begin{equation}
\begin{split}
 \label{Dg_FF_SDC_0pp_33}
\hspace{-1.00cm}
 {\cal D}^{(0^{++})}_g&(z,[3,3]) \,=\, 
 \frac{\pi^{2} \alpha_{s}^{4}(4m_c)}{497664 z(2-z)^{2}(3-z)}\left[186624-430272 z+511072 z^2-425814 z^3\right. \\
 & +\, 217337 z^4-61915 z^5+7466 z^6+42(1-z)(2-z)(3-z)(-144+634 z\\
 & \left.-\, 385 z^2+70 z^3\right) \ln (1-z)+36(2-z)(3-z)\left(144-634 z+749 z^2-364 z^3\right. \\
 & \left.+\, 74 z^4\right) \ln \left(1-\frac{z}{2}\right)+12(2-z)(3-z)\left(72-362 z+361 z^2-136 z^3+23 z^4\right) \\
 & \left.\times\, \ln \left(1-\frac{z}{3}\right)\right]
 \;,
\end{split}
\end{equation}
\\[-0.35cm]
\begin{equation}
\begin{split}
 \label{Dg_FF_SDC_0pp_66}
\hspace{-1.00cm}
 {\cal D}^{(0^{++})}_g&(z,[6,6]) \,=\,  
 \frac{\pi^{2} \alpha_{s}^{4}(4m_c)}{331776 z(2-z)^{2}(3-z)}\left[186624-430272 z+617824 z^2-634902 z^3\right. \\
 & +\, 374489 z^4-115387 z^5+14378 z^6-6(1-z)(2-z)(3-z)(-144-2166 z\\
 & \left.+\, 1015 z^2+70 z^3\right) \ln (1-z)-156(2-z)(3-z)\left(144-1242 z+1693 z^2-876 z^3\right. \\
 & \left.+\, 170 z^4\right) \ln \left(1-\frac{z}{2}\right)+300(2-z)(3-z)\left(72-714 z+953 z^2-472 z^3+87 
 z^4\right) \\
 & \left.\times\, \ln \left(1-\frac{z}{3}\right)\right]
 \;,
\end{split}
\end{equation}
\\[-0.35cm]
\begin{equation}
\begin{split}
 \label{Dg_FF_SDC_0pp_36}
\hspace{-0.00cm}
 {\cal D}^{(0^{++})}_g&(z,[3,6]) \,=\,  
 \frac{\pi^{2} \alpha_{s}^{4}(4m_c)}{165888 z(2-z)^{2}(3-z)}\left[186624-430272 z+490720 z^2-394422 z^3\right. \\
 & +\, 199529 z^4-57547 z^5+7082 z^6+6(1-z)(2-z)(3-z)(-432+3302 z \\
 & \left.-\, 1855 z^2+210 z^3\right) \ln (1-z)-12(2-z)(3-z)\left(720-2258 z+2329 z^2-1052 z^3\right. \\
 & \left.+\, 226 z^4\right) \ln \left(1-\frac{z}{2}\right)+12(2-z)(3-z)\left(936-4882 z+4989 z^2-1936 z^3+331 z^4\right) \\
 & \left.\times\, \ln \left(1-\frac{z}{3}\right)\right]
 \;.
\end{split}
\end{equation}

On the other side, two of the three SDCs describing the perturbative part of the $[g \to \TQcTpp]$ initial-scale FF vanish.
They are the $[6,6]$ term and the $[3,6]$ interference one.
This directly follows from the incompatibility of  the NRQCD operator underlying the $6 \otimes \bar{6}$ configuration with the Fermi--Dirac statistics for a
diquark-antidiquark system in a $2^{++}$ state.
Thus, the only SDC surviving for the $[g \to \TQcTpp]$ channel is
\begin{equation}
\begin{split}
 \label{Dg_FF_SDC_2pp_33}
\hspace{-0.20cm}
 {\cal D}^{(2^{++})}_g&(z,[3,3]) \,=\, 
 \frac{\pi^{2} \alpha_{s}^{4}(4m_c)}{622080 z^2(2-z)^{2}(3-z)}\left[\left(46656-490536 z+1162552 z^2-1156308 z^3\right.\right. \\
 & \left.+\, 595421 z^4-170578 z^5+21212 z^6\right) 2z+3(1-z)(2-z)(3-z)(-20304-31788 z) \\
 & \left.\left.\times\, (1296+1044 z + 73036 z^2-36574 z^3+7975 z^4\right)\right. \\
 & \left.\times\, \ln (1-z)+33(2-z)(3-z)(1296+25)\right] \\
 & \left.\left.\, -9224 z^2+9598 z^3-3943 z^4+725 z^5\right) \ln \left(1-\frac{z}{3}\right)\right]
  \;,
\end{split}
\end{equation}
while ${\cal D}^{(2^{++})}_g(z,[6,6]) = {\cal D}^{(2^{++})}_g(z,[3,6]) \equiv 0$\,.

The remaining ingredients are the color-composite LDMEs, $\langle {\cal O}^{\TQc(J^{PC})}([n]) \rangle$.
As already mentioned, they represent the genuine nonperturbative contribution to the initial-scale FF.
Given that no data are available and lattice simulations for tetraquarks are still at a very early stage, to get a proxy for these matrix elements we need to rely upon potential-model studies.

A suitable strategy consists in computing the radial wave functions at the origin \emph{via} potential models and then relating them to LDMEs \emph{via} the vacuum saturation approximation~\cite{Feng:2020qee}.
In Ref.~\cite{Feng:2020riv} three models were proposed~\cite{Zhao:2020nwy,Lu:2020cns,liu:2020eha}, all of them assuming a Cornell-like potential~\cite{Eichten:1974af,Eichten:1978tg} and accounting for some spin-dependent features.
The first model~\cite{Zhao:2020nwy} and the third one~\cite{liu:2020eha} are based on nonrelativistic quark fields, whereas the second one~\cite{Lu:2020cns} embodies relativistic corrections.

As discussed in Ref.~\cite{Feng:2020riv}, the first model appears to significantly overestimate the cross section when compared with data for $\Jpsi$ production at 13~TeV~CMS~\cite{CMS:2017dju} which, on the other hand, are expected to stay fairly above the $\TQc$ rate.
Then, numeric checks not shown in the present study have highlighted that the resulting FFs built on the basis of LDMEs from the third model are heavily unstable under even a minimal variation of their value, say of the order of 0.1\%.
Therefore, to build the gluon fragmentation channel of our {\tt TQ4Q1.0} functions we will rely upon the second proposed model~\cite{Lu:2020cns}, whose predicted LDMEs read (see Table~I of the published version of Ref.~\cite{Feng:2020riv} for a comparison with values obtained from the other two models) 
\begin{equation}
\begin{split}
\label{LDMEs}
 {\cal O}^{\TQcZpp}([3,3]) &\,=\, 0.0347\mbox{ GeV}^9 \;, \qquad\quad
 {\cal O}^{\TQcTpp}([3,3]) \,=\, 0.072\mbox{ GeV}^9 \;,
 \\
 {\cal O}^{\TQcZpp}([6,6]) &\,=\, 0.0128\mbox{ GeV}^9 \;, \qquad\quad
 {\cal O}^{\TQcTpp}([6,6]) \,=\, 0 \;,
 \\
 {\cal O}^{\TQcZpp}([3,6]) &\,=\, 0.0211\mbox{ GeV}^9 \;, \qquad\quad
 {\cal O}^{\TQcTpp}([3,6]) \,=\, 0 \;.
\end{split}
\end{equation}

To benchmark the numeric implementation of our $[g \to \TQc]$ FFs and compare them with the $\TQcZpp$ ($\TQcTpp$) ones presented in the left (right) panel of Fig.~2 of Ref.~\cite{Feng:2020riv}, we follow the strategy adopted in Section~V of that work.
We evolve the initial-scale input of Eq.~\eqref{Dg_FF_initial-scale} \emph{via} DGLAP and by considering the gluon-to-gluon splitting only
\begin{equation}
\label{Dg_FF_DGLAP_evo_eq}
\frac{\partial}{\partial \ln \mu}
D^{\TQc(J^{PC})}_g(z,\mu)
= \int_{z}^{1} \frac{\drv \xi}{\xi} 
\, \alpha_s \, P_{gg} \left( \frac{z}{\xi} \right) D^{\TQc(J^{PC})}_g(\xi,\mu) 
\;,
\end{equation}
where
\begin{equation}
\label{Pgg_LO}
P_{gg}(y)=\frac{2 C_A}{\pi}\left[\frac{y}{(1-y)_+}+\frac{1-y}{y}+y(1-y)\right]
+\left( \frac{11 N_c - 2 n_f}{6 \pi} \right) \delta(1-y)
\end{equation}
is the LO Altarelli-Parisi gluon-gluon time-like kernel, with $C_A \equiv N_c$ the Casimir factor connected to a gluon emission from a gluon, $N_c$ the number of colors, and $n_f$ the number of active light flavors.
Here we introduced the well-known \emph{plus-prescription}, whose action on a generic function regular behaved at $y=1$ reads
\begin{equation}
\label{plus-prescription}
\int^1_z \drv y \, \frac{f(y)}{(1-y)_+}
=\int^1_z \drv y \, \frac{f(y)-f(1)}{1-y}
-\int^z_0 \drv y \, \frac{f(1)}{1-y}
\;.
\end{equation}
Discarding all quark splittings permits us to analytically solve Eq.~\eqref{Dg_FF_DGLAP_evo_eq}, thus having
\begin{equation}
\begin{split}
\label{Dg_FF_DGLAP_evo_sol}
&\hspace{-0.00cm}
 D^{\TQc(J^{PC})}_g(z,\mu_F) \,=\, 
 D^{\TQc(J^{PC})}_g(z,\mu_{F,0}) +
 \alpha_s \ln \frac{\mu_F}{\mu_{F,0}}
 \left\{
 \int_z^1 \frac{\drv y}{y}
 \left[
 \tilde{P}_{gg}(y) \,
 D^{\TQc(J^{PC})}_g\left(\frac{z}{y},\mu_{F,0}\right)
 \right.\right.
 \\[0.20cm]
 &\left.\left. \!\! -\, 
 \frac{C_A}{\pi} \,\frac{y}{1-y} \,
 D^{\TQc(J^{PC})}_g(z,\mu_{F,0})
 \right]
 + \frac{1}{\pi} \left[
 \frac{11 N_c}{12} - \frac{n_f}{6} + C_A \ln (1-z)
 \right]
 D^{\TQc(J^{PC})}_g(z,\mu_{F,0})
 \right\}
 \;,
\end{split}
\end{equation}
where the actions of the plus-prescription and the delta function of Eq.~\eqref{Dg_FF_DGLAP_evo_eq} have been made explicit, and
\begin{equation}
\label{Pgg_LO_tilde}
\tilde{P}_{gg}(y)=\frac{C_A}{\pi}\left[\frac{y}{1-y}+\frac{1-y}{y}+y(1-y)\right] \;.
\end{equation}

We could not reproduce, however, results of Fig.~2 of Ref.~\cite{Feng:2020riv}.
We note that the very last term of the r.h.s. of Eq.~\eqref{Dg_FF_DGLAP_evo_sol}, coming from the action of the plus-prescription, is proportional to
\begin{equation}
 \label{Dg_FF_DGLAP_evo_sol_div}
 \ln \frac{\mu_F}{\mu_{F,0}}
 \ln  (1-z)
 D^{\TQc(J^{PC})}_g(z,\mu_{F,0})
 \;.
\end{equation}
For initial-scale gluon FFs, $D^{\TQc(J^{PC})}_g(z,\mu_{F,0})$, going to a finite (nonzero) value when $z \to 1^-$, as it happens for our functions, the term in Eq.~\eqref{Dg_FF_DGLAP_evo_sol_div} negatively diverges for $\mu_F > \mu_{F,0}$ and positively diverges in the opposite case.\footnote{The fact that a given FF does not go to zero when $z \to 1$ is rather surprising and one may argue that this is not compatible with the collinear factorization.
Indeed, at leading twist, only one parton fragments into the observed hadron.
Thus, the probability that this parton transfers its entire momentum (as $z$ reaches one) to the hadron should go to zero.
On the other side, dealing with nonzero FFs at the $z$ endpoint is not uncommon in the context of NRQCD.
As an example, the color-singlet $(g \to \eta_{c,b})$ FFs at LO grow with $z$ up to reach their maximum when $z \to 1$~\cite{Braaten:1993rw}.
The same functions negatively diverge at NLO.
Some Authors argue that this does not pose a problem, as the collinear convolution between the divergent FF and the rest of the cross section is well behaved~\cite{Artoisenet:2014lpa}.
Other Authors interpret the singularity as a sign of perturbative instability and suggest resummation(s) as a possible solution~\cite{Zhang:2018mlo}.
We believe that further studies are needed to restore the correct behavior of NRQCD-based FFs at the $z$ endpoint.
This might require an extension or generalization of the NRQCD factorization itself when applied to fragmentation, which clearly goes beyond the scope of our exploratory analysis, but certainly deserves attention in the future.
}

Conversely, functions plotted in Fig.~2 of Ref.~\cite{Feng:2020riv} exhibit regular behavior as $z$ approaches one in the range of $4 m_c < \mu_F < 40\mbox{ GeV}$, as mentioned there.
Given that we exactly reproduced plots of that figure by artificially setting to zero the $\ln (1-z)$ factor of Eq.~\eqref{Dg_FF_DGLAP_evo_sol_div} (for the sake of brevity, we do not show the result of this test here), we suspect that some sort of numeric instability connected to the aforementioned logarithm might have affected the DGLAP-evolution study of Ref.~\cite{Feng:2020riv}.

\begin{figure*}[!t]
\centering

\includegraphics[scale=0.46,clip]{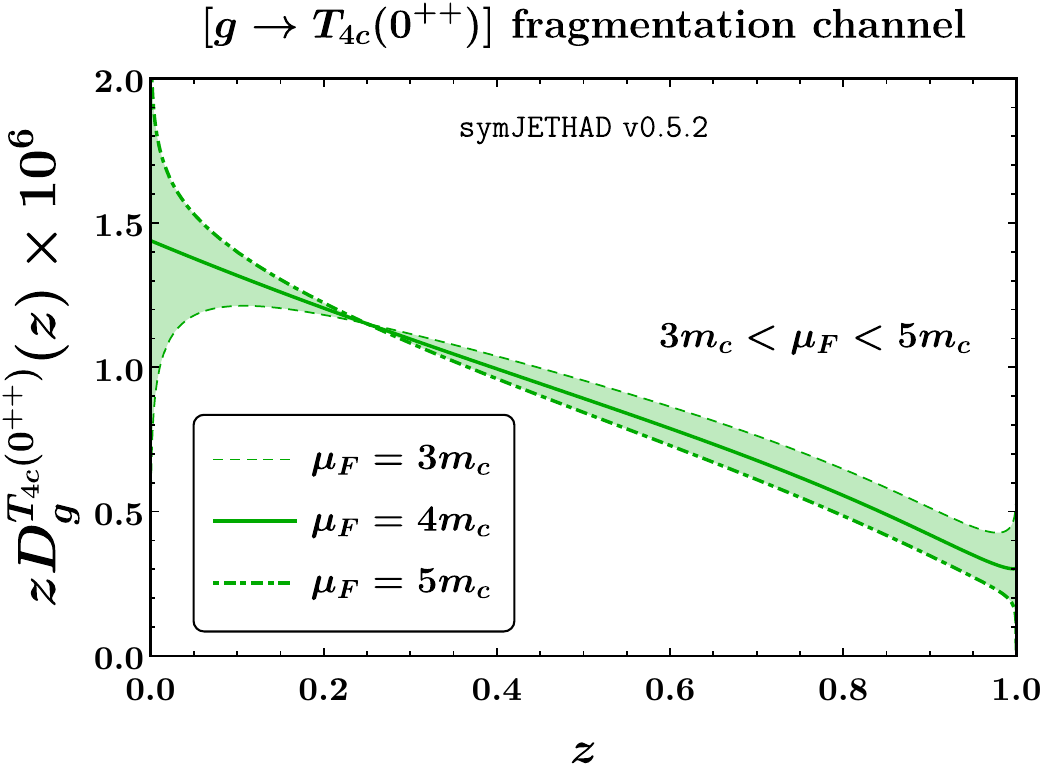}
\hspace{0.50cm}
\includegraphics[scale=0.46,clip]{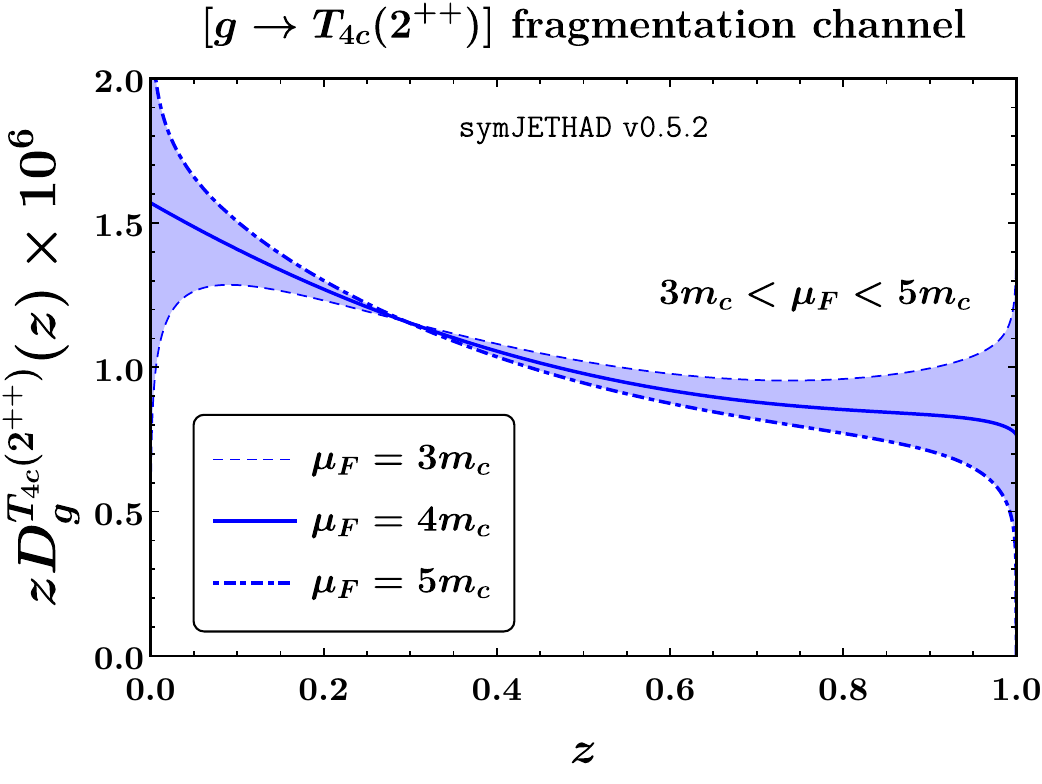}

\caption{Initial-scale inputs for the gluon fragmentation channel to the $\TQcZpp$ state (left) and the $\TQcTpp$ radial excitation (right).
Starting from $\mu_F=4m_c$, the function is evolved within an expanded and decoupled approach in the range $3m_c$ to $5m_c \equiv Q_0$. The latter is the \emph{evolution-ready} scale $Q_0$, namely the threshold to activate the charm-quark fragmentation channel.}
\label{fig:FF_initial-scale_gluon}
\end{figure*}

For clarity, in Fig.~\ref{fig:FF_initial-scale_gluon} we show the $z$-dependence of our $D^{\TQc(0^{++})}_g(z,\mu_{F})$ (left panel) and $D^{\TQc(2^{++})}_g(z,\mu_{F})$ (right panel) functions with uncertainty bands built by setting $\mu_{F,0} = 4 m_c$, and then letting $\mu_F$ run from $3 m_c$ to $5 m_c$, with the DGLAP evolution controlled by the $P_{gg}$ LO splitting kernel only (see Eq.~\eqref{Dg_FF_DGLAP_evo_sol}).
As discussed later (see Section~\ref{ssec:FFs-TQ4Q10}), $\mu_{F,0} = 4 m_c$ will represent the initial-scale for the gluon fragmentation channel, while we will identify $5 m_c$ with the value of the \emph{evolution-ready} scale, $Q_0$, at which the all-order DGLAP evolution for all parton channels will be numerically switched on.

\subsection{Constituent heavy-quark fragmentation channel}
\label{ssec:FFs-Q}

Our proxy for the initial-scale input of the charm-quark fragmentation to a color-singlet $S$-wave $\TQc$ is a model calculation, originally proposed in Ref.~\cite{Nejad:2021mmp} for the study of heavy-light tetraquark, and well-adapted here to fully charmed states.
It relies upon a spin-physics inspired Suzuki approach~\cite{Suzuki:1977km,Suzuki:1985up,Amiri:1986zv}, which accounts for transverse-momentum dependence. 
The collinear limit is obtained by discarding the relative motion of the constituent quarks inside the produced hadron~\cite{Lepage:1980fj,Brodsky:1985cr,Amiri:1986zv}.

In some sense, this treatment follows an analogous factorization scheme as the NRQCD one, where the $(Q\bar{Q})$ pair is produced perturbatively and its subsequent hadronization is encoded in the corresponding LDMEs.
Here, in analogy with NRQCD, a $(cc\bar{c}\bar{c})$ system is first emitted from an outgoing (anti)charm \emph{via} above-threshold perturbative splittings (Fig.~\ref{fig:FF_diagram}, right diagram). 
Then, its production amplitude is convoluted with a bound-state wave function embodying the nonperturbative dynamics of tetraquark hadronization, according to Suzuki.

Assuming complete symmetry between charm and anticharm fragmentation channels, the expression of our initial-scale $[c \to \TQc]$ FF reads
\begin{equation}
 \label{Dc_FF_initial-scale}
 D^{\TQc}_c(z,\mu_{F,0}) \, \equiv \,
 D^{\TQc}_{\bar c}(z,\mu_{F,0}) \, = \,
 {\cal N}_{c} \;
 \frac{(1 - z)^5}{\Delta^2(z)} 
 \sum\limits_{k=0}^3 \gamma_k^{(c)}(z) \, z^{2(k+2)} \left\{ \frac{\vqTTa}{m_c^2} \right\}^{\!\!k}
 \;,
\end{equation}
where
\begin{equation}
 \label{Dc_FF_initial-scale_N}
 {\cal N}_{c} \, = \, \left[ 512 \, \pi^2 \, C_F  \, f_{\cal B} \, \left( \alpha_s(2 m_c) \right)^2 \right]^2
 \;,
\end{equation}
with $C_F \equiv (N_c^2-1)/(2N_c)$ the usual Casimir factor connected to a gluon emission from a quark, $f_{\cal B} = 0.25$~GeV the hadron decay constant~\cite{ParticleDataGroup:2020ssz}, $m_c = 1.5$~GeV the charm-quark mass, and
\begin{equation}
 \label{Dc_FF_initial-scale_Delta}
 \frac{1}{\Delta(z)} = 
 \frac{m_c^8}
 {\bigl[ (4 - 3 z)^2 m_c^2 + z^2 \vqTTa \bigr]^2
  \bigl[ (4 - z)^2 m_c^2 + z^2 \vqTTa \bigr]
  \bigl[ (1-z) M_{\TQc}^2 + z^2 (m_c^2 + \vqTTa) \bigr]}
 \;.
\end{equation}
The $\gamma_k^{(c)}(z)$ coefficients in Eq.~\eqref{Dc_FF_initial-scale} are given by
\begin{equation}
\begin{split}
 \label{Dc_FF_initial-scale_gamma_k}
 \gamma_0^{(c)}(z) &= (4 - z)^2 (256 - 512 z + 416 z^2 - 160 z^3 + 33 z^4) \;, \\
 \gamma_1^{(c)}(z) &= 768 - 1280 z + 1248 z^2 - 464 z^3 + 67 z^4 \;, \\
 \gamma_2^{(c)}(z) &= 48 - 40 z + 35 z^2 \;, \\
 \gamma_3^{(c)}(z) &= 1 \;.
\end{split}
\end{equation}

Formul{\ae} for our $\TQc$ fragmentation channel given above fairly match the ones corresponding to heavy-light states, $\XQq$, obtained in Ref.~\cite{Nejad:2021mmp} and then employed in recent phenomenological studies~\cite{Celiberto:2023rzw}.
A simple analytic change was made by adapting the calculation to the fully charmed case, namely setting the mass of all the constituent quarks to $m_c$.
Then, two important aspects have been addressed.

First, Authors of Ref.~\cite{Nejad:2021mmp} treated the ${\cal N}_{c}$ constant in Eq.~\eqref{Dc_FF_initial-scale_N} as a normalization factor, to be fixed by certain normalization conditions (see Refs.~\cite{Amiri:1986zv,Suzuki:1985up,Chang:1991bp,Chang:1992bb}).
This is needed in the $\XQq$ fragmentation case, since the QCD running coupling in the expression of ${\cal N}_{c}$ is called at $\mu_R = 2m_q$, namely at two times the mass of a constituent light quark (see text right below Eq.~(18) of Ref.~\cite{Nejad:2021mmp}).
In our case however, the presence of four constituent heavy quarks permits to properly calculate $\alpha_s(2m_c)$, and thus ${\cal N}_{c}$.

Second, the choice of the $\vqTTa$ parameter in Eq.~\eqref{Dc_FF_initial-scale} and~\eqref{Dc_FF_initial-scale_Delta} deserves a discussion.
As already mentioned, the original approach by Suzuki accounts for spin correlations.
\emph{De facto}, it comes out as a model for a TMD fragmentation function.
To get the collinear limit, instead of integrating over the squared modulus of the outgoing charm-quark transverse momentum, one can simply replace it by its average value, $\vqTTa$.
In this way, $\vqTTa$ becomes a free parameter to be fixed by reasonable criteria, suggested by phenomenology.
According to a discussion in Ref.~\cite{GomshiNobary:1994eq}, larger and larger values of $\vqTTa$ push the function peak down and down to the low-$z$ region.
Moreover, the function bulk decreases as $\vqTTa$ grows.
The heavy-light tetraquark FF of Ref.~\cite{Nejad:2021mmp} was obtained by setting $\vqTTa = 1\mbox{ GeV}^2$, which was considered as an extreme value for the average squared transverse momentum.

In our case, however, the value of $\vqTTa$ should be selected with more care.
Since no direct indication comes from $\TQc$ phenomenology, we opt for a reasonable choice that aligns with the spirit of our exploratory analysis.
Some recent studies on collinear-fragmentation production at hadron colliders have indicated that the average value of $z$ at which FFs are typically probed is generally confined between 0.4 and 0.6.
This holds as much for light-flavored species~\cite{Celiberto:2016hae,Celiberto:2017ptm,Bolognino:2018oth,Celiberto:2020wpk} as for heavy-flavored ones~\cite{Celiberto:2021dzy,Celiberto:2021fdp,Celiberto:2022dyf,Celiberto:2022keu}.
Therefore, a suitable choice for $\vqTTa$ should not lead to a value of $\langle z \rangle_{D_c}$ smaller than $0.4$.
Here, the ${D_c}$ subscript refers to the fact the we are explicitly considering the charm fragmentation channel.

On the other hand, it is reasonable to request that the initial-scale inputs for gluon and constituent-quark fragmentation channels possess roughly the same order of magnitude.
This is supported by an analogy between our $\TQc$ state and the simplest case in the quarkonium sector: a scalar, color-singlet $S$-wave charmonium state, \emph{i.e.} the $\eta_c$ meson.
The $\eta_c$ fragmentation production can be modeled within NRQCD, with the SDCs being calculated up to ${\cal O}(\alpha_s^3)$ for gluon~\cite{Braaten:1993rw,Artoisenet:2014lpa,Zhang:2018mlo}, charm~\cite{Braaten:1993mp,Zheng:2021ylc}, and nonconstituent quark~\cite{Zheng:2021mqr} channels.
In particular, in the $0.4 \lesssim z \lesssim 0.6$ range, both the LO gluon~\cite{Braaten:1993rw} and charm~\cite{Braaten:1993mp} functions have a magnitude of $\sim 10^{-4}$. 

\begin{figure*}[!t]
\centering

\includegraphics[scale=0.46,clip]{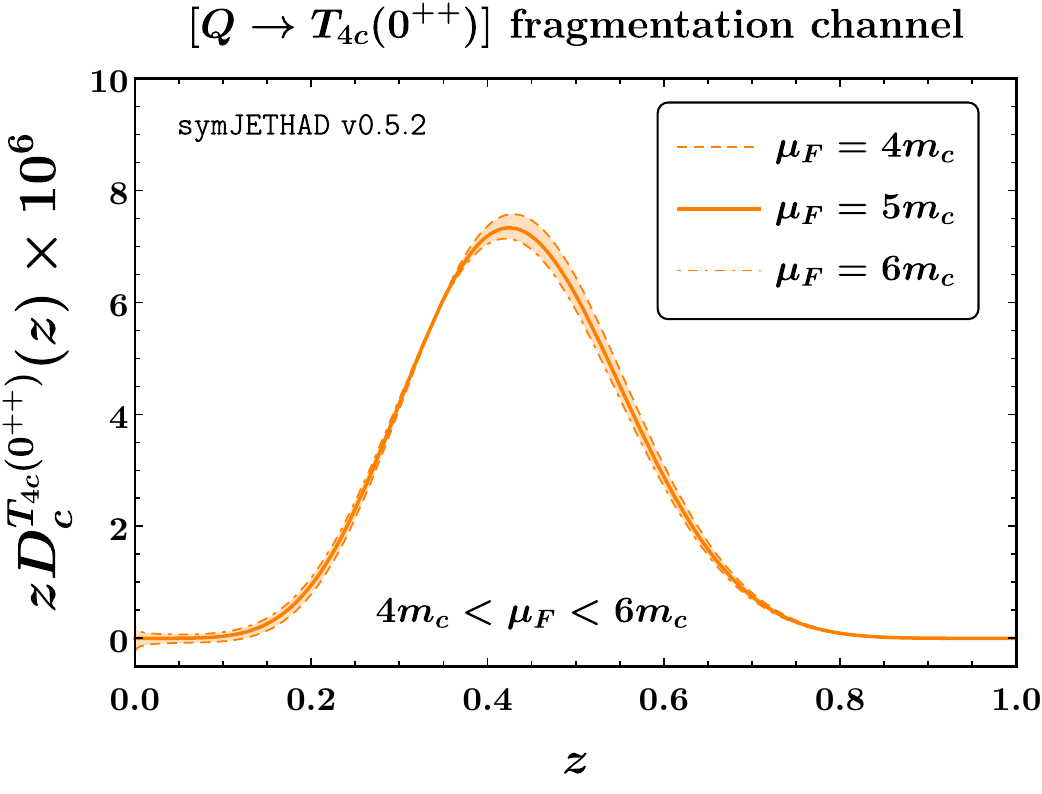}
\hspace{0.50cm}
\includegraphics[scale=0.46,clip]{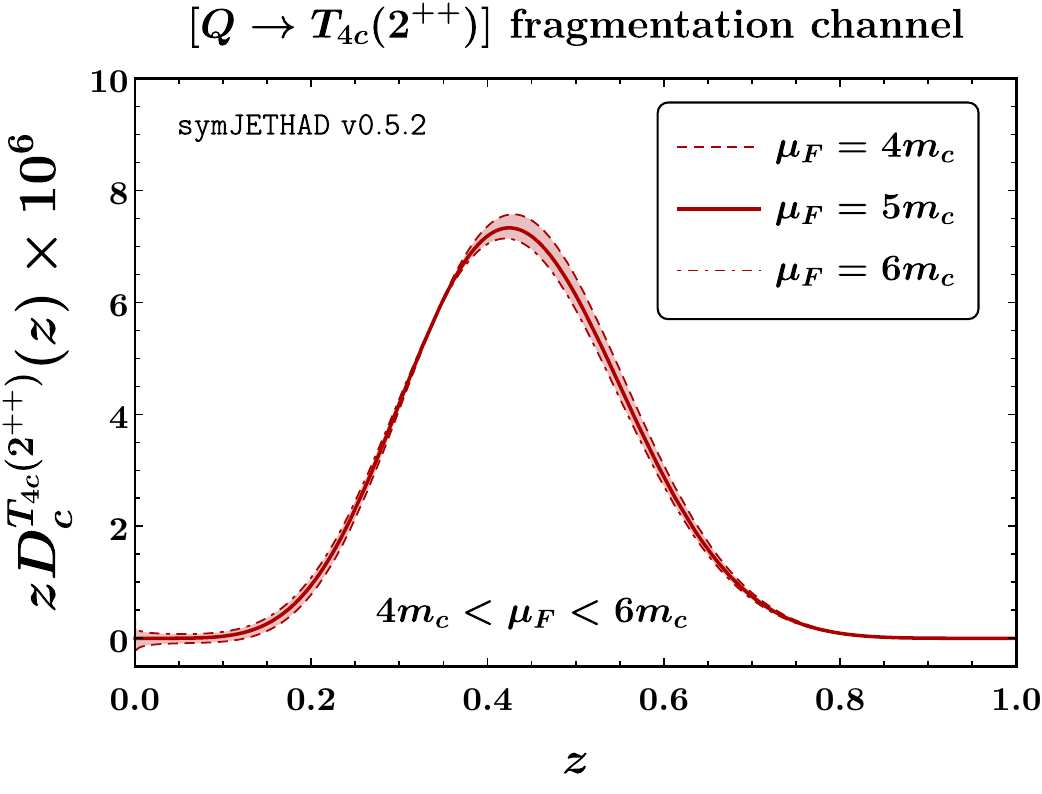}

\caption{Initial-scale inputs for the charm-quark fragmentation channel to the $\TQcZpp$ state (left) and the $\TQcTpp$ radial excitation (right).
For the sake of illustration, an expanded DGLAP evolution is performed around the \emph{evolution-ready} scale, $\mu_F = 5m_c \equiv Q_0$.}
\label{fig:FF_initial-scale_charm}
\end{figure*}

We performed a numeric scan of the $\vqTTa$ parameter range to find the best compromise between the two requirements.
We found that setting $\vqTTa = 70\mbox{ GeV}^2$ leads to $\langle z \rangle_{D_c} = 0.40173$.
It also makes the charm FF be roughly of the same order of the gluon one, as shown by comparing plots of Fig.~\ref{fig:FF_initial-scale_charm} with those of Fig.~\ref{fig:FF_initial-scale_gluon}.
Thus, we adopt $\vqTTa = 70\mbox{ GeV}^2$ as a given parameter of our model determination of the initial-scale charm-quark FF.
As expected, our $D^{\TQc}_c(z,\mu_{F,0})$ functions of Fig.~\ref{fig:FF_initial-scale_gluon} are much lower than the $D^{\Xcq}_c(z,\mu_{F,0})$ ones (see Fig.~2 of Ref.~\cite{Nejad:2021mmp}) and, as anticipated, the peak is shifted backwards of about $0.25$ units of $z$.

One might naively argue that such a backward shift for the charm FF peak could be not in line with general observations on the same function for other heavy-flavored species, say $|Q{\bar q}\rangle$ systems with mass $M$ and momentum $P$.
In that case, indeed, one requires that the constituent heavy quark and the light antiquark (up, down or strange) must have roughly the same velocity, $v_Q \simeq v_q \equiv v$, so that their momenta are $p_Q \equiv z P = m_Q v$ and $p_q = \Lambda_q v$, with $m_Q$ the mass of the heavy quark and $\Lambda_q$ a hadronic mass scale of the order of $\LQCD$.

Given that $M \approx m_Q$ for singly heavy-flavored mesons, we have $m_Q v \approx P = p_Q + p_q = z m_Q v + \Lambda_q v$, and therefore $\langle z \rangle_{D_c} \approx 1 - \Lambda_q/m_Q$.
Hence, one may expect that heavy-quark FFs are peaked in the large-$z$ range and that binding effects scale linearly in the heavy-quark mass~\cite{Suzuki:1977km,Bjorken:1977md}.
This feature does not necessary hold, however, for multiply heavy-flavored hadrons, such as quarkonia and our $\TQc$ tetraquarks. 
Here, no soft scale exists since no light constituent quarks enter the lowest Fock state.
Furthermore, it is reasonable to expect that mutual hadronic interactions among the four constituent charms hinder any possibility of predicting the position of the FF peak solely based on simple kinematic calculations.

In analogy with the gluon case (see Fig.~\ref{fig:FF_initial-scale_gluon}), plots of Fig.~\ref{fig:FF_initial-scale_charm} show the sensitivity of the $D^{\TQc}_c(z,\mu_{F,0})$ FFs on factorization-scale variation.
By using {\symJethad}, here an expanded DGLAP evolution was analytically performed in the $4m_c < \mu_F < 6m_c$ range.
The function shape remains almost unaltered when passing from the $\TQcZpp$ case (left plot) to the $\TQcTpp$ one (right plot).
This is expected, since the $D^{\TQc}_c(z,\mu_{F,0})$ function in Eq.~\eqref{Dc_FF_initial-scale} does not discriminate between the $0^{++}$ state and the $2^{++}$ radial excitation.
A minimal effect, not visible in the plots, is generated by the coupled DGLAP evolution with the $D^{\TQc(J^{PJ})}_g(z,\mu_{F,0})$ input (Eq.~\eqref{Dg_FF_initial-scale}), which does instead depend on the produced tetraquark.

We stress, however, that the study on $\mu_F$ variation in Fig.~\ref{fig:FF_initial-scale_charm} has been shown just for illustrative scopes and for qualitative comparison with Fig.~\ref{fig:FF_initial-scale_gluon}, while a detailed description of our methodology to address the DGLAP evolution of our FFs will be given in the next subsection.

\subsection{The {\tt TQ4Q1.0} functions}
\label{ssec:FFs-TQ4Q10}

The last step to build our {\tt TQ4Q1.0} collinear FFs for fully charmed tetraquarks is performing a proper DGLAP evolution of the initial-scale inputs presented before.
The striking difference with respect to light-hadron fragmentation is that, in our case, not only the (anti)charm channel but also the gluon one has a threshold for evolution.
This directly comes from the $[g \to (cc\bar{c}\bar{c})]$ and $[c,\bar{c} \to (cc\bar{c}\bar{c}) + c,\bar{c}]$ perturbative splittings, respectively encoded in the left and right initial-scale inputs of Fig.~\ref{fig:FF_diagram} and mathematically described by the corresponding SDCs.
It follows from kinematics that the minimal invariant mass for the first splitting is $\mu_{F,0}(g \to \TQc) = 4 m_c$, which we take as the threshold for the gluon fragmentation.
On the other side, the minimal invariant mass for the second splitting is $\mu_{F,0}(c \to \TQc) = 5 m_c$, and we designate this as the threshold for the (anti)charm fragmentation.
These scales are summarized in Table~\ref{tab:muF0_Q0}.

 \begin{table*}[!b]
 \begin{center}
 \begin{tabular}[c]{||c|c||c||}
 \hline
 $\mu_{F,0}(g \to \TQc)$ & $\mu_{F,0}(c \to \TQc)$ &  $Q_0 \equiv \max \left( \{ \mu_{F,0} \} \right)$ \\
 \hline
 $4m_c$ & $5m_c$ & \textcolor{urlblue}{$\boldsymbol{5m_c}$} \\
 \hline
  \end{tabular}
 \caption{First (second) column: initial scale, $\mu_{F,0}$, for the fragmentation of a gluon (charm quark) to a color-singlet $S$-wave $\TQc$ tetraquark.
 Rightmost column: \emph{evolution-ready} scale, $Q_0$, set to the maximum of $\mu_{F,0}$ values (in bold blue font).}
 \label{tab:muF0_Q0}
 \end{center}
 \end{table*}

Now we need to develop a strategy that addresses the DGLAP evolution by properly accounting for these thresholds.
To this extent, we make use of features of the novel {\HFNRevo} method~\cite{Celiberto:2024mex,Celiberto:2024bxu}, designed to portray the DGLAP evolution of heavy-hadron fragmentation from nonrelativistic inputs.
It essentially builds upon three fundamental aspects: interpretation, evolution, and uncertainties.

The first aspect permits us to decipher the short-distance mechanism, dominant at low-$|\vec q_T|$, as a fixed-flavor number-scheme (FFNS) two-parton fragmentation (see also Section~\ref{ssec:HF_fragmentation} of this article).
This lays the ground for a subsequent matching between FFNS and VFNS calculations.
The third aspect allows us to gauge the impact of missing higher-order uncertainties (MHOUs) coming from scale variations connected to DGLAP-evolution thresholds.
Concerning our currently exploratory study on $\TQc$ fragmentation, we postpone those points to the future while, for the time being, we just focus on aspects related to evolution.

\begin{figure*}[!t]
\centering

   \includegraphics[scale=0.35,clip]{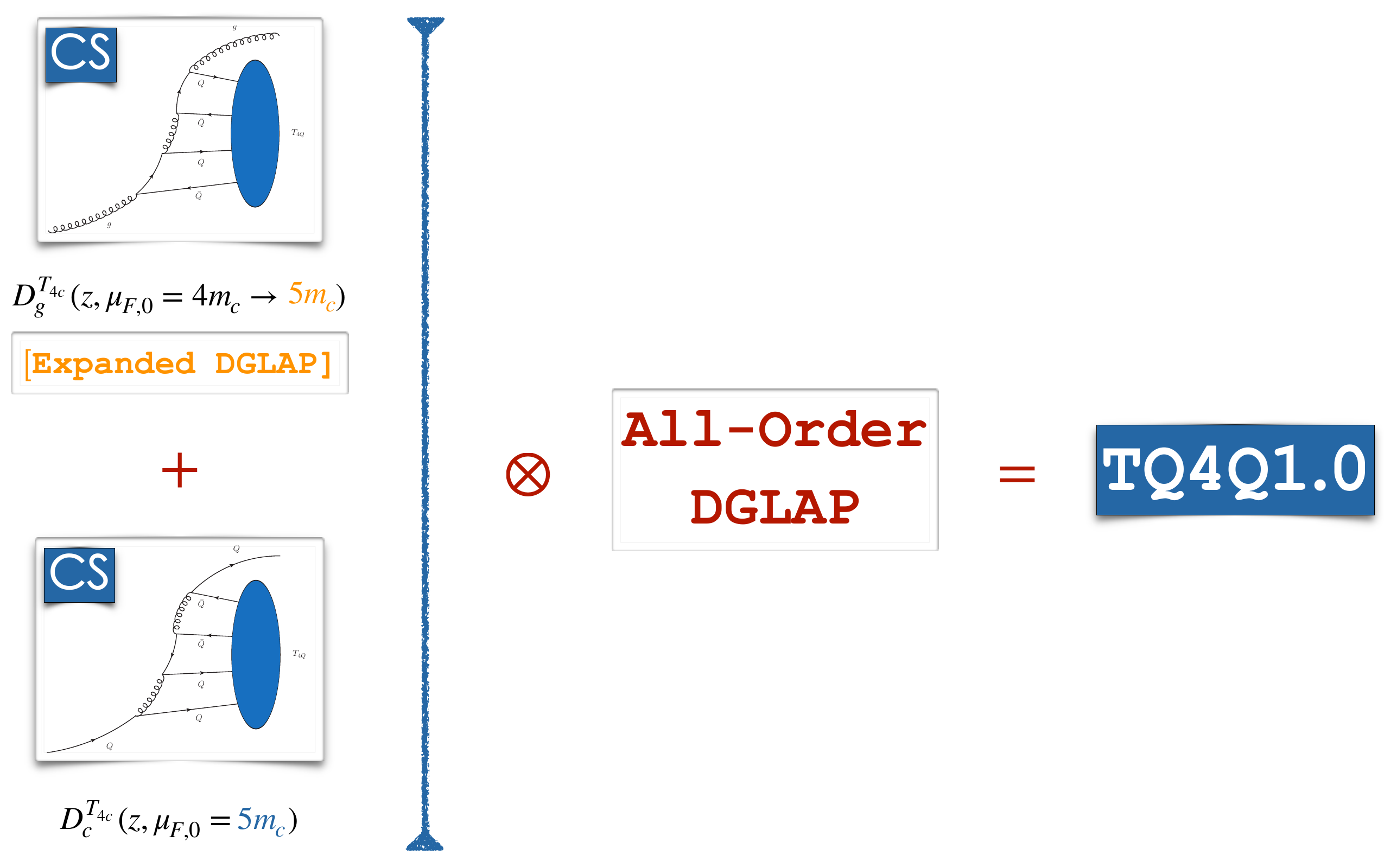}

\caption{A sketch of our methodology to depict $\TQc$ tetraquark fragmentation in the color-singlet (CS) configuration. The initial-scale gluon FF is analytically evolved from $4m_c$ to $Q_0 = 5m_c$ with {\symJethad}. Then, starting from $Q_0$, the {\tt TQ4QF.10} FF sets are obtained \emph{via} the standard all-order DGLAP evolution, performed numerically with {\APFELpp}. Light flavors and the bottom one are generated by evolution.}
\label{fig:FF_evolution_sketch}
\end{figure*}

According to {\HFNRevo}, the DGLAP evolution of a set of heavy-hadron FFs can be split in two steps, as depicted in the sketch of Fig.~\ref{fig:FF_evolution_sketch}.
Contextualizing it to the $\TQc$ case, we start from the fragmentation channel featuring the lowest threshold, namely the gluon one (see Table~\ref{tab:muF0_Q0}).
We take the calculation described in Section~\ref{ssec:FFs-g} as the gluon FF at the initial scale of $\mu_{F,0}(g \to \TQc) = 4 m_c$ and perform a DGLAP evolution regulated by $P_{gg}$ only (see Eqs.~\eqref{Dg_FF_DGLAP_evo_eq} to~\eqref{Dg_FF_DGLAP_evo_sol}) up the initial scale of the charm, $\mu_{F,0}(c \to \TQc) = 5 m_c$.
In this way we generate collinear gluons, and only gluons, between the two scales.
Moreover, being this evolution \emph{expanded} in powers of $\alpha_s$ and \emph{decoupled} from any other quark channel, it can be performed analytically \emph{via} the {\symJethad} plugin.

Now we proceed with the second step.
We combine the gluon FF evolved at $Q_0 \equiv 5 m_c$ with the (anti)charm FF input given in Section~\ref{ssec:FFs-Q} and, starting from this scale, we numerically perform an all-order DGLAP evolution to generate our NLO {\tt TQ4Q1.0} functions released in {\tt LHAPDF} format.
We refer to $Q_0$ as the \emph{evolution-ready} scale, which genuinely corresponds to the maximum between the evolution thresholds of all the parton species (gluon and charm, in our case).
Furthermore, as mentioned above, it represents the initial scale at which the numeric DGLAP-evolution code is called.
In our study we employ {\tt APFEL++}~\cite{Bertone:2013vaa,Carrazza:2014gfa,Bertone:2017gds} while, in the future, we plan to link our technology also to {\tt EKO}~\cite{Candido:2022tld,Hekhorn:2023gul}.

One might argue that our treatment is not complete due the absence of the light- and the bottom-quark channels.
To our knowledge, a calculation for the collinear fragmentation of a nonconstituent quark into a fully charmed tetraquark has not yet been performed.
Consequently, in our two-step evolution setup (see Fig.~\ref{fig:FF_evolution_sketch}) light quarks and the bottom are simply generated by evolution, with no corresponding initial-scale inputs.
However, in analogy with NRQCD studies on color-singlet pseudoscalar~\cite{Braaten:1993rw,Braaten:1993mp,Artoisenet:2014lpa,Zhang:2018mlo,Zheng:2021mqr,Zheng:2021ylc} and vector~\cite{Braaten:1993rw,Braaten:1993mp,Zheng:2019dfk} charmonia, these channels are expected to be heavily suppressed with respect to gluon and charm ones.

\begin{figure*}[!t]
\centering

   \hspace{-0.00cm}
   \includegraphics[scale=0.41,clip]{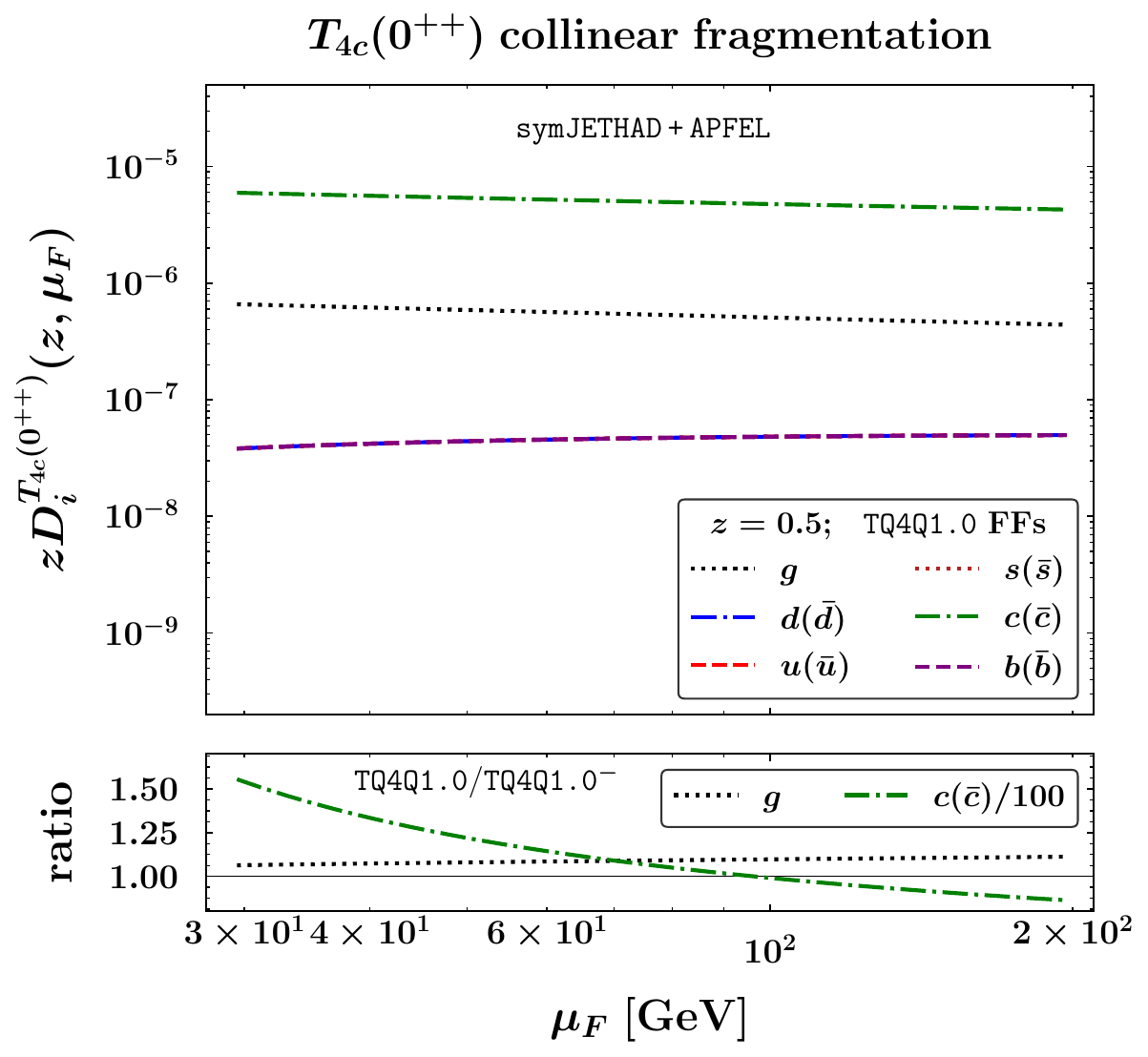}
   \hspace{-0.35cm}
   \includegraphics[scale=0.41,clip]{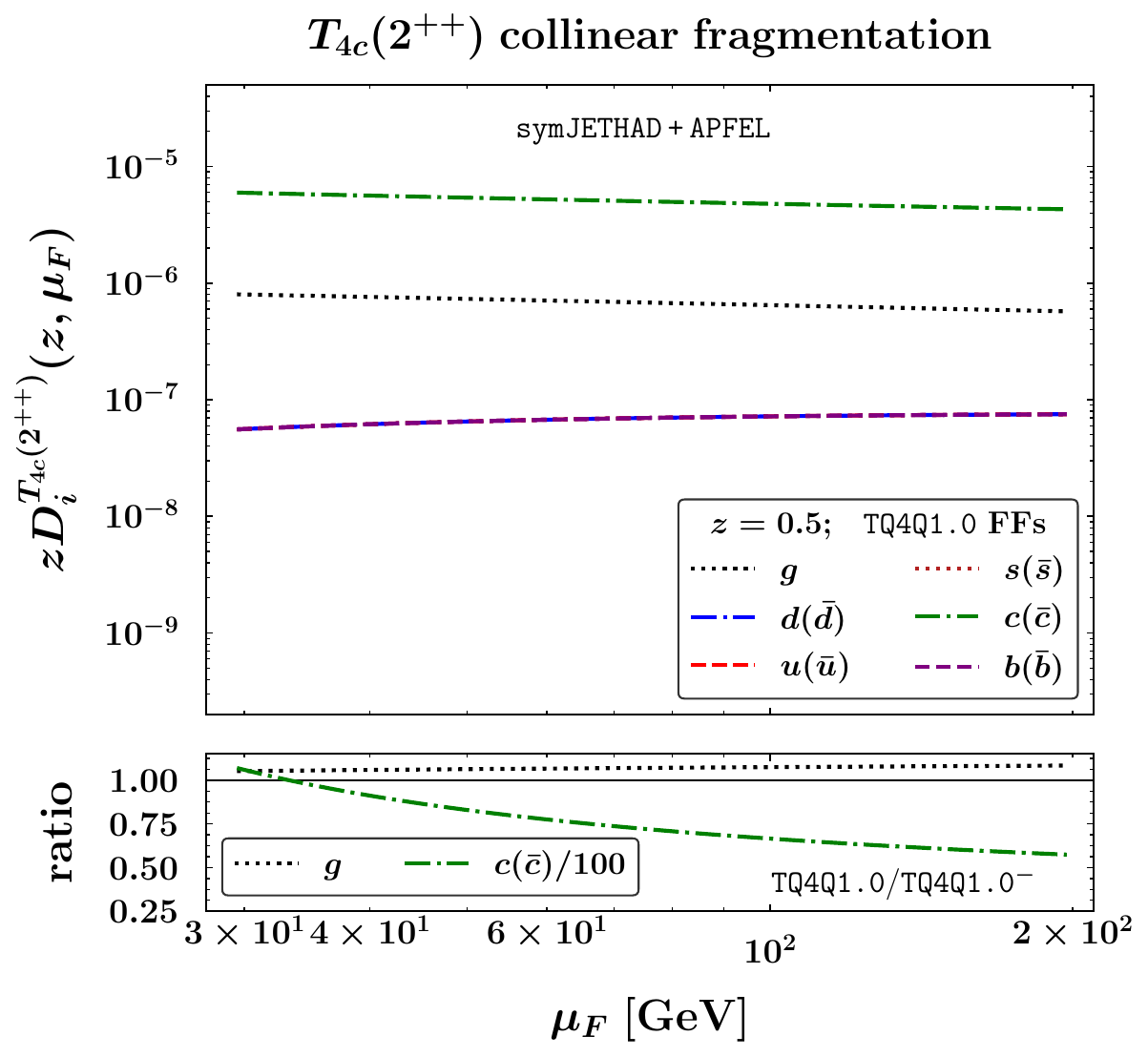}

\caption{Factorization-scale dependence of {\tt TQ4Q1.0} collinear FFs describing $\TQcZpp$ (left) and $\TQcTpp$ (right) production, at $z = 0.50 \simeq \langle z \rangle$.
Ancillary panels below primary plots show the ratio between {\tt TQ4Q1.0} and {\tt TQ4Q1.0}$^-$ functions.
For comparison purposes, the charm ratio has been scaled down by a factor of 100.}
\label{fig:NLO_FFs_Tc0_Tc2}
\end{figure*}

\begin{figure*}[!t]
\centering

   \hspace{-0.25cm} \includegraphics[scale=0.41,clip]{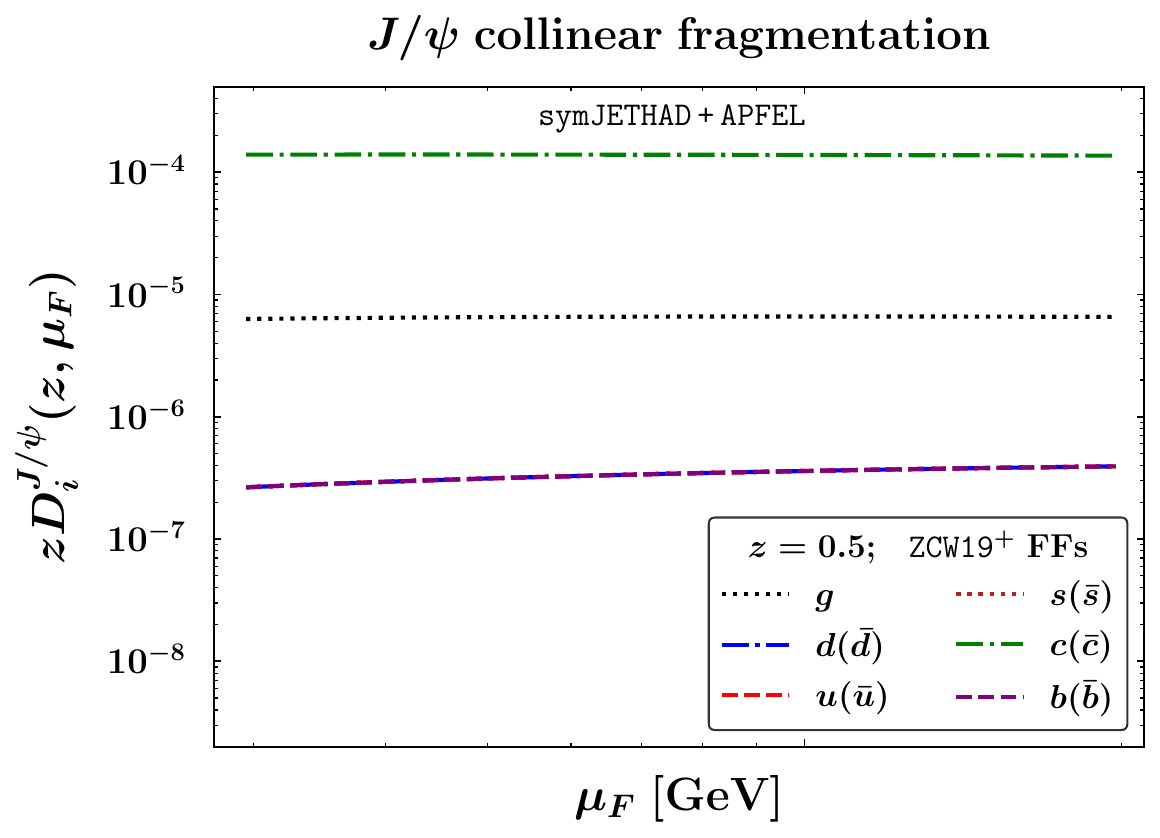}
   \includegraphics[scale=0.41,clip]{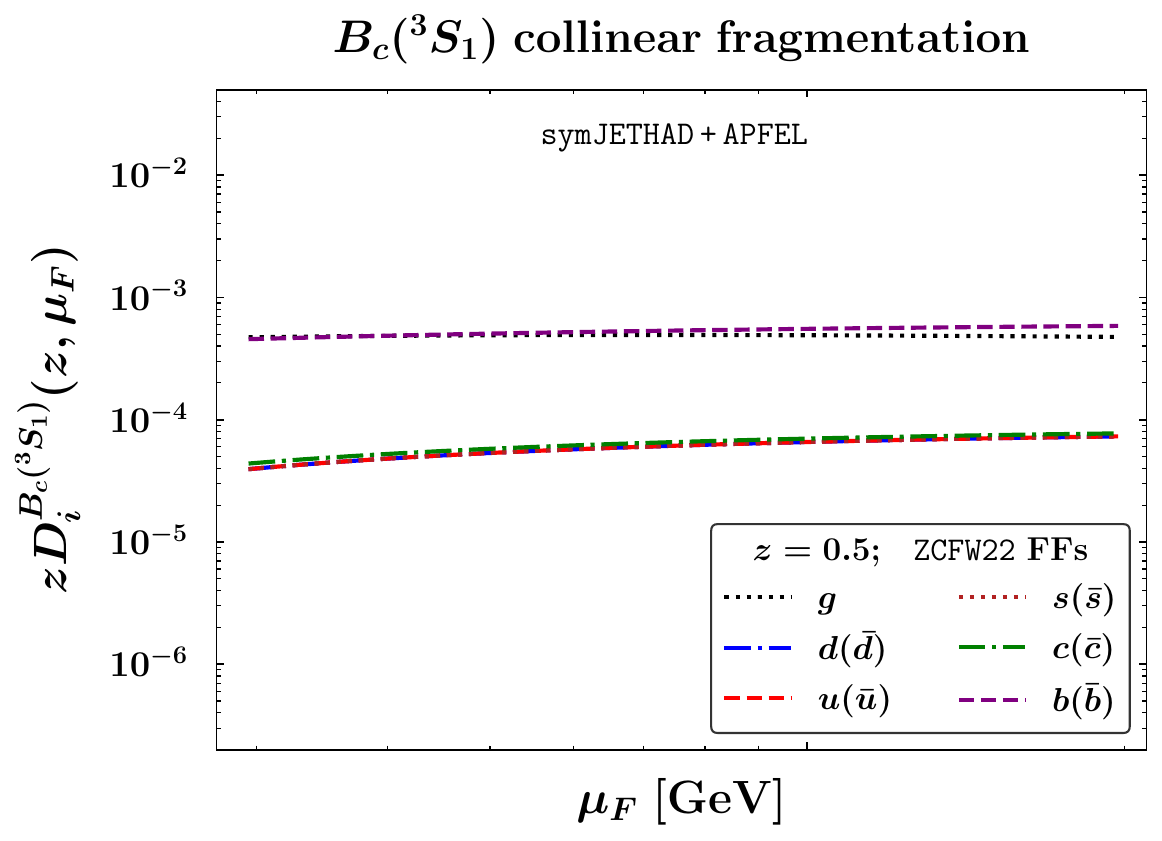}

\caption{Factorization-scale dependence of {\tt ZCW19$^+$}~\cite{Celiberto:2022dyf,Celiberto:2023fzz} and {\tt ZCFW22}~\cite{Celiberto:2022keu,Celiberto:2024omj} collinear FFs respectively describing $\Jpsi$ (left) and $\Bss$ meson (right) production, at $z = 0.50 \simeq \langle z \rangle$.}
\label{fig:NLO_FFs_Jps_Bss}
\end{figure*}

In left and right panels of Fig.~\ref{fig:NLO_FFs_Tc0_Tc2} we show the factorization-scale dependence of {\tt TQ4Q1.0} NLO functions describing $\TQcZpp$ and $\TQcTpp$ collinear fragmentation, respectively.
For comparison, in left and right panels of Fig.~\ref{fig:NLO_FFs_Jps_Bss} we present the $\mu_F$-behavior of {\tt ZCW19$^+$}~\cite{Celiberto:2022dyf,Celiberto:2023fzz} and {\tt ZCFW22}~\cite{Celiberto:2022keu,Celiberto:2024omj} NLO functions for $\Jpsi$ and $\Bss$ fragmentation, respectively.
For brevity, in both figures we consider just one value of the momentum fraction, say $z = 0.5 \simeq \langle z \rangle$.
It roughly represents the mean value at which FFs are generally sounded in semi-hard final states (see Section~\ref{sec:results}).

From the inspection of results in Fig.~\ref{fig:NLO_FFs_Tc0_Tc2}, the (anti)charm to $\TQc$ fragmentation channel heavily dominates on light-parton and bottom ones.
It also stays roughly one order of magnitude above the gluon channel in the whole $\mu_F$-range considered.
This pattern fairly matches the charmonium case ($\Jpsi$, left panel of Fig.~\ref{fig:NLO_FFs_Jps_Bss}),
whereas it differs from the charmed $B$ meson one ($\Bss$, right panel of Fig.~\ref{fig:NLO_FFs_Jps_Bss}).
Indeed, being the latter a ``generalized" quarkonium state (see Section~\ref{ssec:HF_fragmentation}) with two heavy-flavored quarks composing its lowest Fock state, a more involved dynamics is expected to be encoded in its dynamical production mechanism.
Furthermore, the simultaneous presence of both gluon-to-charm and gluon-to-bottom perturbative splittings, accounted for in corresponding SDCs~\cite{Braaten:1993jn,Zheng:2019gnb,Zheng:2021sdo}, qualitatively explains the larger size of the gluon channel on the other partons, with respect to what happens for $\TQc$ and $\Jpsi$.

Despite being the contribution of charm-to-tetraquark fragmentation substantially larger than that of gluon fragmentation, it is the latter that plays a crucial role in describing production rates ad hadron colliders.
This is due to the strong prevalence of the gluon parton density function (PDF) over the quark ones, which makes the $[gg \to gg]$ partonic channel be much more relevant than the $[gg \to c\bar{c}]$ one.
This is the reason why Authors of Ref.~\cite{Feng:2020riv} focused on the NRQCD calculation of the $[g \to \TQc]$ FF.

Since our {\tt TQ4Q1.0} NLO functions also contain the charm FF at the initial scale, we can estimate its impact on corresponding DGLAP-evolved sets.
For this reason, we derived the {\tt TQ4Q1.0$^-$} NLO functions, which we adopt as supplementary sets for testing purposes.
These sets were constructed using the same method as the {\tt TQ4Q1.0} functions, but with the charm input of Section~\ref{ssec:FFs-Q} switched off.
This results in only the gluon FF existing at the initial scale, while the charm is generated solely through evolution, as is the case for the other quark species.

Below the primary plots of Fig.~\ref{fig:NLO_FFs_Tc0_Tc2}, ancillary panels show the ratio between {\tt TQ4Q1.0} and {\tt TQ4Q1.0$^-$} FFs for the gluon and the charm quark.
We observe that the {\tt TQ4Q1.0} gluon FF is approximately 5\% higher than the {\tt TQ4Q1.0$^-$} counterpart, indicating that the presence of the charm initial-scale input has a noticeable impact on the gluon evolved one.
Conversely, as expected, the {\tt TQ4Q1.0} charm FF is significantly higher than the {\tt TQ4Q1.0$^-$} one, by around two orders of magnitudes.
This discrepancy can play a decisive role in accurately describing $\TQc$ production rates at lepton and lepton-hadron colliders, where $[\gamma^{(*)}\gamma^{(*)} \to c\bar{c}]$ and $[\gamma^{(*)}g \to c\bar{c}]$ partonic subprocesses are relevant.
Taken together, these observations suggest that, since our {\tt TQ4Q1.0} determinations contain both the gluon and charm at the initial scale, they can be universally applied in the description of a broad range of processes, spanning from hadron to lepton and lepton-hadron colliders.

Finally, we note that the gluon-to-tetraquark evolved FFs of Fig.~\ref{fig:NLO_FFs_Tc0_Tc2} exhibit a slow decrease with $\mu_F$.
This behavior is partially shared by the corresponding gluon-to-charmonium and gluon-to-$B_c$ channels of Fig.~\ref{fig:NLO_FFs_Jps_Bss}, whose trend is either flat or slowly increasing with $\mu_F$.
Recently, it was pointed out that gluon FFs exhibiting a smooth $\mu_F$-pattern act as ``stabilizers" of high-energy resummed distributions sensitive to the semi-inclusive production of singly~\cite{Celiberto:2021dzy,Celiberto:2021fdp} as well as multiply~\cite{Celiberto:2022dyf,Celiberto:2022keu,Celiberto:2023rzw} heavy-flavored hadrons.
We refer to this remarkable feature as the \emph{natural stability}~\cite{Celiberto:2022grc} of the high-energy resummation (see Section~\ref{ssec:HE_resummation}). 
Natural stability as resulting from heavy-flavor fragmentation will represent a key ingredient for our next phenomenological analysis (see Section~\ref{sec:results}).

\section{$\boldsymbol{\NLLp}$ hybrid factorization}
\label{sec:hybrid_factorization}

In the first part of this Section (\ref{ssec:HE_resummation}) we present a brief recap of recent phenomenological advancements in the study of the semi-hard regime of QCD. 
In the second part (\ref{ssec:NLL_cross_section}) we give details on the way the $\NLLp$ hybrid factorization is built and adapted to the study of the semi-inclusive $\TQc$ plus jet hadroproduction process.

\subsection{High-energy resummation: A brief overview}
\label{ssec:HE_resummation}

Channels involving the production of hadrons with heavy quarks serve as valuable probes for studying high-energy QCD. 
Here, energy logarithms become very large, affecting the running-coupling expansion to all orders and potentially hampering the convergence of perturbative QCD.
The Balitsky--Fadin--Kuraev--Lipatov (BFKL) resummation~\cite{Fadin:1975cb,Kuraev:1977fs,Balitsky:1978ic} provides a framework to account for these logarithms to all orders. 
Its applicability extends up to the leading level (LL), involving the resummation logarithms of the form $[\alpha_s \ln (s)]^n$, and the next-to-leading level (NLL), resumming contributions proportional to $\alpha_s [\alpha_s^n \ln (s)]^n$.

BFKL-resummed cross sections for hadronic final states are expressed as a transverse-momentum convolution between a process-independent Green's function, calculated within the NLO~\cite{Fadin:1998py,Ciafaloni:1998gs,Fadin:2004zq,Fadin:2023roz}, and two process-related singly off-shell emissions functions, also known as forward-production impact factors.
The latters encompass collinear elements, namely PDFs and FFs. 
This nested collinear factorization inside an overall BFKL convolution justifies naming our formalism as \emph{hybrid} high-energy and collinear factorization.

The validity of BFKL has been assessed through various phenomenological studies within a full or partial $\NLLp$ accuracy.
We mention: the Mueller–Navelet~\cite{Mueller:1986ey} di-jet channel~\cite{Colferai:2010wu,Ducloue:2013hia,Ducloue:2013bva,Caporale:2014gpa,Colferai:2015zfa,Celiberto:2015yba,Celiberto:2015mpa,Celiberto:2016ygs,Caporale:2018qnm,deLeon:2021ecb,Celiberto:2022gji}, light di-hadron~\cite{Celiberto:2016hae,Celiberto:2017ptm,Celiberto:2020rxb,Celiberto:2022rfj}, hadron-jet~\cite{Bolognino:2018oth,Bolognino:2019cac,Bolognino:2019yqj,Celiberto:2020wpk,Celiberto:2020rxb,Celiberto:2022kxx}, and multijet~\cite{Caporale:2016soq,Caporale:2016xku,Celiberto:2016vhn,Caporale:2016zkc} tags, then forward Higgs~\cite{Hentschinski:2020tbi,Celiberto:2022fgx,Celiberto:2020tmb,Mohammed:2022gbk,Celiberto:2023rtu,Celiberto:2023uuk,Celiberto:2023eba,Celiberto:2023nym,Celiberto:2023rqp,Celiberto:2022zdg,Nefedov:2019mrg,Fucilla:2022whr,Fucilla:2024cpf}, Drell–Yan~\cite{Celiberto:2018muu,Golec-Biernat:2018kem}, and heavy-hadron detections~\cite{Celiberto:2017nyx,Boussarie:2017oae,Bolognino:2019ouc,Bolognino:2019yls,Celiberto:2021dzy,Celiberto:2021fdp,Celiberto:2022dyf,Celiberto:2023fzz,Celiberto:2022grc,Bolognino:2022paj,Celiberto:2022keu,Celiberto:2022kza,Celiberto:2024omj}.
Furthermore, emissions of single objects in forward directions provide us with a direct avenue for probing the gluon content of the proton at small-$x$ through the unintegrated gluon distribution (UGD). The evolution of the UGD is governed by the BFKL kernel. Phenomenological analyses of the UGD have been conducted through exclusive light vector-meson leptoproduction at HERA~\cite{Anikin:2011sa,Besse:2013muy,Bolognino:2018rhb,Bolognino:2018mlw,Bolognino:2019bko,Bolognino:2019pba,Celiberto:2019slj,Luszczak:2022fkf} and the EIC~\cite{Bolognino:2021niq,Bolognino:2021gjm,Bolognino:2022uty,Bolognino:2022ndh}, as well as by means of the vector-quarkonium photoproduction~\cite{Bautista:2016xnp,Garcia:2019tne,Hentschinski:2020yfm}.

The gluon-content information carried by the UGD has been instrumental in refining the collinear-factorization framework \emph{via} the determination of resummed small-$x$ PDFs~\cite{Ball:2017otu,Abdolmaleki:2018jln,Bonvini:2019wxf}. Additionally, it has been employed to determine low-$x$ improved spin-dependent transverse momentum densities (TMDs) at small $x$~\cite{Bacchetta:2020vty,Bacchetta:2024fci,Celiberto:2021zww,Bacchetta:2021oht,Bacchetta:2021lvw,Bacchetta:2021twk,Bacchetta:2022esb,Bacchetta:2022crh,Bacchetta:2022nyv,Celiberto:2022omz,Bacchetta:2023zir}.
References~\cite{Hentschinski:2021lsh,Mukherjee:2023snp} unravel the interplay between BFKL and TMD physics, while Refs.~\cite{Boroun:2023goy,Boroun:2023ldq} connect color-dipole cross sections with the UGD.

A striking result coming from high-energy emissions of singly heavy-flavored bound states, such as $\Lambda_c$ baryons~\cite{Celiberto:2021dzy} or $b$-hadrons~\cite{Celiberto:2021fdp},
is the possibility of overcome the well-known issues in the description of semi-hard final states at natural scales, 
which become particularly manifest when lighter objects are tagged~\cite{Ducloue:2013bva,Caporale:2014gpa,Bolognino:2018oth,Celiberto:2020wpk}.
Here, large and negative NLL corrections, together with nonresummed \emph{threshold} logarithms, spoil the convergence of the resummed series.
This effect becomes evident when studies on assessing the weight of MHOUs \emph{via} factorization and renormalization scale variations are undertaken. 

On the contrary, a \emph{natural stabilization} pattern~\cite{Celiberto:2022grc} (see also a brief, related discussion at the end of Section~\ref{ssec:FFs-TQ4Q10}) emerges for final states sensitive to the (semi-)inclusive emission of heavy hadrons at large transverse momentum, where the dominant mechanism describing their dynamic production is VFNS collinear fragmentation.
Studies on the natural stability were further extended to doubly heavy-flavored mesons within a collinearly enhanced nonrelativistic fragmentation approach.
To this extent, novel VFNS, DGLAP-evolving FFs were derived on the basis of NRQCD inputs~\cite{Braaten:1993mp,Zheng:2019dfk,Braaten:1993rw,Chang:1992bb,Braaten:1993jn,Ma:1994zt,Zheng:2019gnb,Zheng:2021sdo,Feng:2021qjm,Feng:2018ulg} first to vector quarkonia~\cite{Celiberto:2022dyf,Celiberto:2023fzz}, and then to $\BCs$ and $\Bss$ bound states~\cite{Celiberto:2022keu,Celiberto:2024omj}.

By taking advantage of the natural stability, a first contact point between the high-energy QCD sector and the physics of exotic hadrons was established in Ref.~\cite{Celiberto:2023rzw} (see Ref.~\cite{Celiberto:2024mrq} for a review).
As discussed in Section~\ref{ssec:HF_fragmentation}, novel sets of VFNS FFs, named {\tt TQHL1.0} functions, were obtained in Ref.~\cite{Celiberto:2023rzw} to describe the production of four hidden-flavored, neutral heavy-light tetraquark species: $\Xcu$, $\Xcs$, $\Xbu$, and $\Xbs$.

\subsection{NLL-resummed cross section}
\label{ssec:NLL_cross_section}

\begin{figure*}[!t]
\centering
\includegraphics[width=0.75\textwidth]{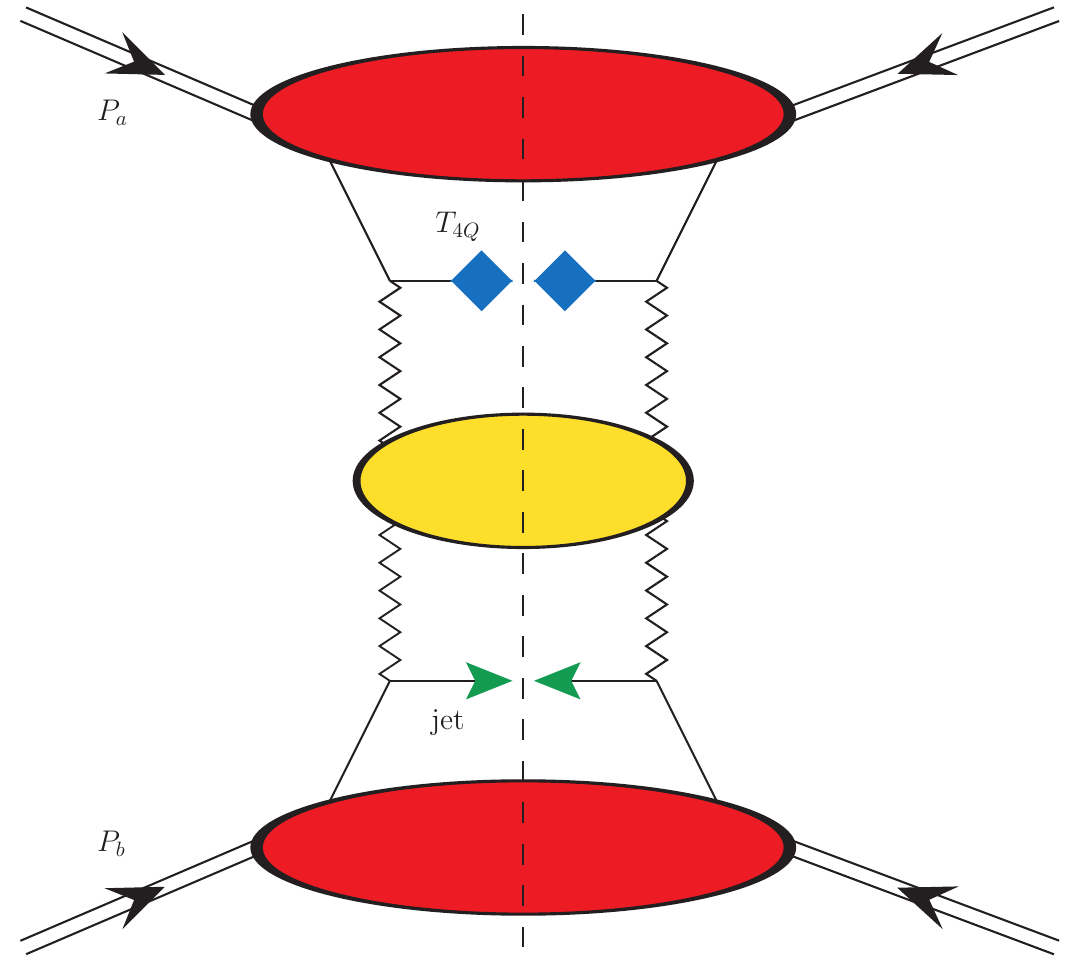}

\caption{Hybrid factorization for the semi-inclusive hadroproduction of a $\TQQ$ tetraquark in association with a jet.
Blue rhombi denote tetraquark
collinear FFs. Green arrows stand for light jets. 
Large red blobs depict proton collinear PDFs.
The BFKL Green’s function (yellow oval) is connected to the two off-shell emission functions \emph{via} Reggeon lines.}
\label{fig:reaction}
\end{figure*}

The process matter of our interest is (see Fig.~\ref{fig:reaction})
\begin{equation}
\label{process}
    {\rm p}(P_a) + {\rm p}(P_b) \, \rightarrow \, \TQQ(\kappa_1, \phi_1, y_1) + {\cal X} + {\rm jet}(\kappa_2,\phi_2, y_2) \; ,
\end{equation}
where ${\rm p}(P_{a,b})$ is a parent proton having momentum $P_{a,b}$.
Then, $\TQQ$ inclusively refers to a fully charmed tetraquark, $\TQcZpp$ or the $\TQcTpp$ resonance, produced with momentum $\kappa_1$, azimuthal angle $\phi_1$ and rapidity $y_1$.
A light jet is simultaneously tagged with momentum $\kappa_2$, azimuthal angle $\phi_2$ and rapidity $y_2$, whereas ${\cal X}$ refers to all the undetected products. 
On the one side, high transverse momenta, $|\vec \kappa_{1,2}|$, together with a large rapidity interval, $\DY = y_1 - y_2$, permit us to deal with diffractive semi-hard final states.
On the other side, large transverse-momentum ranges are required to preserve the validity of a VFNS treatment, as well as to ensure that collinear fragmentation is the prevailing mechanism for the formation of our heavy tetraquark.

Incoming protons' form a Sudakov-like basis with $P_a^2= P_b^2=0$ and $(P_a\cdot P_b) = s/2$.
Thus we have
\begin{equation}\label{sudakov}
\kappa_{1,2} = x_{1,2} P_{a,b} - \frac{ \kappa_{1,2\perp}^2}{x_{1,2} s}P_{b,a} + \kappa_{1,2\perp} \ , \quad
\vec \kappa_{1,2}^{\,2} \equiv -\kappa_{1,2\perp}^2\;,
\end{equation}
and the relations
\begin{equation}\label{y-vs-x}
y_{1,2}=\pm\frac{1}{2}\ln\frac{x_{1,2}^2 s}
{\vec \kappa_{1,2}^2 }
\qquad \mbox{and} \qquad
\drv y_{1,2} = \pm \frac{\drv x_{1,2}}{x_{1,2}}
\end{equation}
hold between longitudinal fractions of final-state particles ($x_{1,2}$) and the corresponding rapidities.
Furthermore, one has 
\begin{equation}
\label{DeltaY}
 \DY = y_1 - y_2 = \ln \left( \frac{x_1 x_2}{|\vec \kappa_1||\vec \kappa_2|} s \right) \;.
\end{equation}

In a pure LO QCD collinear-factorization, the differential cross section for our process comes as a collinear convolution between the on-shell hard-subprocess term, parent-proton PDFs, and tetraquark FFs
\begin{equation}
\label{sigma_collinear}
\begin{split}
\hspace{-0.25cm}
\frac{\drv\sigma^{\rm LO}_{\rm [coll.]}}{\drv x_1\drv x_2\drv ^2\vec \kappa_1\drv ^2\vec \kappa_2}
= \hspace{-0.25cm} \sum_{m,n=q,{\bar q},g}\int_0^1 \hspace{-0.20cm} \drv x_a \int_0^1 \hspace{-0.20cm} \drv x_b\ f_m\left(x_a\right) f_n\left(x_b\right)
\int_{x_1}^1 \hspace{-0.15cm} \frac{\drv \upsilon}{\upsilon}D^{\TQQ}_m\left(\frac{x_1}{\upsilon}\right) 
\frac{\drv {\hat\sigma}_{m,n}\left(\hat s\right)}
{\drv x_1\drv x_2\drv ^2\vec \kappa_1\drv ^2\vec \kappa_2}\;,
\end{split}
\end{equation}
where the $m,n$ indices run over all parton species except for the top quark, which does not fragment.\footnote{For the sake of brevity, the explicit dependence on $\mu_F$ has been dropped from Eq.~\eqref{sigma_collinear}.}
Then, $f_{m,n}\left(x_{a,b}, \mu_F \right)$ depict the proton PDFs, while $D^{\TQQ}_m\left(x_1/\upsilon, \mu_F \right)$ are the tetraquark FFs.
Furthermore, $x_{a,b}$ are incoming partons' longitudinal-momentum fractions, whereas $\upsilon$ denotes the momentum fraction of the outgoing parton fragmenting into the exotic hadron. 
Finally, $\drv\hat\sigma_{m,n}\left(\hat s \right)$ stands for the partonic cross sections and $\hat s = x_a x_b s$ is the  partonic-scattering center-of-mass energy squared.

Conversely, the high-energy resummed differential cross section within the hybrid factorization reads as a transverse-momentum convolution between the BFKL Green's function and two forward-particle, singly off-shell emission functions.
Collinear ingredients (PDFs and FFs) are encoded in the latters.
We can express the cross section as a Fourier series in the azimuthal angle difference, $\phi = \phi_1 - \phi_2 - \pi$, to have
\begin{equation}
 \label{dsigma_Fourier}
 \frac{\drv \sigma}{\drv y_1 \drv y_2 \drv \vec \kappa_1 \drv \vec \kappa_2 \drv \phi_1 \drv \phi_2} =
 \frac{1}{(2\pi)^2} \left[{\cal C}_0 + 2 \sum_{l=1}^\infty \cos (l \phi)\,
 {\cal C}_l \right]\, .
\end{equation}
The first building block, namely the NLL Green's function, reads
\begin{equation}
\label{G_BFKL_NLL}
 {\cal G}_{\rm NLL}(\DY,l,\nu,\mu_R) = e^{{\DY} \bar \alpha_s(\mu_R) \,
 \chi^{\rm NLO}(l,\nu)} \; ,
\end{equation}
where $\bar \alpha_s(\mu_R) \equiv \alpha_s(\mu_R) N_c/\pi$, with $\beta_0 = 11N_c/3 - 2 n_f/3$ the QCD $\beta$-function first coefficient.
The main function entering the exponent of Eq.~\eqref{G_BFKL_NLL} is the BFKL kernel in the Mellin space.
It contains the resummation of energy logarithms at NLL.
One has
\begin{eqnarray}
 \label{chi}
 \chi^{\rm NLO}(l,\nu) = \chi(l,\nu) + \bar\alpha_s \tilde{\chi}(l,\nu) \;,
\end{eqnarray}
with $\chi(l,\nu)$ being the LO kernel eigenvalues
\begin{eqnarray}
 \label{kernel_LO}
 \chi\left(l,\nu\right) = -2\gamma_{\rm E} - 2 \, {\rm Re} \left\{ \psi\left(\frac{1}{2} + \frac{l}{2} + i \nu \right) \right\} \, ,
\end{eqnarray}
$\gamma_{\rm E}$ the Euler-Mascheroni constant and $\psi(z) \equiv \Gamma^\prime
(z)/\Gamma(z)$ the logarithmic derivative of the Gamma function. 
The $\tilde{\chi}(l,\nu)$ function in Eq.\eref{chi} is the NLO correction to the kernel
\begin{equation}
\label{chi_NLO}
\tilde{\chi} \left(l,\nu\right) = \bar\chi(l,\nu)+\frac{\beta_0}{2 N_c}\chi(l,\nu)
\left( - \frac{1}{4}\chi(l,\nu) + \frac{5}{6} + \ln\frac{\mu_R}{\sqrt{|\vec \kappa_1||\vec \kappa_2|}} \right) \;.
\end{equation}
The $\bar\chi(l,\nu)$ function used here was obtained in Refs.~\cite{Kotikov:2000pm,Kotikov:2002ab}. 

Another core ingredient to construct our resummed differential cross section is the tetraquark NLO emission function projected onto the LO kernel eigenfunctions. We rely on the expression obtained in Ref.~\cite{Ivanov:2012iv}, well-adapted to the study of heavy hadrons at large transverse momentum:
\begin{equation}
\label{TIF}
\E_{\TQQ}^{\rm NLO}(l,\nu,|\vec \kappa|,x) =
\E_{\TQQ}(\nu,|\vec \kappa|,x) +
\alpha_s(\mu_R) \, \hat \E_{\TQQ}(l,\nu,|\vec \kappa|,x) \;.
\end{equation}
Its LO term reads
\begin{equation}
\label{LOTIF}
\hspace{-0.30cm}
\E_{\TQQ}(\nu,|\vec \kappa|,x) 
= 2 \sqrt{\frac{C_F}{C_A}}
|\vec \kappa|^{2i\nu-1}
\!\!\!\int_{x}^1\frac{\drv \upsilon}{\upsilon}
\left( \frac{\upsilon}{x} \right)
^{2 i\nu-1} 
 \!\left[\frac{C_A}{C_F}f_g(\upsilon)D_g^{\TQQ}\left(\frac{x}{\upsilon}\right)
 +\!\!\!\sum_{m=q,\bar q}\!f_m(\upsilon)D_m^{\TQQ}\left(\frac{x}{\upsilon}\right)\right] 
\end{equation}
while the NLO correction, $\hat \E_{\TQQ}(l,\nu,|\vec \kappa|,x)$, is given in Eqs.~(4.58) to~(4.65) of Ref.~\cite{Ivanov:2012iv}.

The last building block is represented by the Mellin projection of the light-jet NLO emission function
\begin{equation}
\label{JIF}
\E_J^{\rm NLO}(l,\nu,|\vec \kappa|,x) =
\E_J(\nu,|\vec \kappa|,x) +
\alpha_s(\mu_R) \, \hat \E_J(l,\nu,|\vec \kappa|,x) \;,
\end{equation}
where
\begin{equation}
 \label{LOJIF}
 \E_J(\nu,|\vec \kappa|,x) =  2 \sqrt{\frac{C_F}{C_A}}
 |\vec \kappa|^{2i\nu-1}\left[\frac{C_A}{C_F}f_g(x)
 +\sum_{n=q,\bar q}f_n(x)\right] \;
\end{equation}
is its LO limit.
The NLO correction, $\hat \E_J(l,\nu,|\vec \kappa|,x)$, depends on the jet selection function. 
We rely upon a suitable choice, obtained by combining Eq.~(36) of Ref.~\cite{Caporale:2012ih} with Eqs.~(4.19) and~(4.20) of Ref.~\cite{Colferai:2015zfa}.
It bases on computations in Refs.~\cite{Ivanov:2012ms}, suited to numerical analyses, where a jet selection function was calculated within the small-cone approximation (SCA)~\cite{Furman:1981kf,Aversa:1988vb} and then adapted to the cone-type algorithm (see Ref.~\cite{Colferai:2015zfa} for technical details). 
Following the choice done by recent CMS experimental analyses on forward-jet events~\cite{Khachatryan:2016udy}, we set the jet-cone radius to ${\cal R}_J = 0.5$.

We combine all the ingredients to write our master formula, valid in the $\MSb$ renormalization scheme~\cite{PhysRevD.18.3998}, for $\NLLp$ azimuthal coefficients (see Ref.~\cite{Caporale:2012ih} for more details).

\begin{eqnarray}
\label{Cl_NLLp_MSb}
 \ClNLLp \!\! &=& \!\! 
 \frac{e^{\DY}}{s} 
 \int_{-\infty}^{+\infty} \drv \nu \, 
 {\cal G}_{\rm NLL}(\DY,l,\nu,\mu_R) \,
 \alpha_s^2(\mu_R) 
 \\ \nonumber
 \!\! &\times& \!\! \biggl\{\E_{\TQQ}^{\rm NLO}(l,\nu,|\vec \kappa_1|, x_1) \,
 [\E_J^{\rm NLO}(l,\nu,|\vec \kappa_2|,x_2)]^*
 \\ \nonumber
 \!\! &+& \!\!
 \left.
 \alpha_s^2(\mu_R) \DY \frac{\beta_0}{4 \pi} \,
 \chi(l,\nu)
 \left[\ln\left(|\vec \kappa_1| |\vec \kappa_2|\right) + \frac{i}{2} \, \frac{\drv}{\drv \nu} \ln\frac{\E_{\TQQ}}{\E_J^*}\right]
 \right\}
 \;.
\end{eqnarray}
The label $\NLLp$ indicates a complete NLL resummation of energy logarithms within the NLO perturbative accuracy. The `$+$' superscript denotes that terms beyond the NLL level, arising from the cross product of the NLO emission-function corrections, are included in our representation for azimuthal coefficients.

For the sake of comparison, we will also consider the pure LL limit in the $\MSb$ scheme
\begin{equation}
\label{Cm_LL_MSb}
 \ClLL = 
 \frac{e^{\DY}}{s} 
 \int_{-\infty}^{+\infty} \drv \nu \, 
 e^{{\cal G}_{\rm NLL}^{(0)}(\DY,l,\nu,\mu_R)} 
 \alpha_s^2(\mu_R) \, \E_{\TQQ}(l,\nu,|\vec \kappa_1|, x_1)[\E_J(l,\nu,|\vec \kappa_2|,x_2)]^* \;,
\end{equation}
with $\E_{\TQQ}$ and $\E_J$ respectively being the LO heavy-tetraquarks and light-jet off-shell emission functions, see Eqs.\eref{LOTIF} and\eref{LOJIF}.

Then, a proper comparison between high-energy resummed and DGLAP predictions involves the evaluation of observables using both our hybrid factorization and pure fixed-order computations.
Unfortunately, as of our current knowledge, there is no available numerical code for studying fixed-order distributions in inclusive semi-hard hadron-plus-jet hadroproductions at NLO. 
Therefore, to assess the impact of the high-energy resummation on top of DGLAP predictions, we will compare our BFKL-based results with corresponding ones calculated using a high-energy fixed-order treatment. 
This approach, originally developed for light di-jet~\cite{Celiberto:2015yba,Celiberto:2015mpa} and hadron-jet~\cite{Celiberto:2020wpk} azimuthal correlations, involves truncating the high-energy series up to NLO accuracy. 
This truncation allows us to mimic the high-energy signal from a pure NLO calculation.
From a pragmatic point of view, we cut the expansion of azimuthal coefficients in Eq.~(\ref{Cl_NLLp_MSb}) up to ${\cal O}(\alpha_s^3)$.
This gives us an effective high-energy fixed-order ($\HENLOp$) expression that is suitable for our phenomenological study.

The $\MSb$ expression of our azimuthal coefficients in the $\HENLOp$ case is
\begin{align}
\label{Cn_HENLOp_MSb}
 \ClHENLOp &= 
 \frac{e^{\DY}}{s} 
 \int_{-\infty}^{+\infty} \drv \nu \, 
 \alpha_s^2(\mu_R) \,
 \left[ 1 + {\cal G}_{\rm NLL}^{(0)}(\DY,l,\nu,\mu_R) \right]
 \\ \nonumber
 &\times
 \E_{\TQQ}^{\rm NLO}(l,\nu,|\vec \kappa_1|, x_1)[\E_J^{\rm NLO}(l,\nu,|\vec \kappa_2|,x_2)]^* \;,
\end{align}
where
\begin{equation}
\label{G_BFKL_0}
 {\cal G}_{\rm NLL}^{(0)}(\DY,l,\nu,\mu_R) = \bar \alpha_s(\mu_R) \DY \chi(l,\nu)
\end{equation}
is first term of the expansion in $\alpha_s$ of the BFKL kernel.

We set the factorization scale ($\mu_F$) and renormalization one ($\mu_R$) to \emph{natural} energies provided by the given final state. 
We choose $\mu_F = \mu_R = \mu_{\cal N}$, with $\mu_{\cal N} = M_{\TQQ \perp} + |\vec \kappa_2|$ taken as the process-natural reference scale. 
Here $M_{\TQQ \perp} = \sqrt{M_{\TQQ}^2 + |\vec \kappa_1|^2}$ denotes the transverse mass of the tetraquark, and we set $M_{\TQQ} = 4 m_Q$.
To be precise, any energy scale with an order of magnitude similar to the typical energy of the emitted particle can be seen, in principle, as a \emph{natural} scale. 
In our case, the emission of two particles leads to two potential scales: $M_{\TQQ \perp}$ for the tetraquark and $|\vec \kappa_2|$ for the jet.
To ease the comparison with potential predictions from other approaches, we opt for a simplified choice that combines both scales into a single value, namely the sum of $m_{1 \perp}$ and $m_{2 \perp}$. 
This selection is in line with settings of many other numerical codes suited precision QCD phenomenology (see, for instance, Refs.~\cite{Alioli:2010xd,Campbell:2012am,Hamilton:2012rf}).
To gauge the size of MHOUs, $\mu_F$ and $\mu_R$ scales will be uniformly varied in the range $\mu_{\cal N}/2$ to $2\mu_{\cal N}$ through the $C_\mu$ parameter (see Section~\ref{sec:results}).

\section{Hadron-collider phenomenology}
\label{sec:results}

All numeric results presented here were computed \emph{via} the \textsc{Python}+\textsc{Fortran} {\Jethad} multimodular interface~\cite{Celiberto:2020wpk,Celiberto:2022rfj,Celiberto:2023fzz,Celiberto:2024mrq,Celiberto:2024swu}.
As for proton PDFs, we made use of the {\tt NNPDF4.0} NLO determination~\cite{NNPDF:2021uiq,NNPDF:2021njg} obtained from {\tt LHAPDF v6.5.4}~\cite{Buckley:2014ana}.
The sensitivity of our observables on MHOUs was assessed by letting $\mu_F$ and $\mu_R$ be around the \emph{natural} scale provided by kinematics, up to a factor ranging from 1/2 to two, controlled by the $C_\mu$ scale parameter.
Uncertainty bands entering plots combine the net effect of MHOUs and errors coming from multidimensional numeric integrations. The latters were kept uniformly below 1\% by the {\Jethad} integrators.

\subsection{Rapidity-interval distributions}
\label{ssec:DY}

\begin{figure*}[!t]
\centering

   \hspace{0.00cm}
   \includegraphics[scale=0.395,clip]{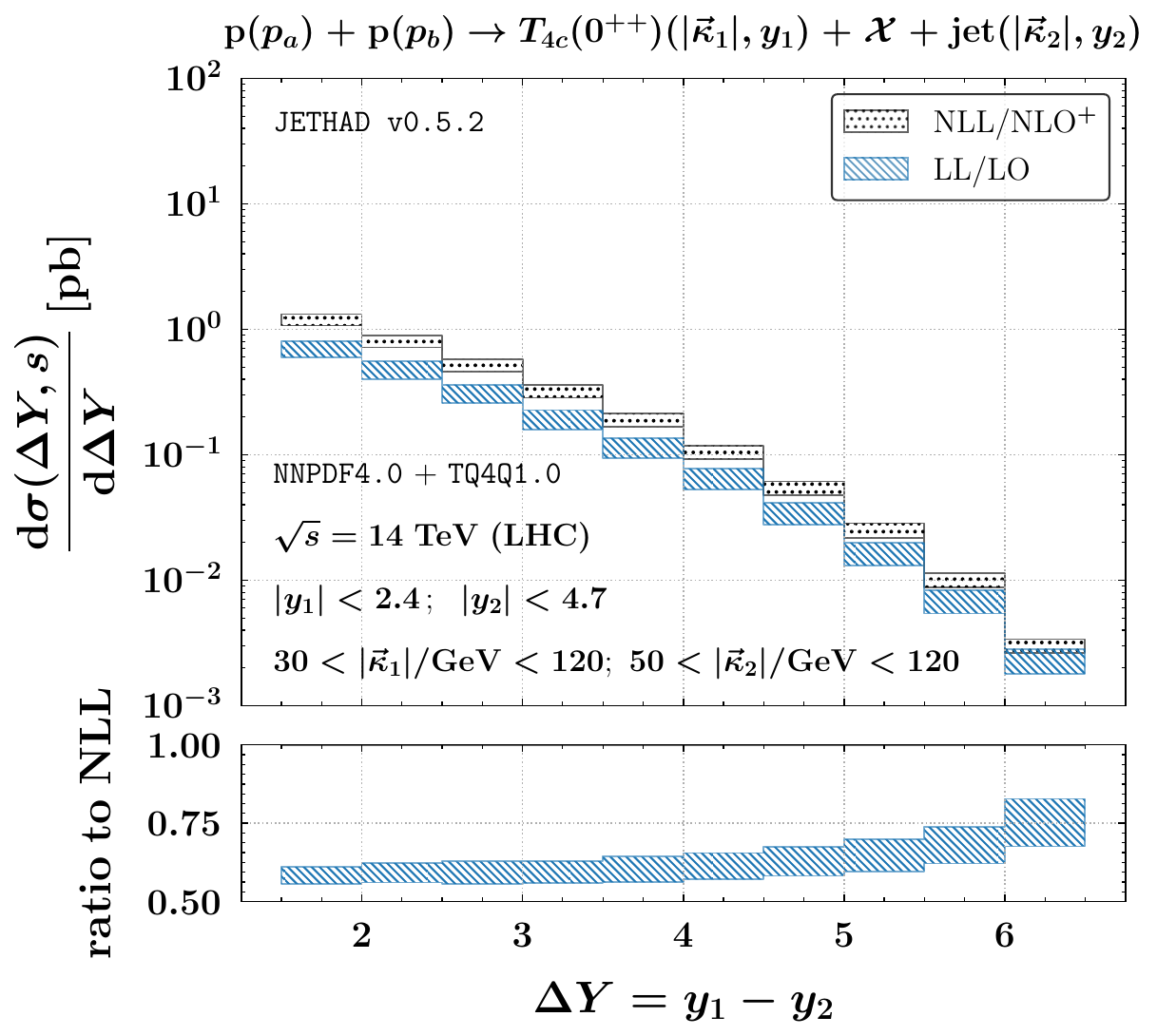}
   \hspace{-0.00cm}
   \includegraphics[scale=0.395,clip]{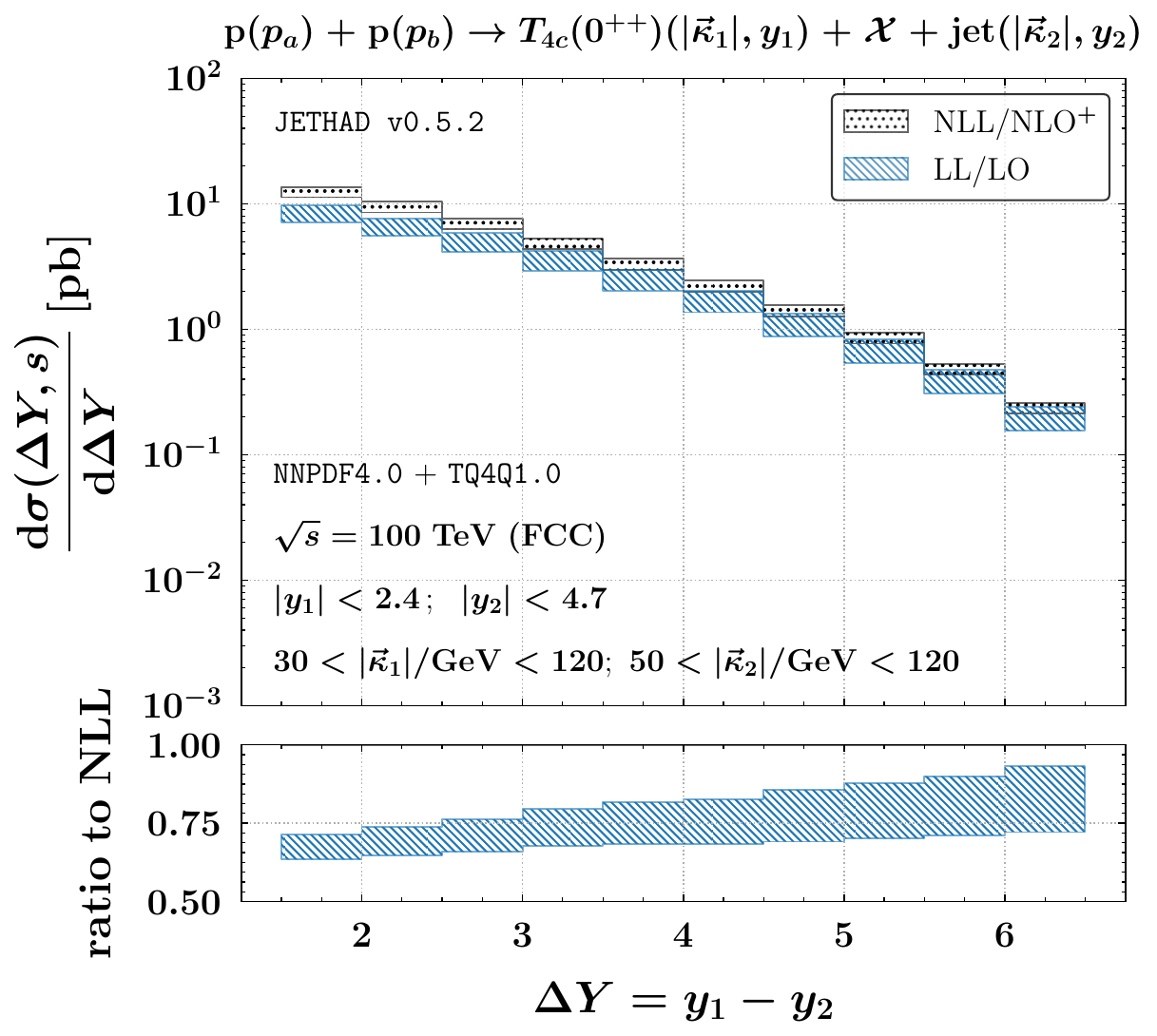}

   \vspace{0.50cm}

   \includegraphics[scale=0.395,clip]{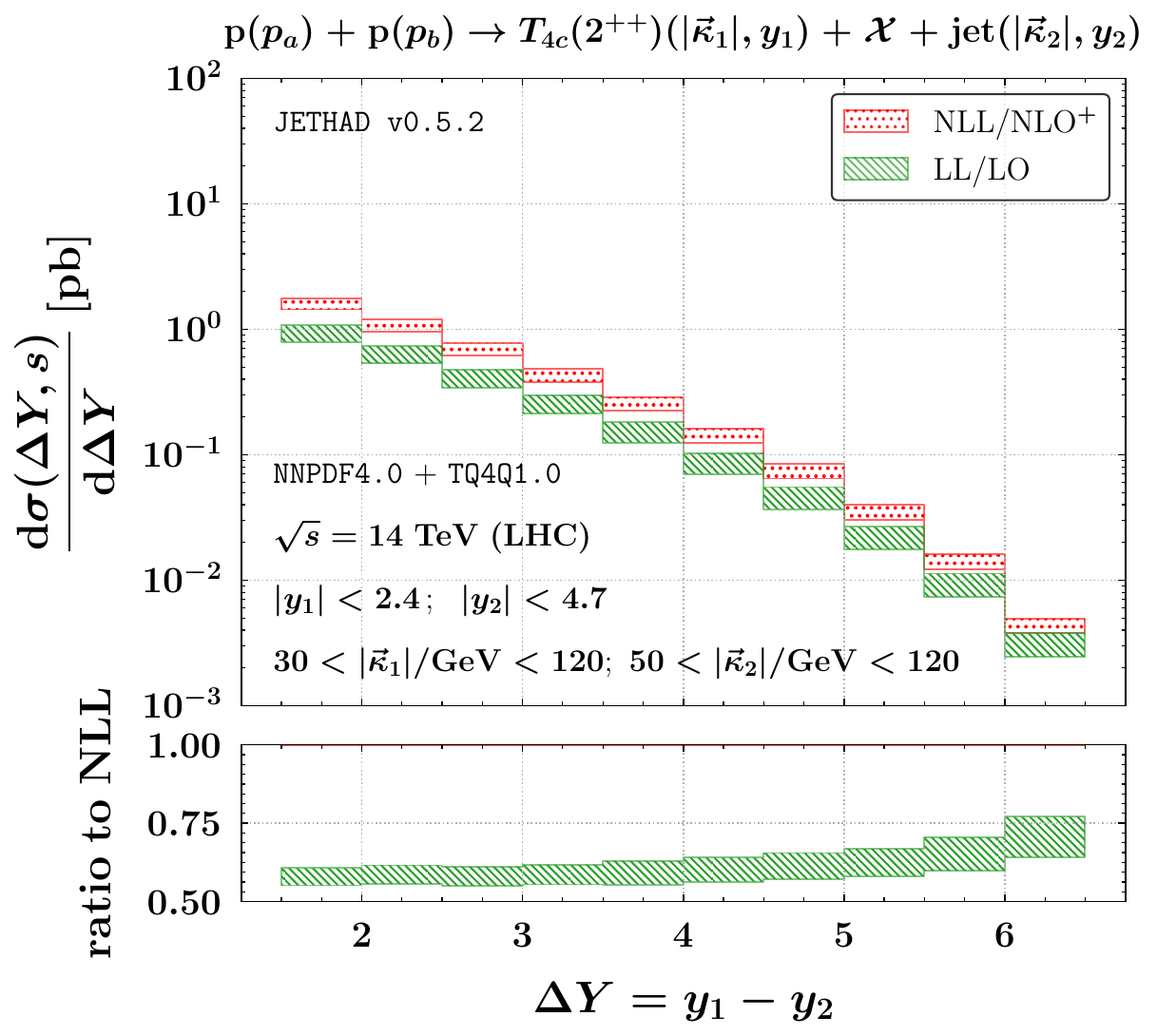}
   \hspace{-0.00cm}
   \includegraphics[scale=0.395,clip]{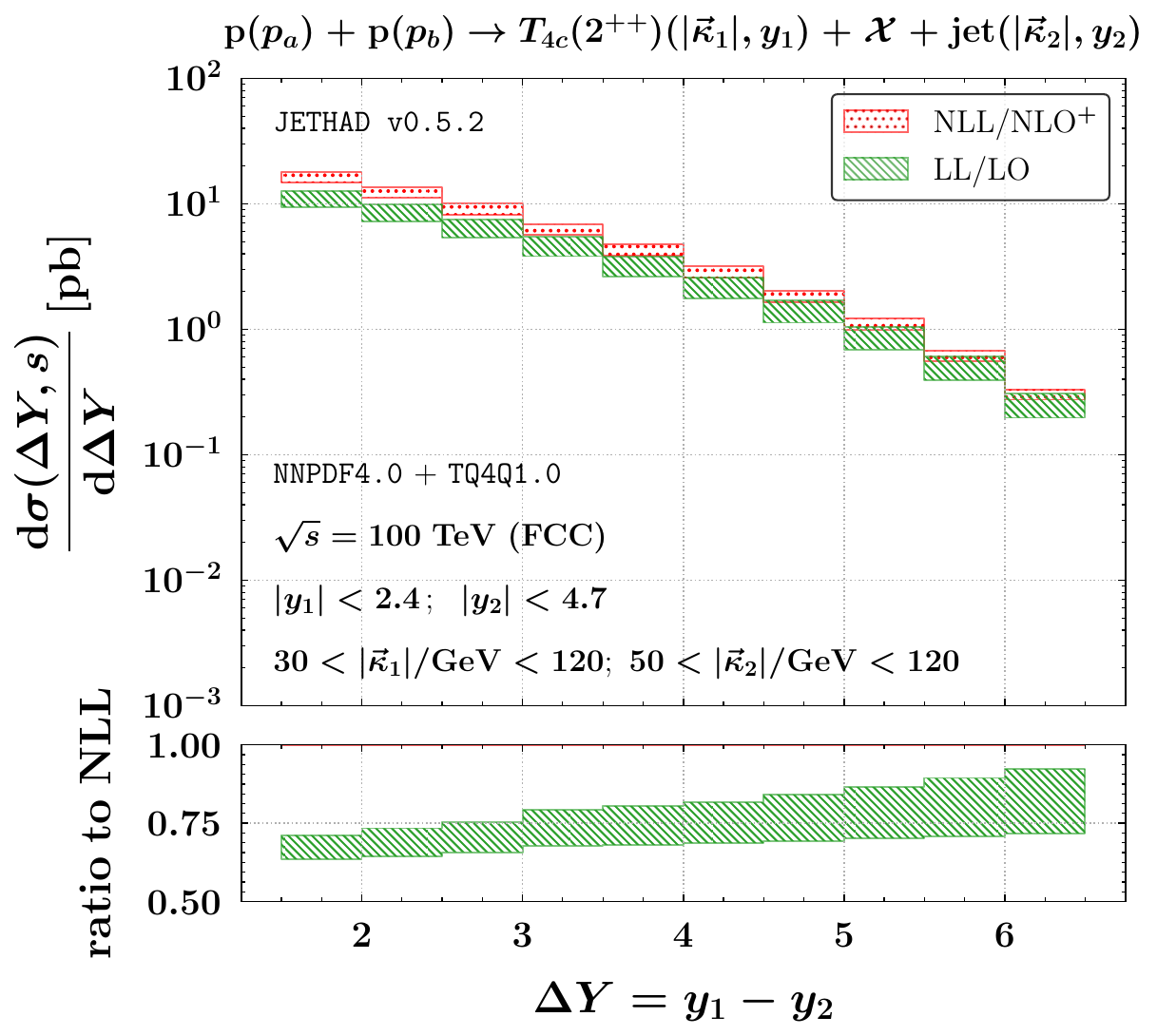}

\caption{Rapidity-interval distributions for $\TQcZpp$ (upper) or $\TQcTpp$ (lower) plus jet detections at $\sqrt{s} = 14$ TeV (LHC, left) or $100$ TeV (nominal FCC, right).
Ancillary panels below primary plots show the ratio of the pure $\LL$ rate to the $\NLLp$ case.
Shaded bands embody the combined effect of MHOUs and phase-space numeric multidimensional integration.}
\label{fig:I}
\end{figure*}

The first observable falling in the viewfinder of our investigations is the rapidity-interval distribution.
It genuinely corresponds to the cross section differential in the rapidity distance, $\DY \equiv y_1 - y_2$, between the tetraquark and the jet. 
One has
\begin{equation}
 \label{DY_distribution}
 \frac{\drv \sigma(\DY, s)}{\drv \DY} =
 \int_{y_1^{\rm min}}^{y_1^{\rm max}} \drv y_1
 \int_{y_2^{\rm min}}^{y_2^{\rm max}} \drv y_2
 \, \,
 \delta (\DY - (y_1 - y_2))
 \int_{|\vec \kappa_1|^{\rm min}}^{|\vec \kappa_1|^{\rm max}} 
 \!\!\drv |\vec \kappa_1|
 \int_{|\vec \kappa_2|^{\rm min}}^{|\vec \kappa_2|^{\rm max}} 
 \!\!\drv |\vec \kappa_2|
 \, \,
 {\cal C}_{l=0}^{\rm [order]}
 \;,
\end{equation}
where ${\cal C}_{l=0}$ is the first azimuthal coefficient (see Section~\ref{ssec:NLL_cross_section}).
Furthermore, the `${\rm [order]}$' label can be: $\NLLp$, $\LL$, or $\HENLOp$.
Transverse momenta of the $\TQc$ particle live in the range $30 < |\vec \kappa_1| /{\rm GeV} < 120$, while jet ones stay in the range $50 < |\vec \kappa_2| /{\rm GeV} < 120$. 
The latters slightly differ, but they are still compatible with current and forthcoming studies at the LHC~\cite{Khachatryan:2016udy}, say $50 < |\vec \kappa_2| / \rm{GeV} < 60$.

The adoption of \emph{asymmetric} transverse-momentum ranges proves beneficial for disentangling the high-energy signal from the fixed-order one~\cite{Celiberto:2015yba,Celiberto:2015mpa,Celiberto:2020wpk}. 
This choice not only mitigates large Sudakov logarithms arising from nearly back-to-back configurations, which would necessitate an additional resummation~\cite{Mueller:2013wwa,Marzani:2015oyb,Mueller:2015ael,Xiao:2018esv,Hatta:2020bgy,Hatta:2021jcd}, but also alleviates instabilities associated with radiative corrections~\cite{Andersen:2001kta,Fontannaz:2001nq}. 
Additionally, it suppresses violations of the energy-momentum conservation relation~\cite{Ducloue:2014koa}.
Our selection aligns with the established criteria current LHC studies. 
Tetraquark detections, limited to the barrel calorimeter akin to the CMS experiment~\cite{Chatrchyan:2012xg}, are confined within the rapidity interval of $-2.4$ to $2.4$. 
In the case of jets, which are also traceable in the endcaps' regions~\cite{Khachatryan:2016udy}, we consider a broader rapidity span, ranging from $-4.7$ to $4.7$.

Our predictions for rapidity-interval distributions are presented in Fig.~\ref{fig:I}. Upper and lower panels refer to the production of a $\TQcZpp$ tetraquark and its $\TQcTpp$ excitation, respectively. The center-of-mass energy is fixed at $14$~TeV (left) or $100$~TeV (right).
To propose realistic setups for direct comparisons with future experimental data, we adopt $\DY$-bins of a uniform size, their length being fixed at 0.5.
Ancillary panels below primary plots show the ratio between pure $\LL$ predictions and $\NLLp$ results.
Statistics ranges from $10^{-3}$~pb to around $2 \times 10^2$~pb.
As expected, it is substantially lower than the one found for heavy-light $\XQq$ tetraquarks plus jet emissions (roughly five orders of magnitude, see left plot of Fig.~3 of Ref.~\cite{Celiberto:2023rzw}), but still very promising.
Overall, rates grow by roughly one order of magnitude when $\sqrt{s}$ goes from typical LHC energies to nominal FCC ones.

We note that all $\DY$-distributions follow a common trend: they fall down as $\DY$ increases.
This comes out as the net effect of two competing aspects.
On the one side, the partonic hard factor would increase with energy (and thus with $\DY$), as predicted by the high-energy resummation. 
On the other side, this growth is significantly suppressed because of the collinear convolution with PDFs and FFs in the emission functions (see Eqs.~\eqref{LOTIF} and~\eqref{LOJIF}).

The main outcome emerging from the inspection of our plots is twofold.
We notice a solid stability under MHOUs, with the uncertainty bands fairly staying below $1.5$ relative size, but also under NLL corrections, with the $\NLLp$ bands being everywhere smaller than the $\LL$ ones and becoming closer and closer to them, up to be partially nested in the large-$\DY$ region.
This feature confirms what has been already observed in studies on heavy-light tetraquarks~\cite{Celiberto:2023rzw} and corroborates the statement that tetraquark production from single-parton fragmentation acts a quit stable channel to access high-energy QCD dynamics.

\subsection{Transverse-momentum rates}
\label{ssec:pT}

In this Section we investigate the high-energy behavior of two kinds of distributions.

The first observable matter of our interest here is the single transverse-momentum rate
\begin{equation}
\label{pT1_distribution}
 \frac{\drv \sigma(|\vec \kappa_1|, s)}{\drv |\vec \kappa_1|} =
 \int_{\DY^{\rm min}}^{\DY^{\rm max}} \drv \DY
 \int_{y_1^{\rm min}}^{y_1^{\rm max}} \drv y_1
 \int_{y_2^{\rm min}}^{y_2^{\rm max}} \drv y_2
 \, \,
 \delta (\DY - (y_1 - y_2))
 \int_{|\vec \kappa_2|^{\rm min}}^{|\vec \kappa_2|^{\rm max}} 
 \!\!\drv |\vec \kappa_2|
 \, \,
 {\cal C}_{l=0}^{\rm [order]}
 \;,
\end{equation}
namely the cross section differential in the $\TQc$ transverse momentum, but integrated over the 40~GeV~$< |\vec \kappa_2| <$~120~GeV jet transverse-momentum range and over a given $\DY$-bin.

This $|\vec \kappa_1|$-distribution permits us to sound common ground for possible connections between the $\NLLp$ factorization and other formalisms. 
Allowing $|\vec \kappa_1|$ to stay from $10$ to $100$~GeV enables the exploration of a wide kinematic sector, where other resummation mechanisms may become relevant and needed. 
Recent findings on high-energy Higgs~\cite{Celiberto:2020tmb} and heavy-jet~\cite{Bolognino:2021mrc} tags highlighted that our hybrid approach is valid in the moderate transverse-momentum regime, say when $|\vec \kappa_1|$ and $|\vec \kappa_2|$ are roughly of the same order.
Conversely, large $|\vec \kappa_1|$ and threshold logarithms are expected to become prominent in the soft ($|\vec \kappa_1| \ll |\vec \kappa_2|^{\rm min}$) and hard ($|\vec \kappa_1| > |\vec \kappa_2|^{\rm max}$) regimes, respectively.
Therefore, we propose the study of $|\vec \kappa_1|$-rates
as a reliability test for our approach, as well as a suitable way whereby setting the stage for the interplay among other resummations.

The second observable considered here is the double transverse momentum rate
\begin{equation}
\label{pT12_distribution}
 \frac{\drv \sigma(|\vec \kappa_1|, |\vec \kappa_2|, s)}{\drv |\vec \kappa_1| \, \drv |\vec \kappa_2|} =
 \int_{\DY^{\rm min}}^{\DY^{\rm max}} \drv \DY
 \int_{y_1^{\rm min}}^{y_1^{\rm max}} \drv y_1
 \int_{y_2^{\rm min}}^{y_2^{\rm max}} \drv y_2
 \, \,
 \delta (\DY - (y_1 - y_2))
 \, \,
 {\cal C}_{l=0}^{\rm [order]}
 \;,
\end{equation}
namely the cross section differential both in the $\TQc$ and the light-jet transverse momenta.
As for the previous observable, also this distribution is integrated over a given $\DY$-bin.

Here, when the separation between $|\vec \kappa_1|$ and $|\vec \kappa_2|$ widens, additional kinematic regimes, adjacent to the semi-hard sector, come into play (see recent studies on bottom-flavor~\cite{Celiberto:2021fdp,Celiberto:2024omj} and cascade-baryon~\cite{Celiberto:2022kxx} detections). 
For completeness, we mention that a joint resummation of transverse-momentum logarithms for two-particle distributions was first undertaken in the context of Higgs-plus-jet hadroproduction~\cite{Monni:2019yyr} through the {\RadISH} momentum-space method~\cite{Bizon:2017rah}.

We opt for a \emph{symmetric} configuration, setting $|\vec \kappa_1| = |\vec \kappa_2|$ and allowing them to range from $10$ to $120$~GeV. 
This almost back-to-back selection permits us to make stringent tests of the BFKL regime.
It also aims for a precise determination of the impact of NLL corrections.

\begin{figure*}[!t]
\centering

   \hspace{0.00cm}
   \includegraphics[scale=0.395,clip]{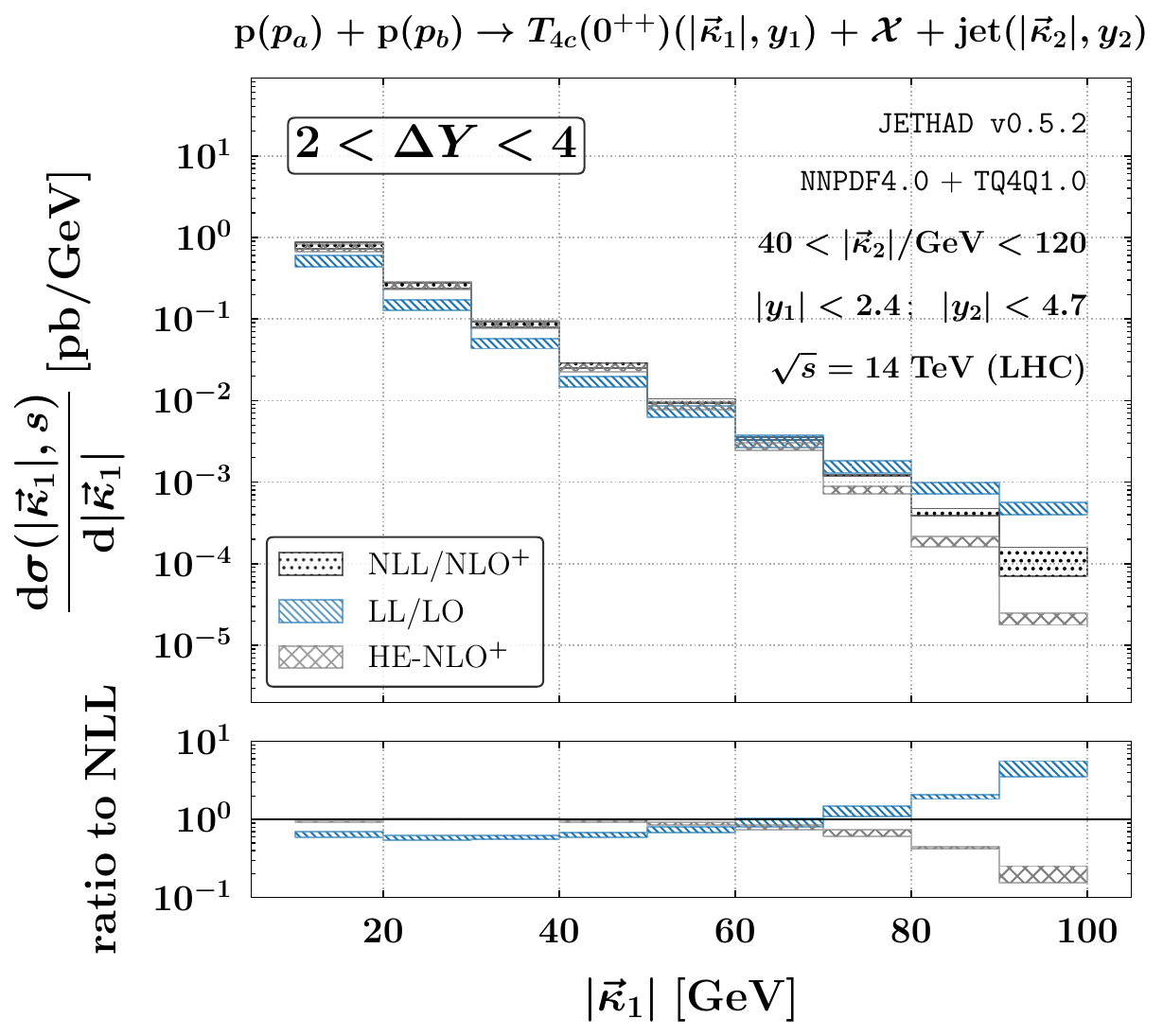}
   \hspace{-0.00cm}
   \includegraphics[scale=0.395,clip]{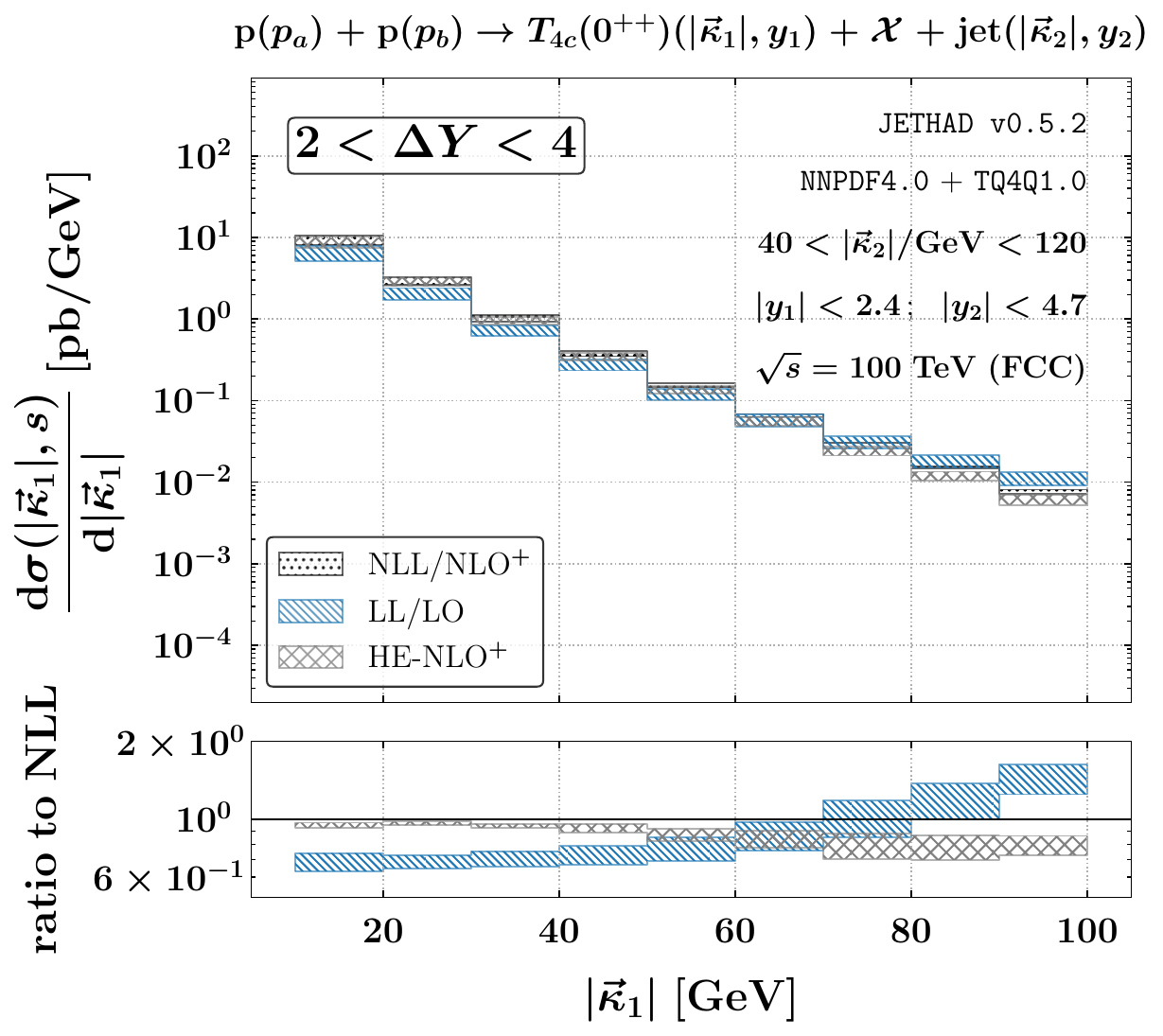}

   \vspace{0.50cm}

   \includegraphics[scale=0.395,clip]{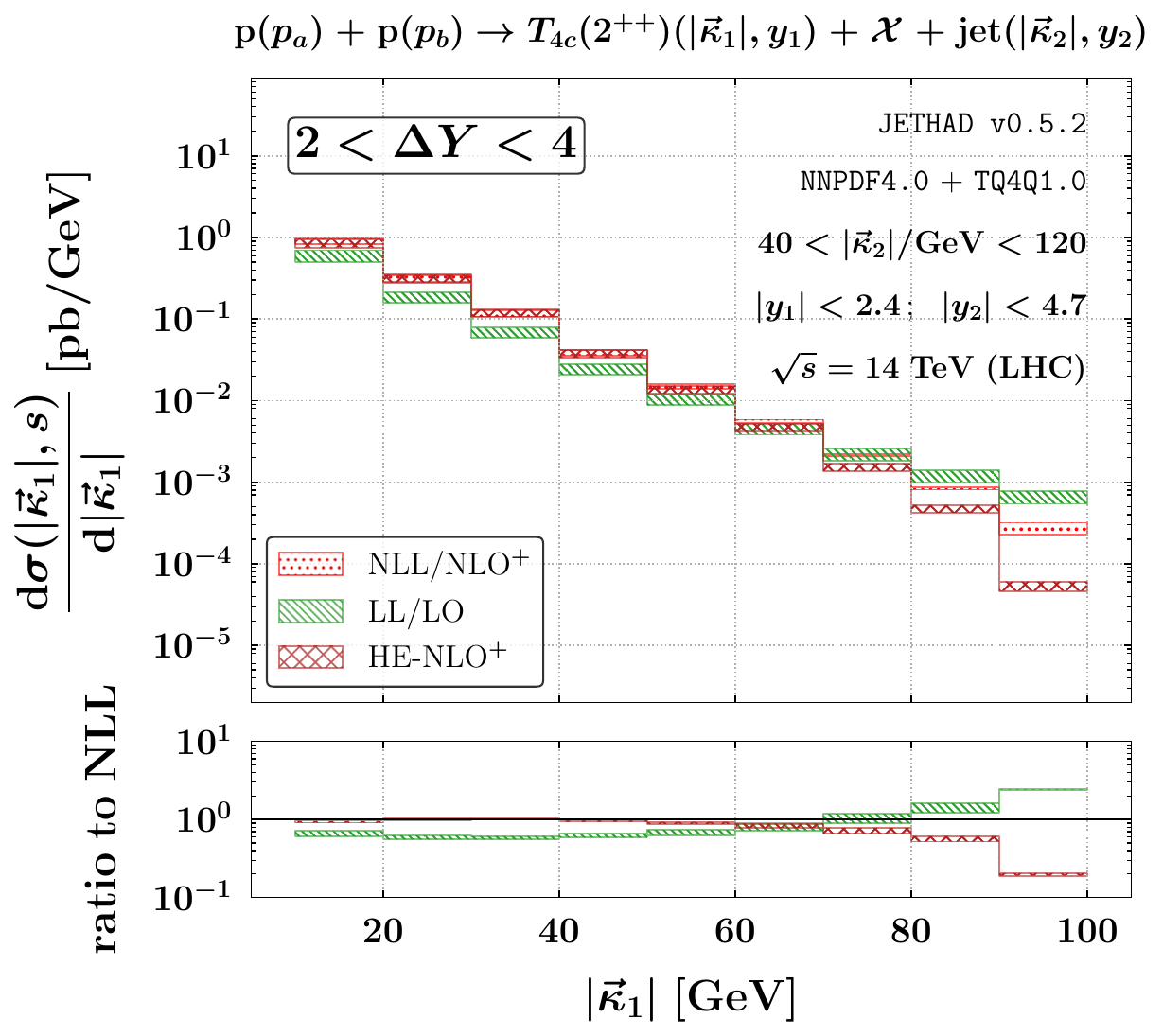}
   \hspace{-0.00cm}
   \includegraphics[scale=0.395,clip]{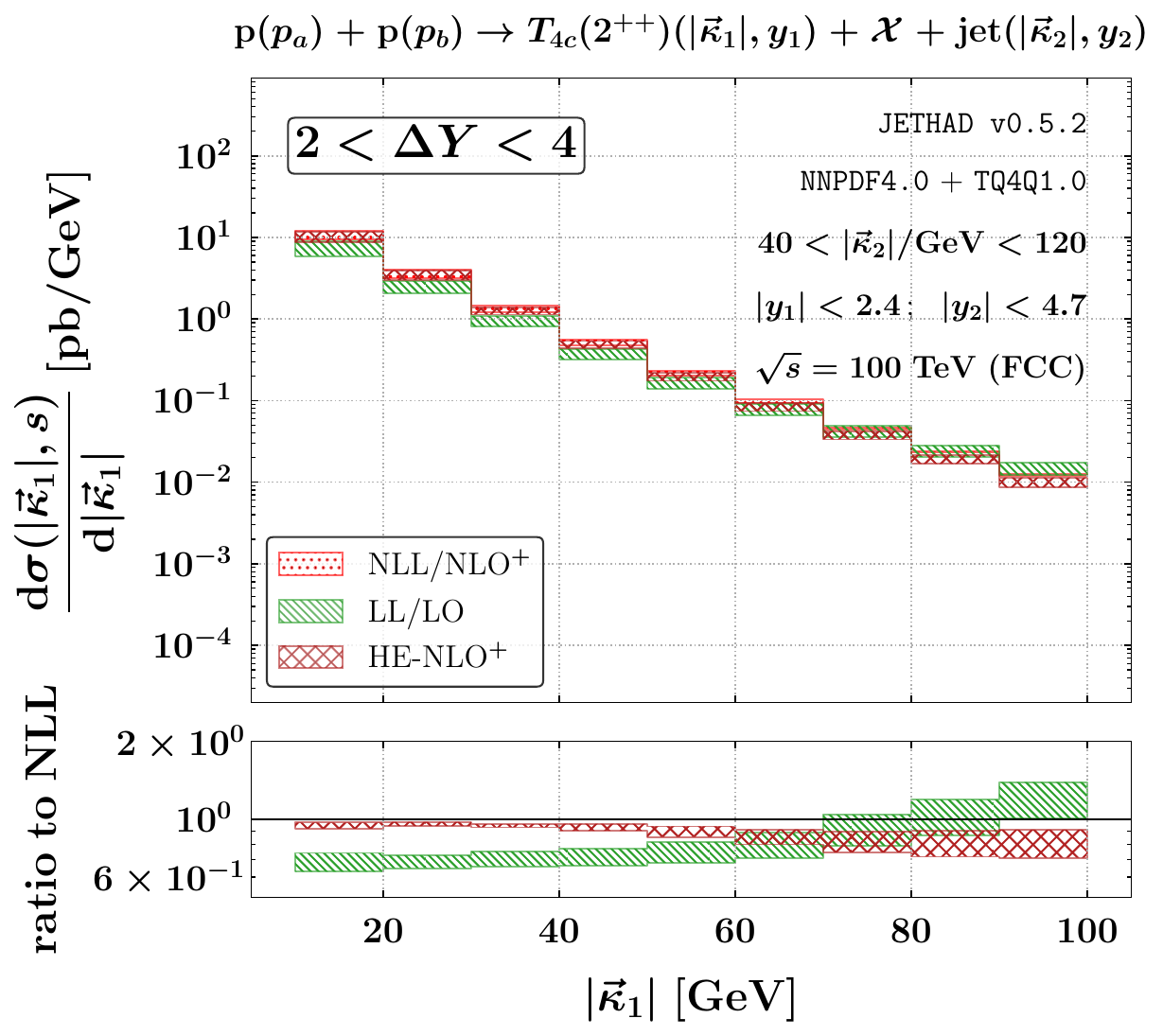}

\caption{Single transverse-momentum rate for $\TQcZpp$ (upper) or $\TQcTpp$ (lower) plus jet detections at $\sqrt{s} = 14$ TeV (LHC, left) or $100$ TeV (nominal FCC, right), and for $2 < \DY < 4$.
Ancillary panels below primary plots show the ratio of a given distribution to the $\NLLp$ case.
Shaded bands embody the combined effect of MHOUs and phase-space numeric multidimensional integration.}
\label{fig:I-k1b-S}
\end{figure*}

\begin{figure*}[!t]
\centering

   \hspace{0.00cm}
   \includegraphics[scale=0.395,clip]{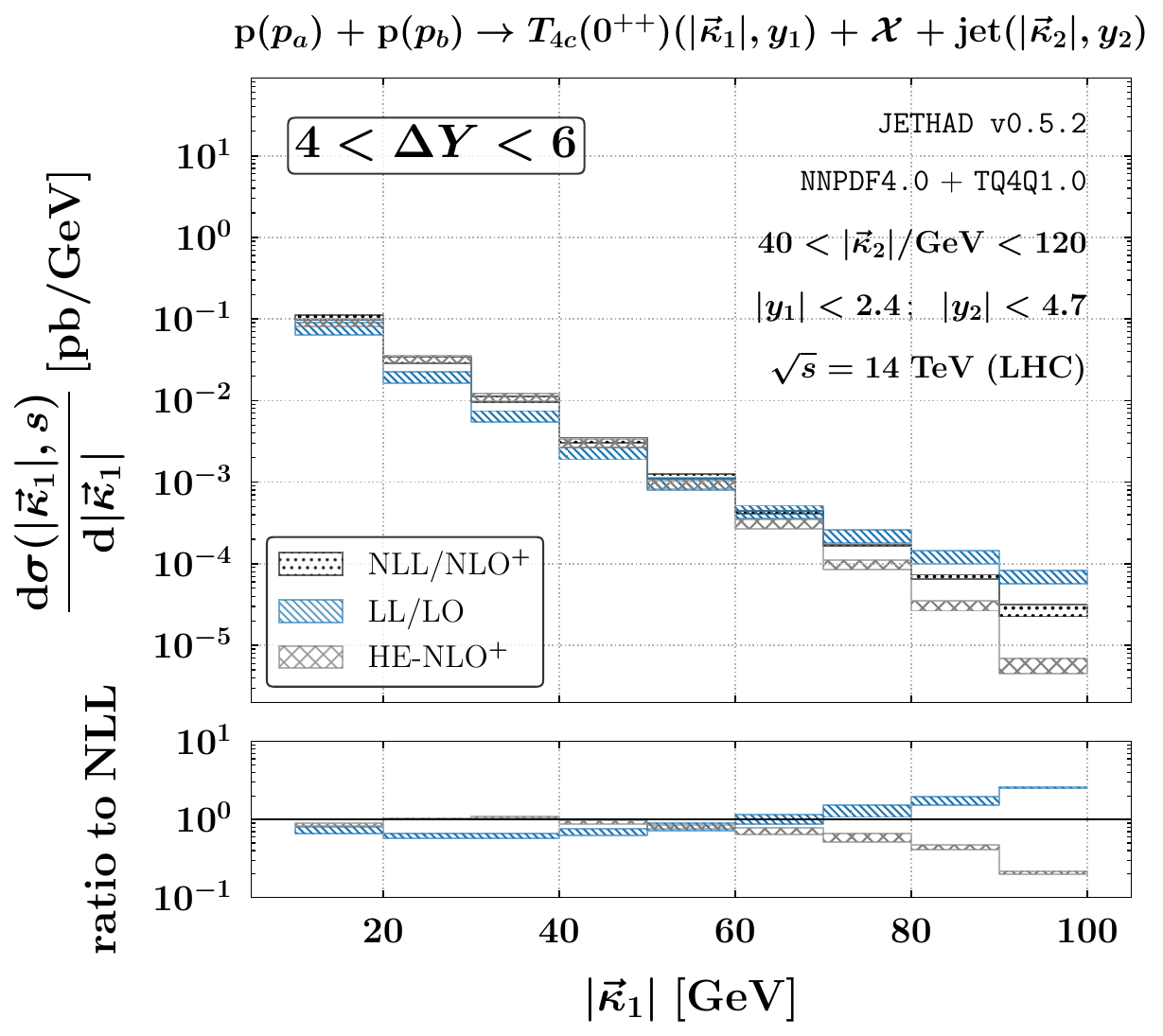}
   \hspace{-0.00cm}
   \includegraphics[scale=0.395,clip]{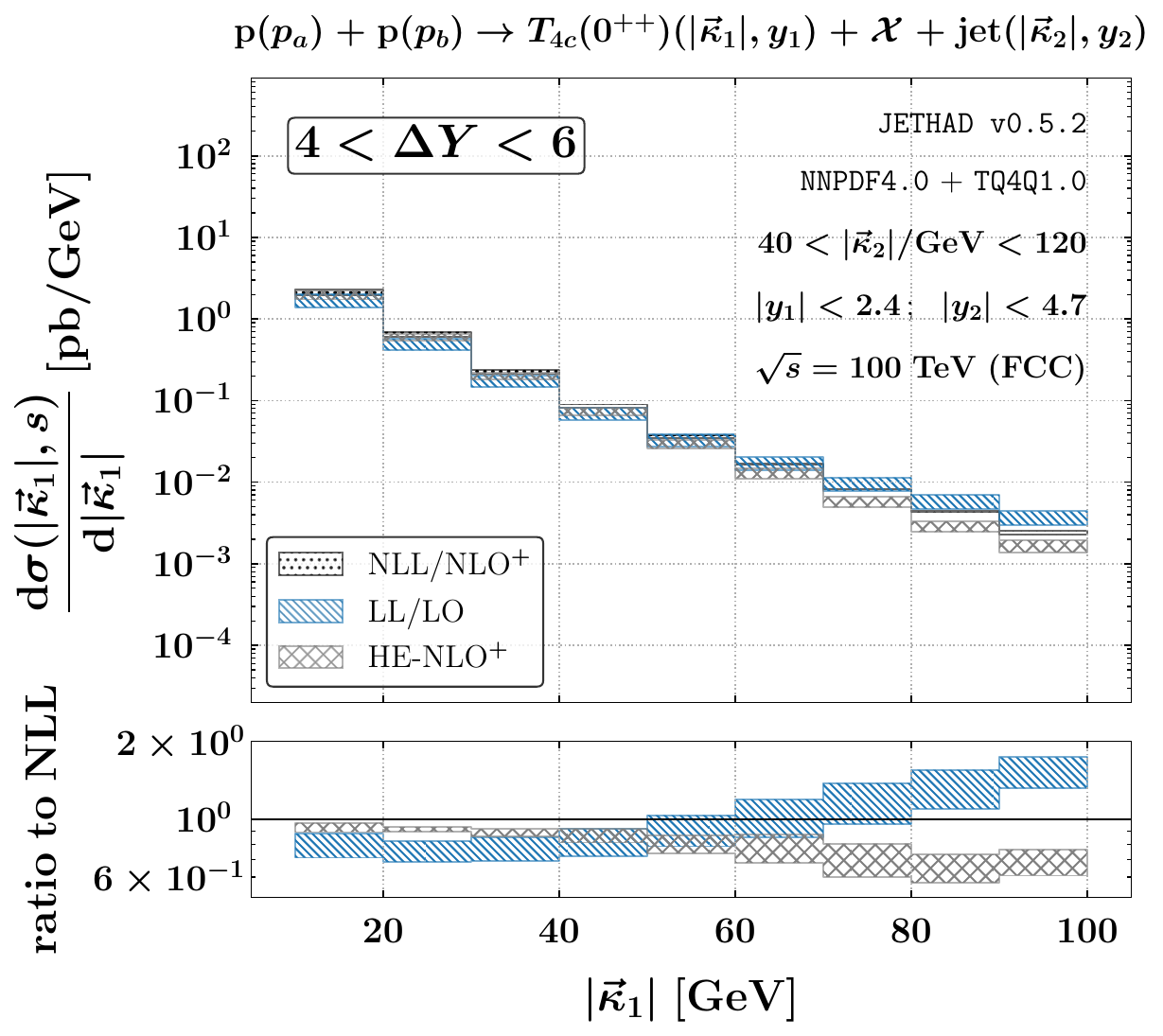}

   \vspace{0.50cm}

   \includegraphics[scale=0.395,clip]{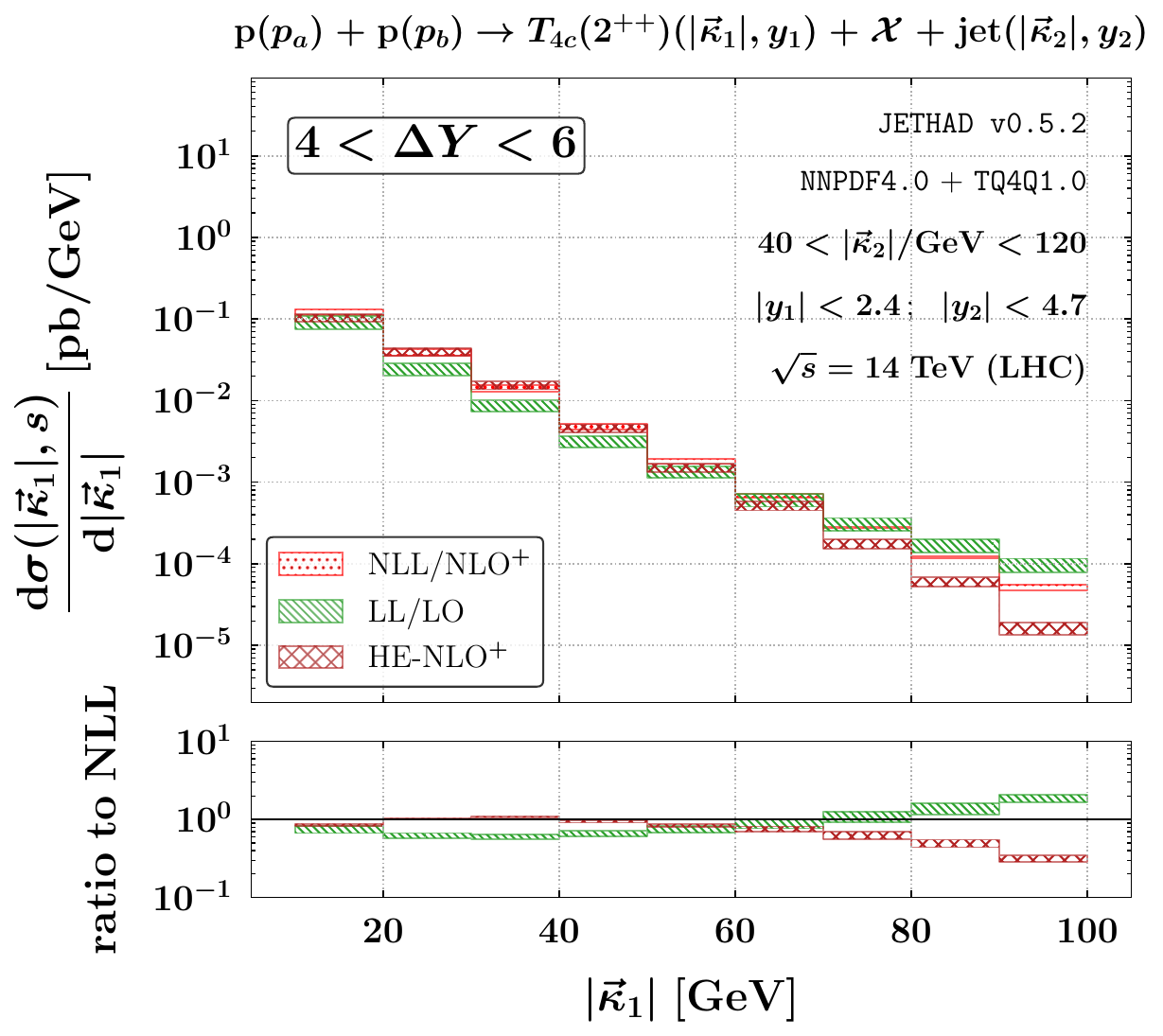}
   \hspace{-0.00cm}
   \includegraphics[scale=0.395,clip]{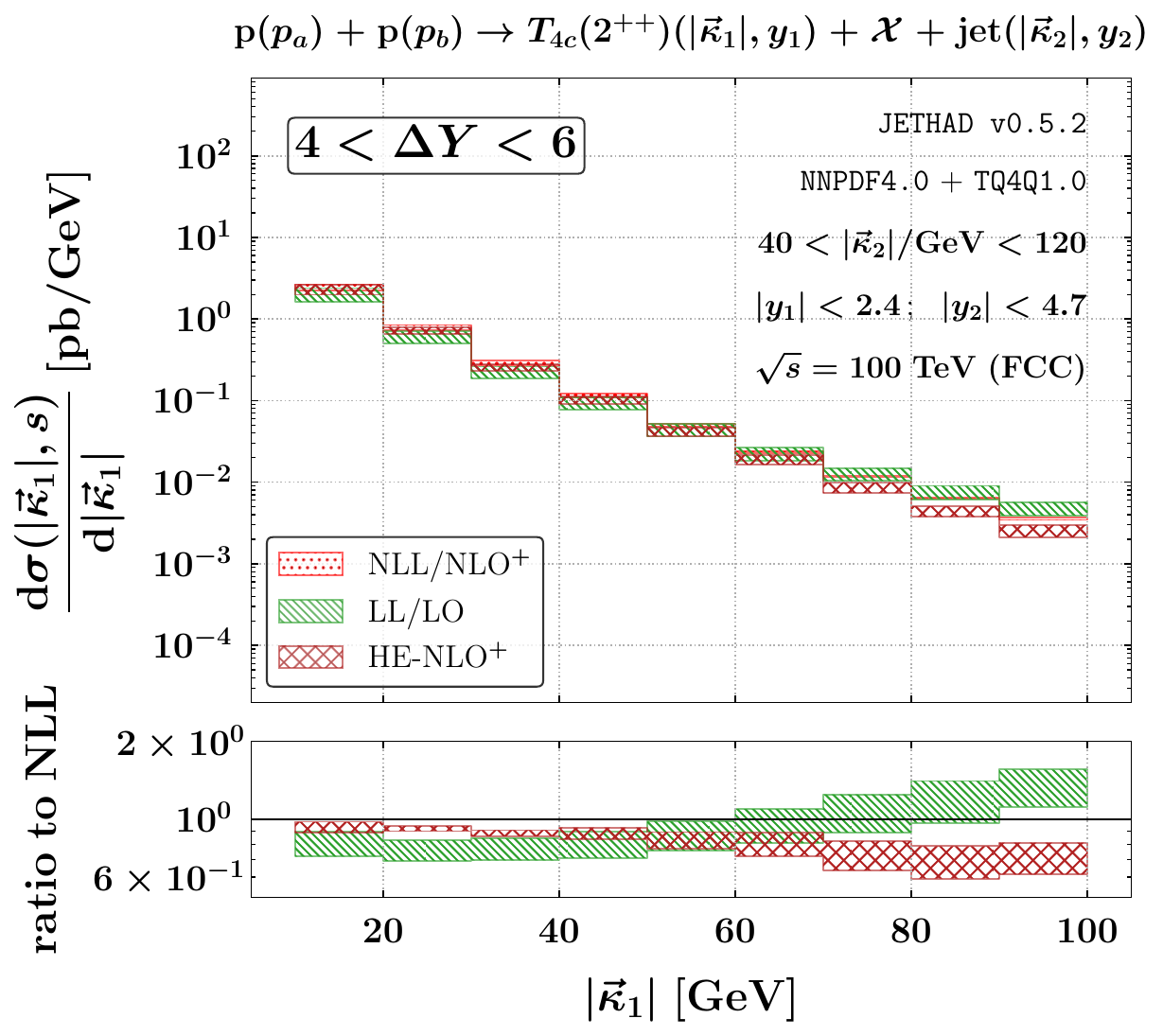}

\caption{Single transverse-momentum rate for $\TQcZpp$ (upper) or $\TQcTpp$ (lower) plus jet detections at $\sqrt{s} = 14$ TeV (LHC, left) or $100$ TeV (nominal FCC, right), and for $4 < \DY < 6$.
Ancillary panels below primary plots show the ratio of a given distribution to the $\NLLp$ case.
Shaded bands embody the combined effect of MHOUs and phase-space numeric multidimensional integration.}
\label{fig:I-k1b-M}
\end{figure*}

\begin{figure*}[!t]
\centering

   \hspace{0.00cm}
   \includegraphics[scale=0.395,clip]{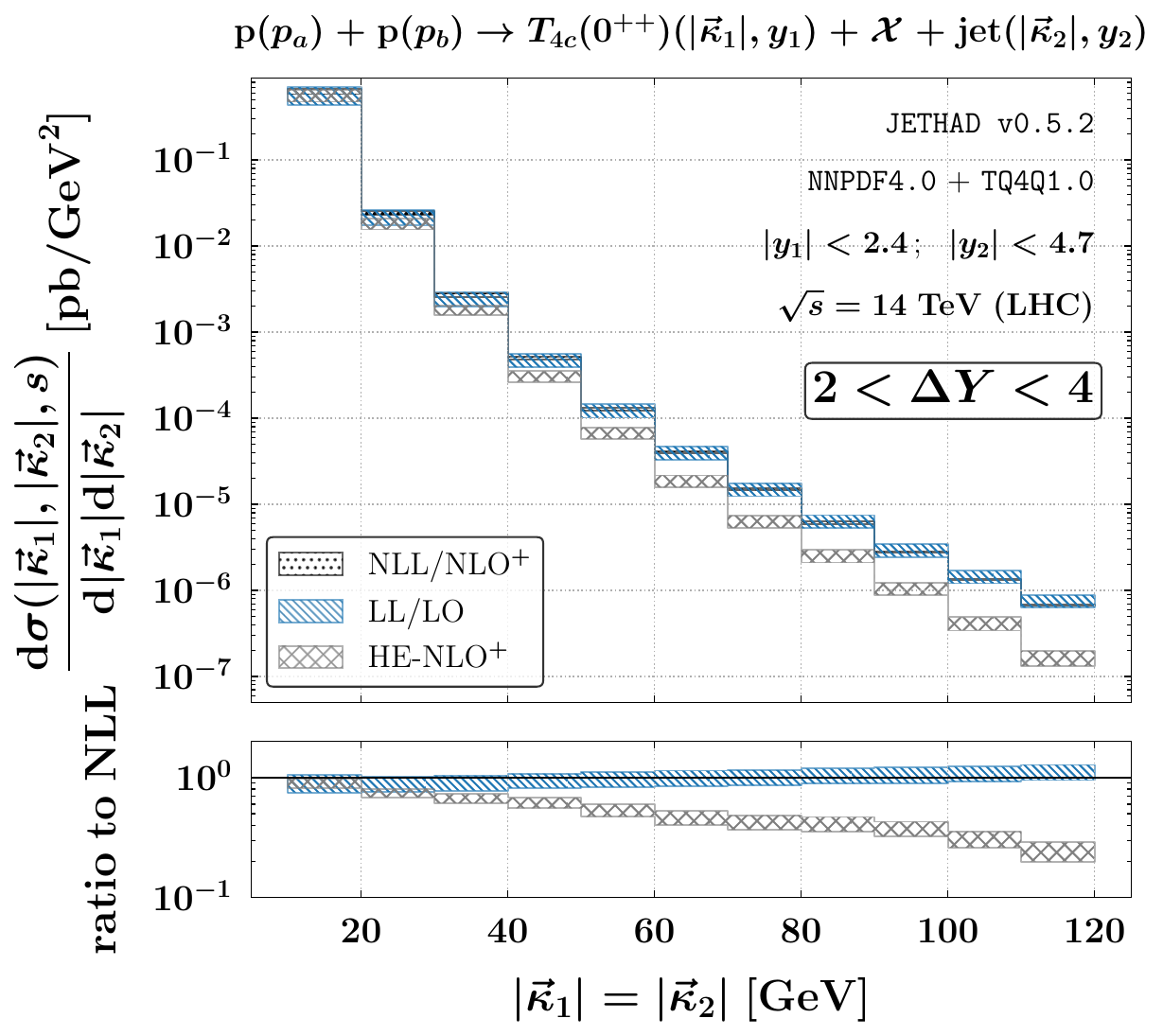}
   \hspace{-0.00cm}
   \includegraphics[scale=0.395,clip]{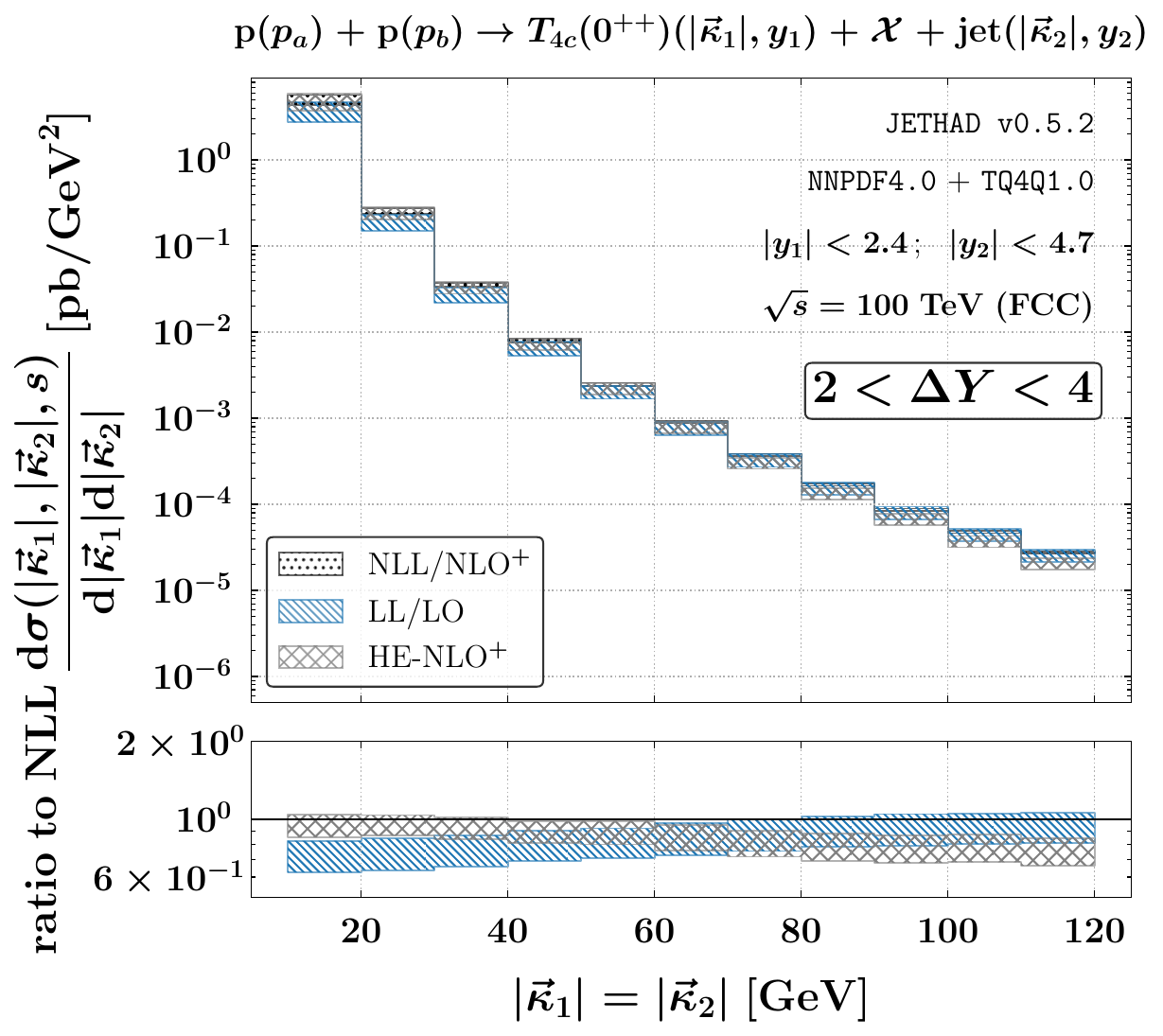}

   \vspace{0.50cm}

   \includegraphics[scale=0.395,clip]{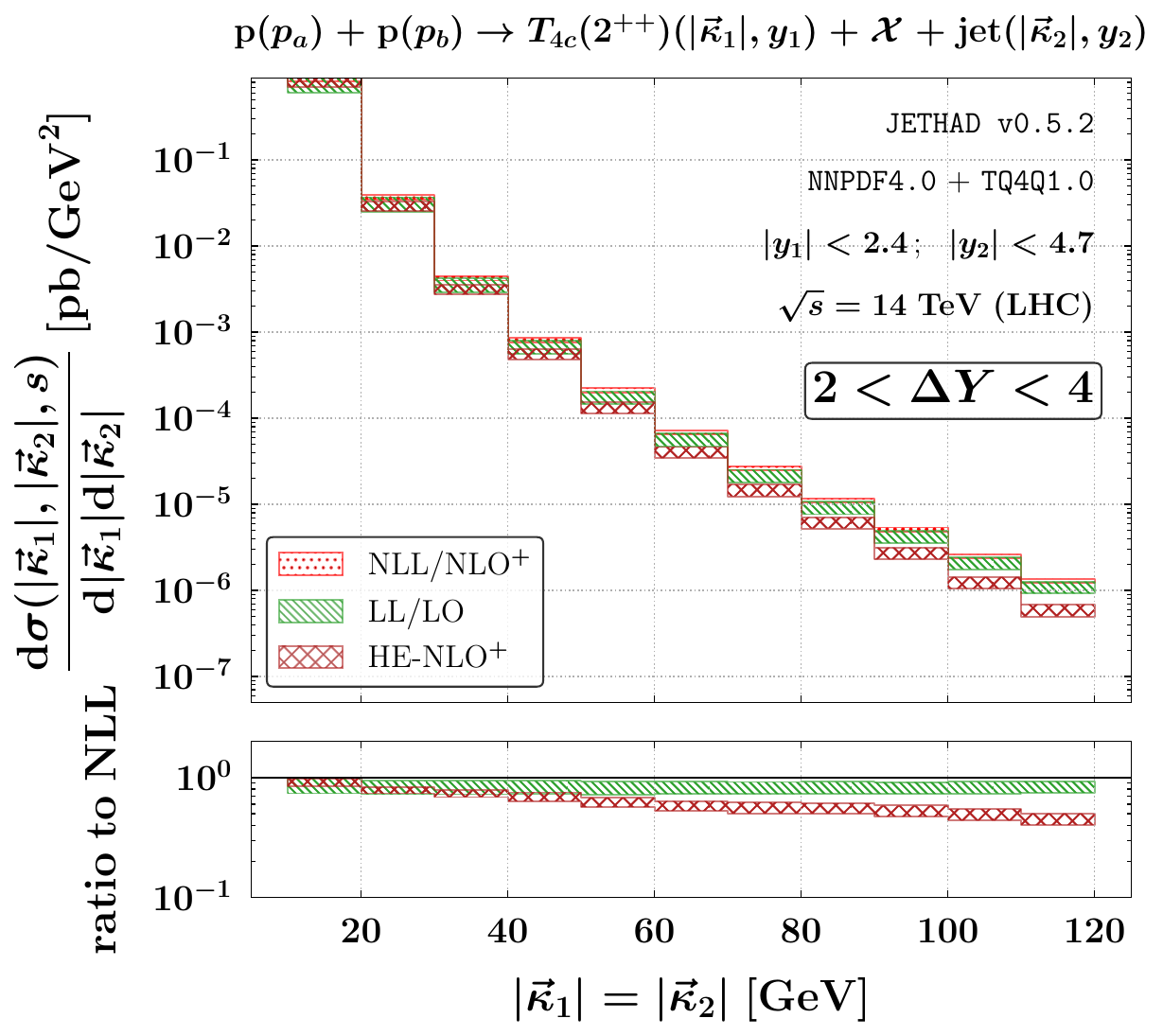}
   \hspace{-0.00cm}
   \includegraphics[scale=0.395,clip]{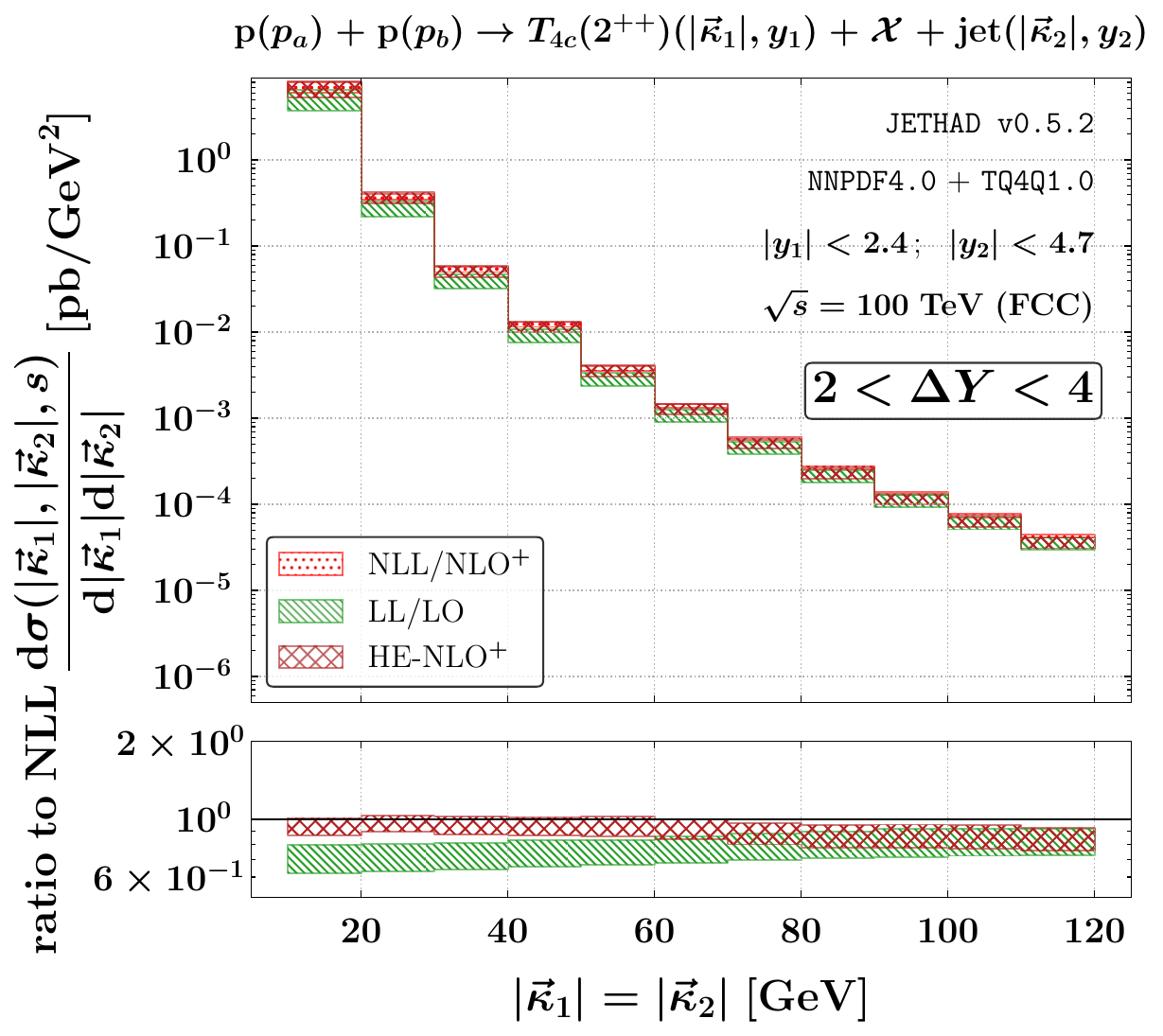}

\caption{Symmetric double transverse-momentum rate for $\TQcZpp$ (upper) or $\TQcTpp$ (lower) plus jet detections at $\sqrt{s} = 14$ TeV (LHC, left) or $100$ TeV (nominal FCC, right), and for $2 < \DY < 4$.
Ancillary panels below primary plots show the ratio of a given distribution to the $\NLLp$ case.
Shaded bands embody the combined effect of MHOUs and phase-space numeric multidimensional integration.}
\label{fig:I-k12b-S}
\end{figure*}

\begin{figure*}[!t]
\centering

   \hspace{0.00cm}
   \includegraphics[scale=0.395,clip]{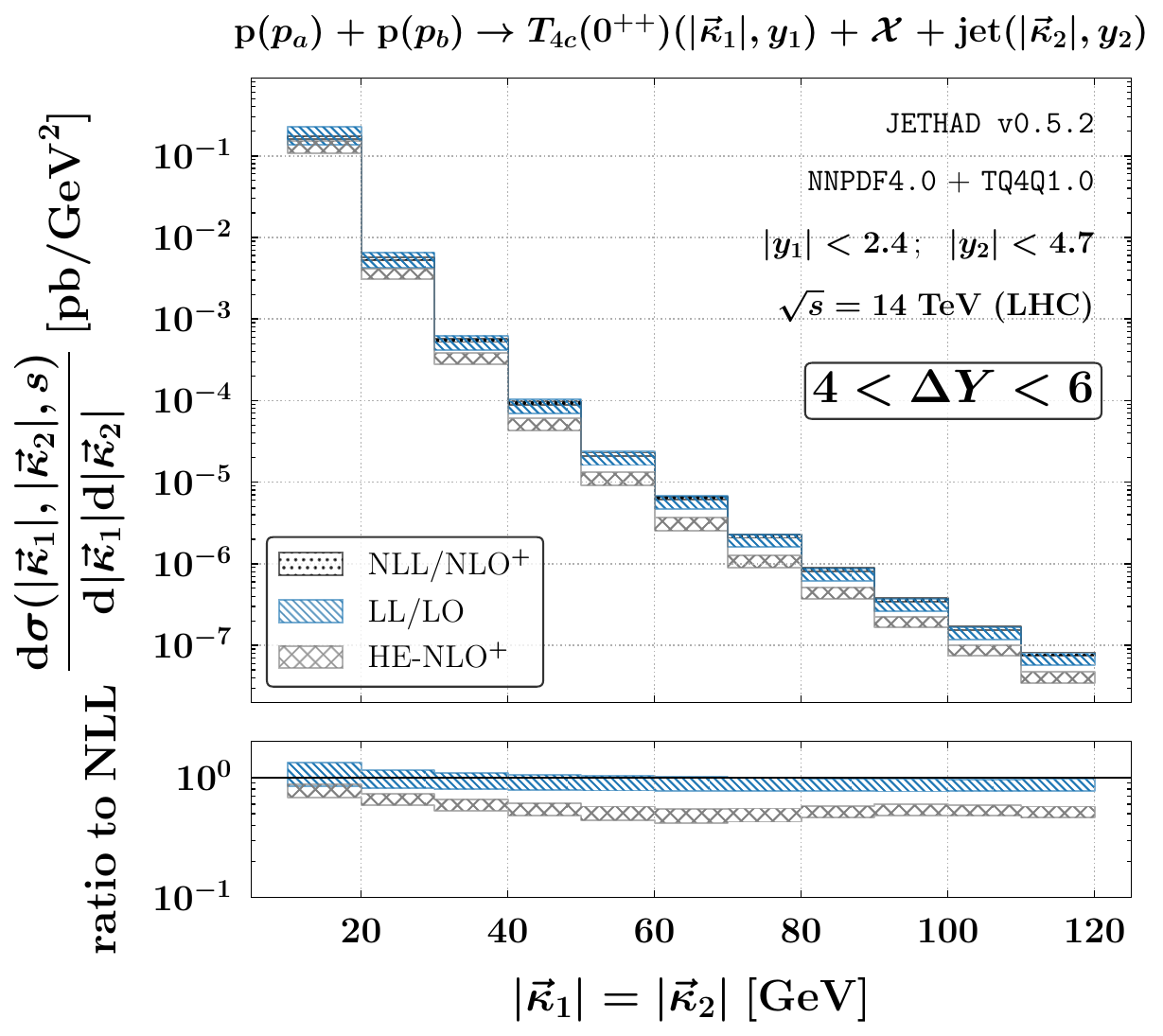}
   \hspace{-0.00cm}
   \includegraphics[scale=0.395,clip]{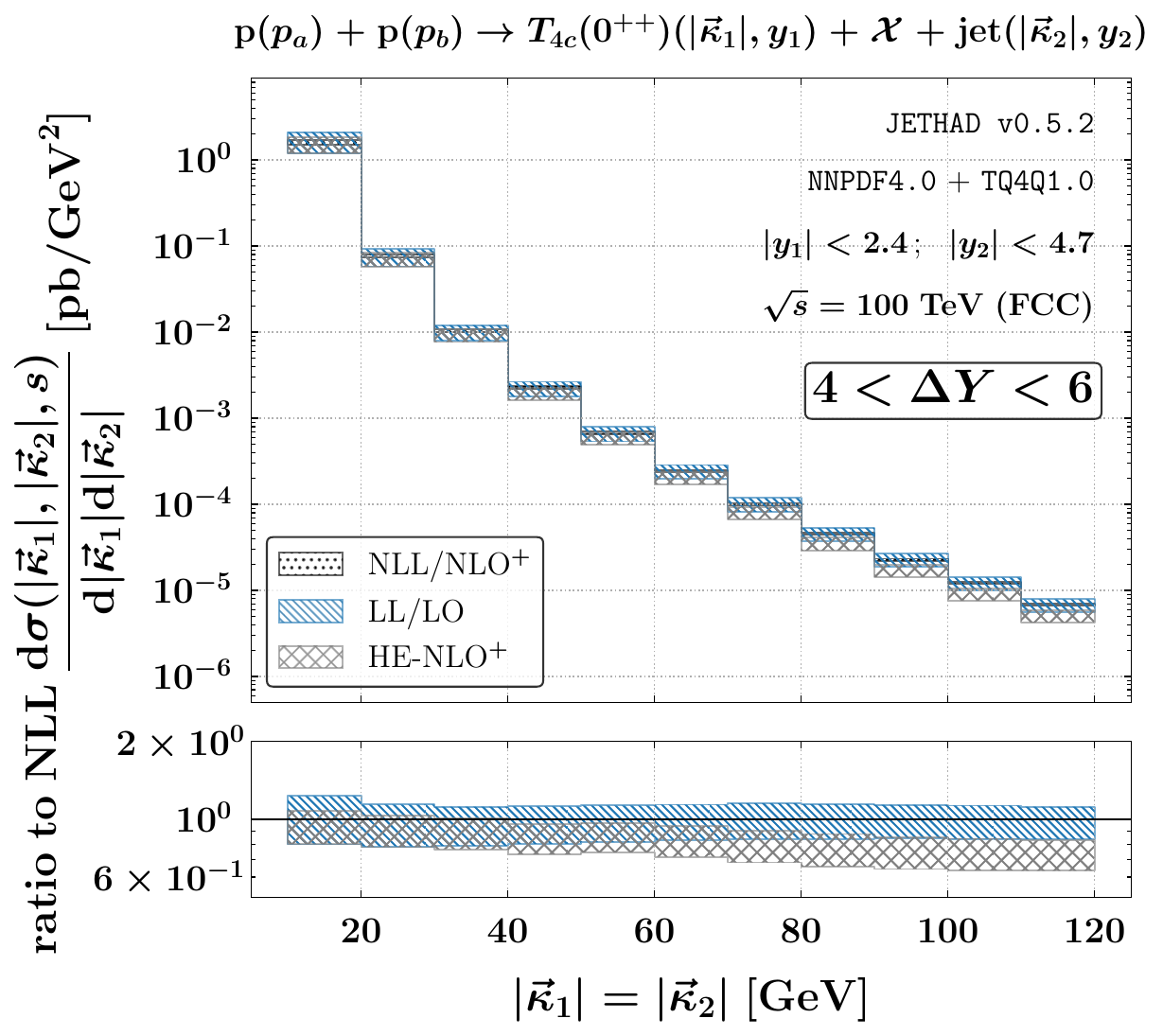}

   \vspace{0.50cm}

   \includegraphics[scale=0.395,clip]{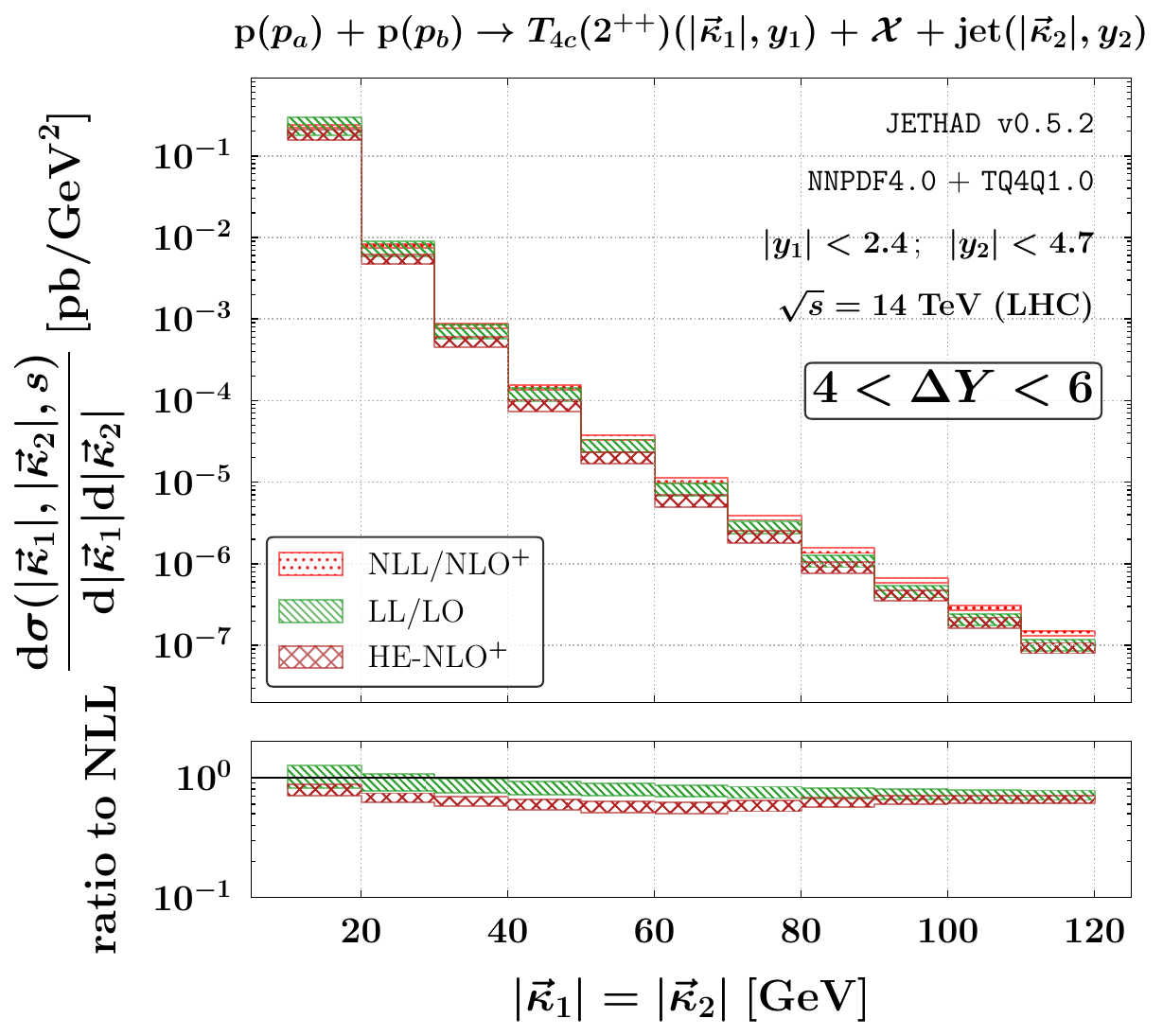}
   \hspace{-0.00cm}
   \includegraphics[scale=0.395,clip]{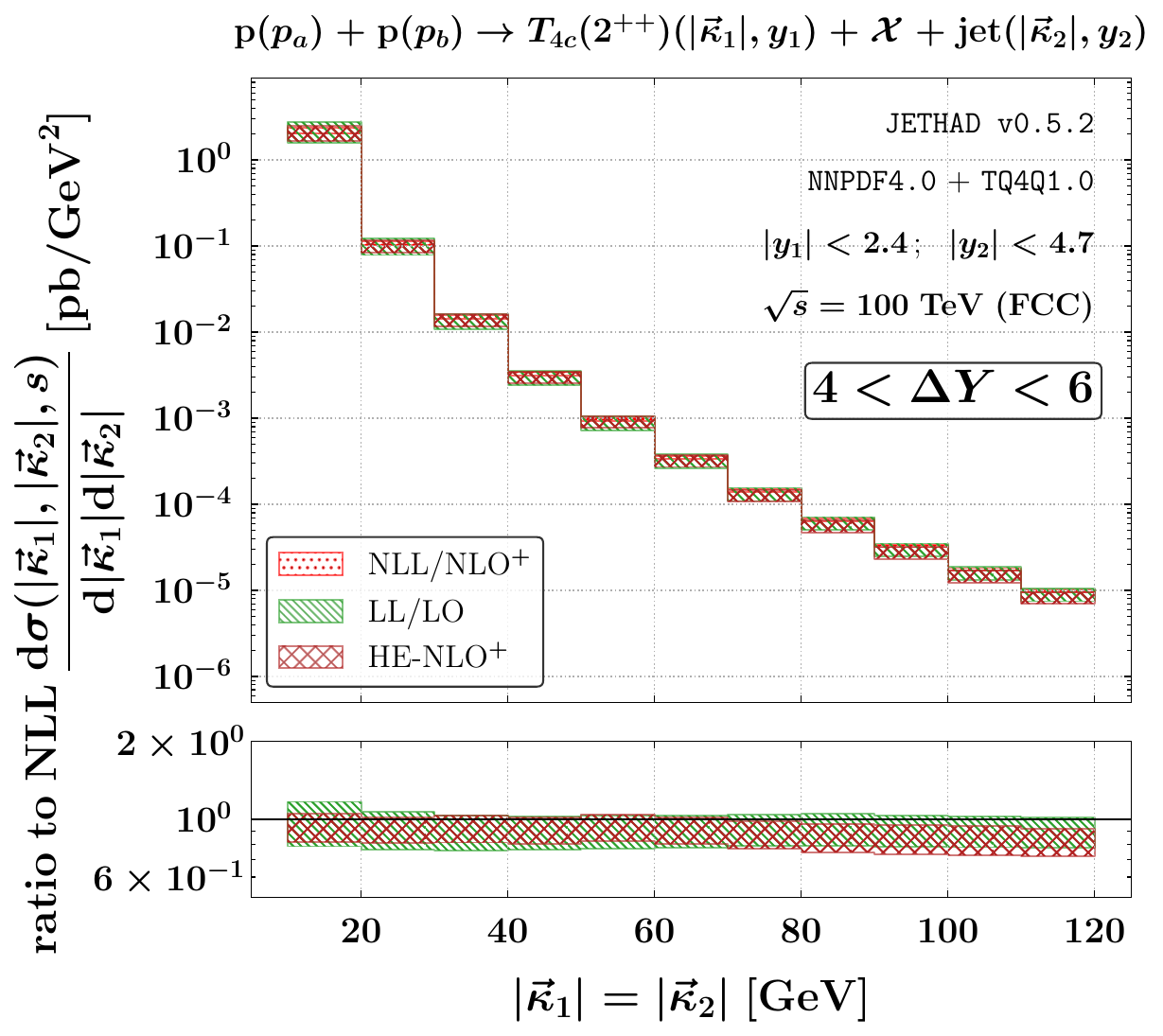}

\caption{Symmetric double transverse-momentum rate for $\TQcZpp$ (upper) or $\TQcTpp$ (lower) plus jet detections at $\sqrt{s} = 14$ TeV (LHC, left) or $100$ TeV (nominal FCC, right), and for $4 < \DY < 6$.
Ancillary panels below primary plots show the ratio of a given distribution to the $\NLLp$ case.
Shaded bands embody the combined effect of MHOUs and phase-space numeric multidimensional integration.}
\label{fig:I-k12b-M}
\end{figure*}

Upper (lower) plots of Fig.~\ref{fig:I-k1b-S} are for predictions of the single transverse-momentum rate for the $\TQcZpp$ ($\TQcTpp$) plus jet production at $14$~TeV~LHC (left) and $100$~TeV~FCC~(right), and with $\DY$ integrated in a ``lower'' bin, $2 < \DY < 4$.
To ease the comparison with future experimental data, we adopt transverse-momentum bins of a uniform size, their length being fixed at 10~GeV.
Plots of Fig.~\ref{fig:I-k1b-M} are organized in the same way of Fig.~\ref{fig:I-k1b-S}, but this time $\DY$ lies on an ``upper'' bin, $4 < \DY < 6$, contiguous to the previous one.
Then, plots in Figs.~\ref{fig:I-k12b-S} and~\ref{fig:I-k12b-M} correspond to the same configurations, but for the symmetric double transverse-momentum rate.
Ancillary panels below primary plots are for the ratio between $\LL$ or $\HENLOp$ cases and $\NLLp$ predictions.
The general behavior of both single and double transverse-momentum distributions is a sharp downtrend with $|\vec \kappa_1|$.
Results are very stable under energy-scale variations, with error bands featuring a 20\% width at most.

Concerning the single transverse-momentum rate, a very distinctive pattern comes out from the inspection of ancillary panels of Figs.~\ref{fig:I-k1b-S} and~\ref{fig:I-k1b-M}.
Here we note that the $\HENLOp$ over $\NLLp$ ratio stays almost everywhere below one, and it decreases with $\vec \kappa_1$. 
This falloff is less pronounced at nominal FCC energies ($100$~TeV, right plots) than at typical LHC ones ($14$~TeV, left plots).
Conversely, the $\LL$ over $\NLLp$ ratio exhibits an almost opposite trend.
It is smaller than one in the low-$|\vec \kappa_1|$ region, but it grows and grows as $|\vec \kappa_1|$ increases, up to reach $1.5$ to $2$ as a maximum  value.
Explaining these two features is not straightforward, since they encode a combination of multiple effects.

On the one hand, previous studies on other semi-hard two-particle hadroproductions have highlighted how the behavior of the NLL-resumed signal with respect to its NLO high-energy background for single transverse-momentum distributions generally depends on the considered process.
While the $\HENLOp$ over $\NLLp$ ratio is always larger than one in $\Xi^-/\bar\Xi^+$ baryon plus jet detections (see Fig.~7 of Ref.~\cite{Celiberto:2022kxx}), preliminary analyses on Higgs plus jet distribution within a partial NLL-to-NLO matched accuracy have shown a more intricate trend~\cite{Celiberto:2023dkr}.
Therefore, we must conclude that $\HENLOp$ over $\NLLp$ ratios smaller than one come out as a peculiar feature of $\TQc$ plus jet emissions.

On the other hand, also the behavior of the $\LL$ over $\NLLp$ ratio depends on a complicated interplay between distinct effects.
Indeed, it is known that NLO corrections to the jet emission function are negative~\cite{Bartels:2001ge,Ivanov:2012ms,Colferai:2015zfa}, whereas the hadron emission function has positive NLO corrections coming from the $C_{gg}$ perturbative coefficient function and negative from the other ones~\cite{Ivanov:2012iv}.
These two features could (partially) balance each other in some sectors of the transverse-momentum phase space. 
For the sake of qualitative comparison, we report that the $\LL$ over $\NLLp$ ratio is larger than one in the $\Xi^-/\bar\Xi^+$ plus jet case~\cite{Celiberto:2022kxx}, but its behavior is less pronounced in other processes, such as the heavy-light tetraquark plus jet reaction~\cite{Celiberto:2023rzw}.

We now present and discuss our results for the double transverse-momentum rate (see Figs.~\ref{fig:I-k1b-S} and~\ref{fig:I-k1b-M}).
Apart from the expected downtrend with $|\vec \kappa_1| \equiv |\vec \kappa_2|$, we observe that $\NLLp$ uncertainty bands are almost always overlapped with $\LL$ ones, if not completely nested for some bins of the transverse momentum.
This represents a clear signal of a reached stability of our hybrid factorization for this observable.
It is directly connected to the symmetric selection made for $|\vec \kappa_1|$ and $|\vec \kappa_2|$ ranges.
Although this symmetric configuration should not favor a clear disentanglement of the core high-energy signal from the fixed-order background (see a related discussion in Section~\ref{ssec:DY}), we surprisingly report that $\HENLOp$ bands visibly decouple from $\NLLp$ ones as the transverse momentum grows, the $\HENLOp$ over $\NLLp$ ratio becoming smaller and smaller, particularly at 14~TeV~LHC.

We conclude this Section by making some overall statements about the transverse-momentum analysis proposed for our process.
The solid stability under MHOUs and NLL corrections shown by predictions of Figs.~\ref{fig:I-k1b-S} to~\ref{fig:I-k12b-M} makes the tetraquark-plus-jet transverse-momentum spectrum one of the most promising set of observables, among those proposed to date, for the study of the core high-energy QCD dynamics.
The found stabilizing pattern, directly connected to the $\TQc$ fragmentation in a VFNS, is evident at LHC energies and persists also at nominal FCC ones.\footnote{We remark that the apparent larger size of ancillary panels of FCC plots in Figs.~\ref{fig:I-k1b-S} to~\ref{fig:I-k12b-M} (right) is simply due to a different scale of the $y$-axis with respect to the LHC one (left).}
As a striking feature, large transverse-momenta seem to offer us a favorable way to clearly discriminate BFKL from fixed-order results.

Finally, in our previous work on heavy-light $\XQq$ tetraquarks it was highlighted how our $\NLLp$ factorization suffers from a loss of stability in the very last transverse-momentum bins. That issue is less pronounced when a $\Xbs$ tetraquark is emitted instead of a $\Xcu$ one~\cite{Celiberto:2023rzw}. 
This is in agreement with the fact that VFNS FFs describing bottomed particles leads to a better stabilization than charmed ones~\cite{Celiberto:2021fdp}.
In the present study we notice how this problem is completely overcome when a fully charmed tetraquark is produced.
Remarkably, because of the much lower value of the charm mass with respect to the bottom one, emissions of $\TQc$ also permit to better restrain instabilities rising in the moderate-to-low transverse-momentum regimes, when energy scales rapidly approach thresholds for DGLAP evolution connected to the heavy-quark masses. 

\subsection{Azimuthal-angle multiplicities}
\label{ssec:phi}

We complement out phenomenological analysis by addressing the azimuthal spectrum of our reaction.
More in particular, we focus on the following multiplicities
\begin{equation}
 \label{azimuthal_multiplicity}
 \frac{1}{\sigma} \frac{\drv \sigma(\phi, s)}{\drv \phi} = \frac{1}{2 \pi} + \frac{1}{\pi} \sum_{l=1}^\infty
 \langle \cos(l \phi) \rangle \, \cos (l \phi)\;,
\end{equation}
where $\phi \equiv \phi_1 - \phi_2 - \pi$ and the mean values $\langle \cos(l \phi) \rangle \equiv C_l^{\rm [order]}/C_0^{\rm [order]}$ are the azimuthal-correlation moments, with $C_{l\geq0}^{\rm [order]}$ the azimuthal coefficients integrated over rapidity and trans\-ver\-se-momentum ranges as indicated in Section~\ref{ssec:DY}, and taken at fixed bins of $\DY$.

Originally introduced to access the light di-jet azimuthal spectrum~\cite{Marquet:2007xx,Ducloue:2013hia}, these multiplicities stand out as particularly promising observables for probing high-energy QCD dynamics.
They encapsulate signals from all azimuthal modes.
Furthermore, their differential nature in $\phi$ simplifies the comparison with experimental data, given that detectors often do not cover the entire $(2\pi)$ range.
A recent analysis on di-jet multiplicities has unveiled two significant advantages~\cite{Celiberto:2022gji}.
First, it allows us to overcome the well-known challenges associated with describing light-flavored final states at natural energy scales. Second, it enhances the agreement with experimental data collected at 7 TeV by CMS~\cite{Khachatryan:2016udy}.

\begin{figure*}[!t]
\centering

   \hspace{0.00cm}
   \includegraphics[scale=0.395,clip]{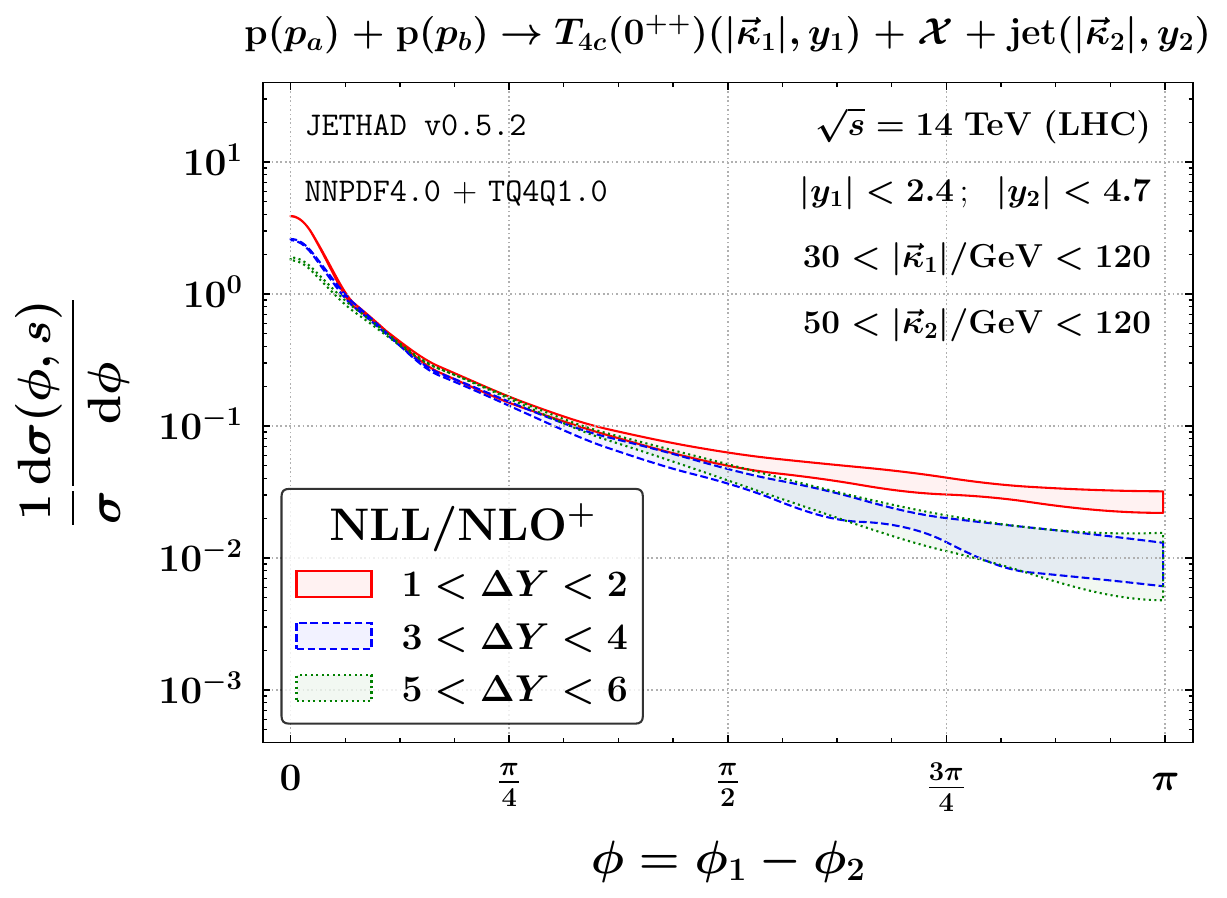}
   \hspace{-0.00cm}
   \includegraphics[scale=0.395,clip]{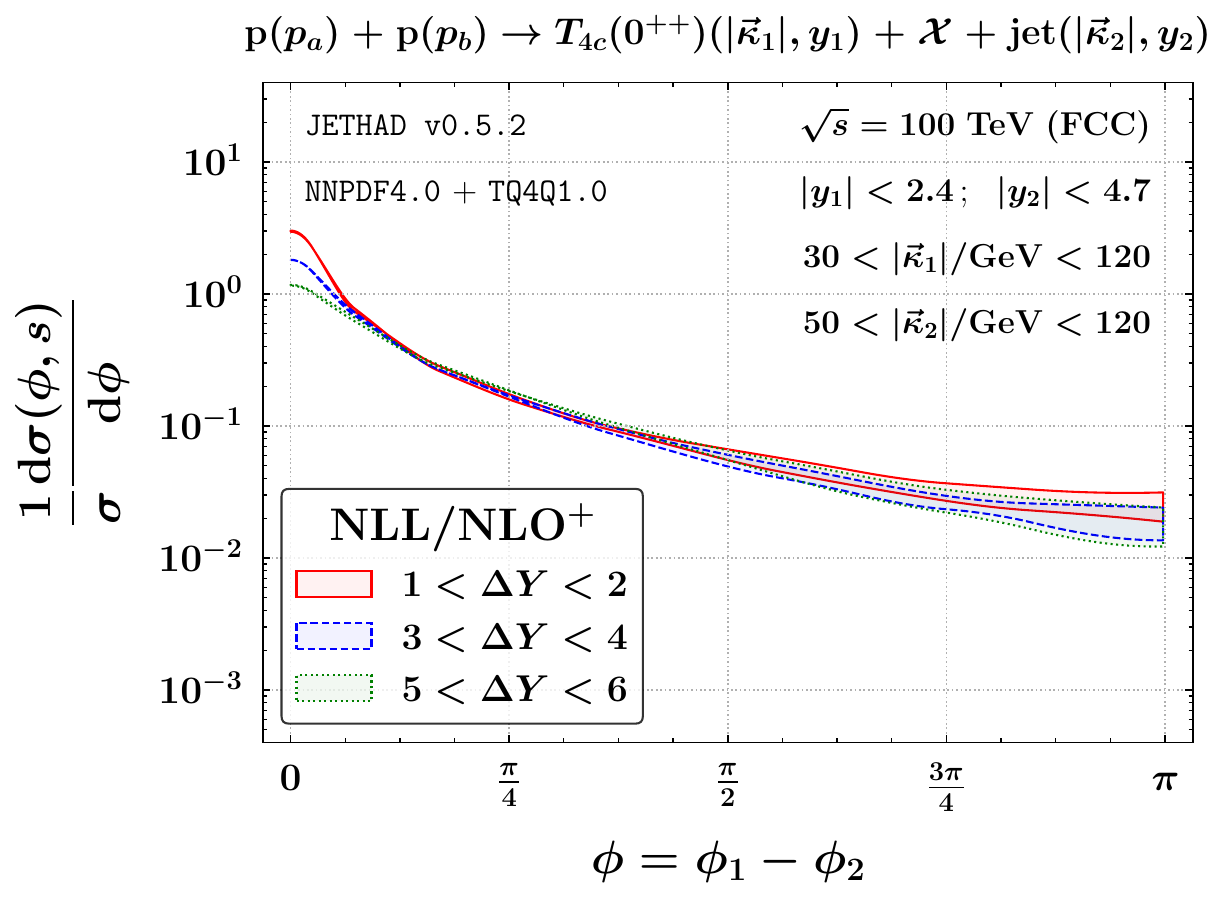}

   \vspace{0.50cm}

   \includegraphics[scale=0.395,clip]{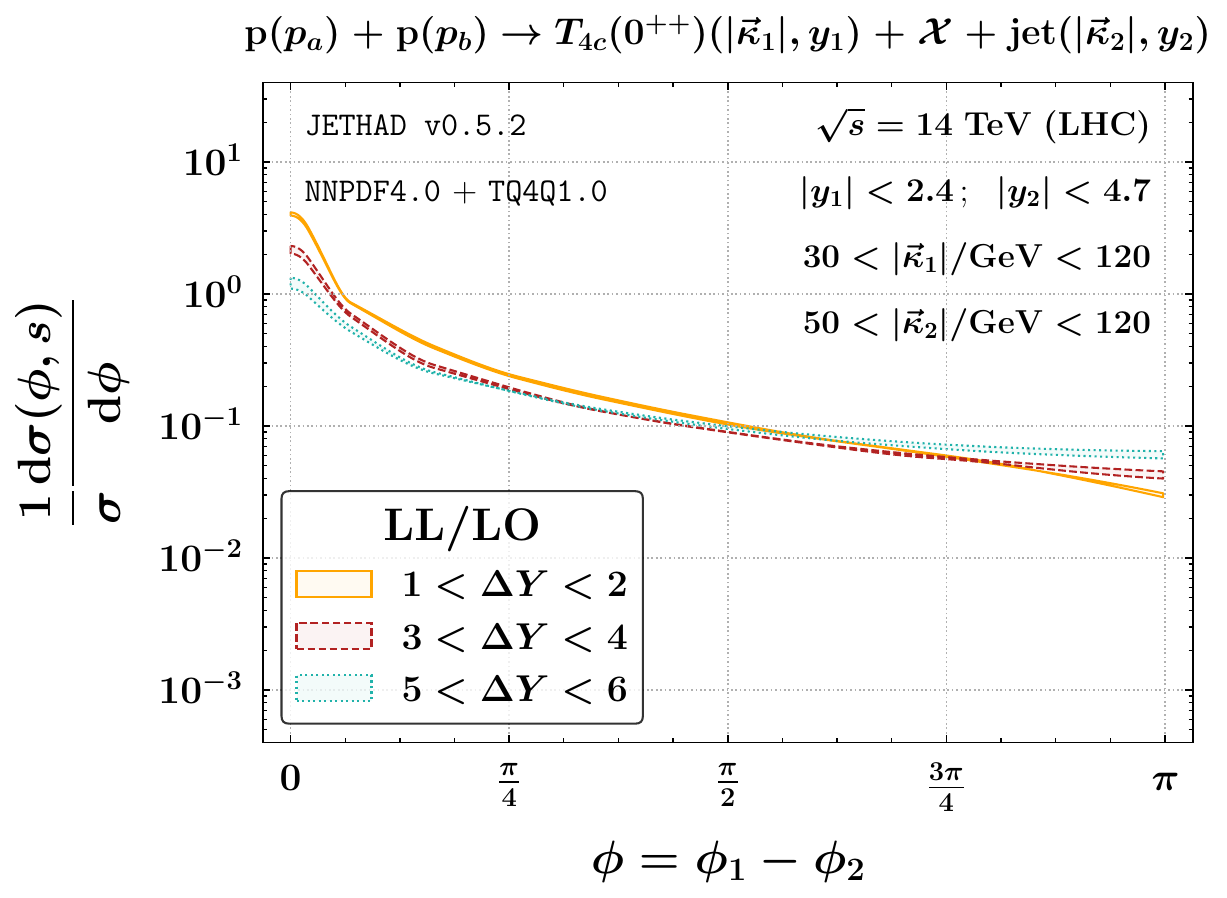}
   \hspace{-0.00cm}
   \includegraphics[scale=0.395,clip]{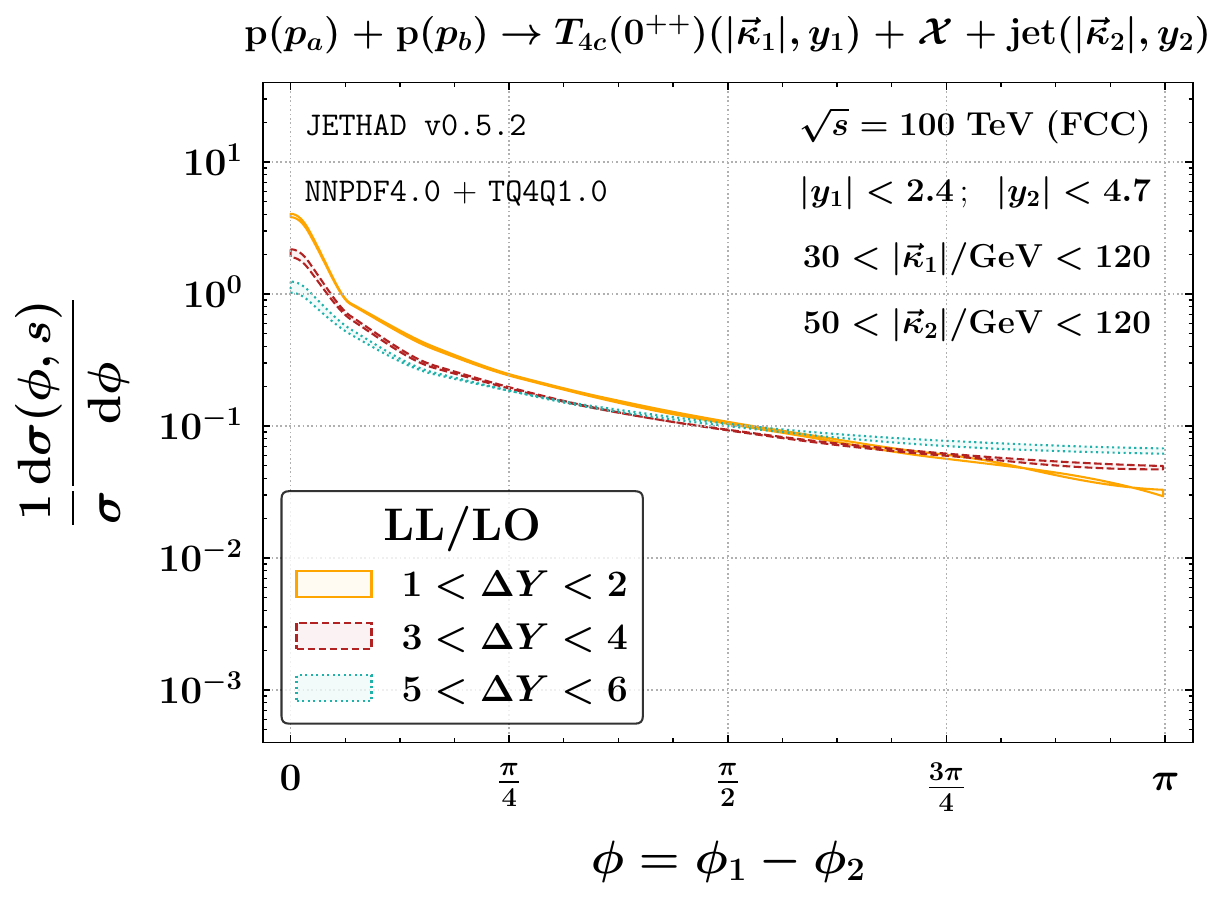}

\caption{Azimuthal multiplicities for the $\TQcZpp$ plus jet detection at $\sqrt{s} = 14$ TeV (LHC, left) or $100$ TeV (nominal FCC, right), and for three different bins of $\DY$.
Upper (lower) plots contain predictions calculated within the $\NLLp$ ($\LL$) hybrid factorization accuracy.
Shaded bands embody the combined effect of MHOUs and phase-space numeric multidimensional integration.}
\label{fig:I-phi}
\end{figure*}

For the sake of simplicity, we show results just for the $\TQcZpp$ channel.
Indeed, being our multiplicities \emph{de facto} normalized cross sections, possible differences connected to the emission of the $\TQcTpp$ resonance are suppressed and shapes for the two channels look almost identical.
Predictions within $\NLLp$ accuracy (Fig.~\ref{fig:I-phi}, upper plots) are compared with the ones at $\LL$ (Fig.~\ref{fig:I-phi}, lower plots).
The range of $\DY$ spans from one to six units, with three selected bins of unit length.
By construction, azimuthal distributions in Eq.~\eqref{azimuthal_multiplicity} are symmetric under $\phi \to -\phi$.
For this reason, we plot them just in the range $0 < \phi < \pi$.

Our multiplicities exhibit a maximum around $\phi = 0$, corresponding kinematically to (almost) back-to-back detections.
The largest maximum occurs in the lowest rapidity bin, $1 < \DY < 2$.
As $\DY$ increases, the maximum height decreases.
This trend indicates that high-energy dynamics comes more and more into play as we approach the large-$\DY$ sector.
Here, the weight of secondary gluons accounted for by the BFKL resummation grows. 
Consequently, the number of back-to-back events diminishes, reducing the azimuthal correlation between the two outgoing objects.

As a noteworthy observation, the computed $\phi$-distributions consistently remain positive. The stabilizing effect originating from tetraquark fragmentation surpasses oscillating instabilities typically affecting the large-$\phi$ tail of multiplicities sensitive to other final states (see, Refs.~\cite{Celiberto:2022dyf,Celiberto:2022zdg,Celiberto:2022keu}).
Those oscillations are connected to \emph{threshold} logarithmic contaminations~\cite{Celiberto:2022kxx}. 
These soft double logarithms, not genuinely accounted for by our resummation, may still survive and contribute to the increasing width of uncertainty bands with $\phi$ observed at $\sqrt{s} = 14$~TeV (LHC).
Conversely, bands for predictions taken at $\sqrt{s} = 14$~TeV do not enlarge when $\phi$ grows.
This is directly connected to kinematics.
Indeed, if one reverts the left-hand side Eq.~\eqref{y-vs-x} to isolate the longitudinal fractions $x_{1,2}$, it becomes evident that, at fixed values of rapidities and transverse momenta, the larger the $\sqrt{s}$, the smaller the values of $x_{1,2}$.
Thus, FCC energies make us drift away from the threshold region, $x_{1,2} \lesssim 1$, where soft-gluon radiation becomes important.

\section{Closing remarks and Outlook}
\label{sec:conclusions}

We investigated the semi-inclusive detection of fully charmed tetraquarks, including $T_{4c}(0^{++})$ bound states and their $T_{4c}(2^{++})$ radial excitations, in high-energy proton collisions.
Our study was based on the single-parton collinear fragmentation in a VFNS, suited for describing the tetraquark production mechanism at moderate and large transverse momenta.
To this end, we derived a new set of collinear FFs, the {\tt TQ4Q1.0} functions, which incorporate initial-scale inputs from both gluon and charm fragmentation channels, respectively defined in the context of quark-potential NRQCD and spin-physics inspired models.
By making use of basic features of the novel {\HFNRevo} method~\cite{Celiberto:2024mex,Celiberto:2024bxu}, we performed a proper DGLAP evolution of these inputs, which consistently accounts for thresholds of the gluon and the charm quark.

As an application to phenomenology, we worked within the $\NLLp$ hybrid factorization scheme and used the {\Jethad} numeric interface along with the {\symJethad} symbolic calculation plugin~\cite{Celiberto:2020wpk,Celiberto:2022rfj,Celiberto:2023fzz,Celiberto:2024mrq,Celiberto:2024swu}.
With these tools, we provided predictions for high-energy observables sensitive to $T_{4c}$ plus jet emissions at center-of-mass energies ranging from 14~TeV~LHC to 100~TeV nominal energy of~FCC.
In this context, the VFNS collinear fragmentation mechanism describing the production of a $\TQc$ state acted as a high-energy stabilizer that shields the hybrid factorization from eventual instabilities arising from genuine NLL corrections as well as from nonresummed threshold logarithms.
The resulting \emph{natural stability} protected us form these issues, which might otherwise hamper the convergence of the resummed series.
Moreover, it preserved the validity of our formalism across a wide range of center-of-mass energies.

Concerning future studies on the exotic sector within high-energy QCD, a significant step forward will be achieved by exploring single inclusive tetraquark detections in forward rapidity directions within our $\NLLp$ formalism. 
These channels would provide direct access to the small-$x$ UGD in the proton, which is currently understood only qualitatively and heavily relies on models.

Being based on both gluon and charm-quark initial-scale inputs, our {\tt TQ4Q1.0} sets can serve as useful guidance for exploratory studies on the production of $\TQc$ states across a wide range of processes, spanning from semi-inclusive emissions at hadron colliders to detections at lepton and lepton-hadron machines.
As a prospect, we plan to enhance our description of tetraquark fragmentation \emph{via} a proper determination of uncertainty, possibly connected to MHOU effects~\cite{Kassabov:2022orn,Harland-Lang:2018bxd,Ball:2021icz,McGowan:2022nag,NNPDF:2024dpb,Pasquini:2023aaf}.
Another possible advancement would be to compare our {\tt TQ4Q1.0} functions with new ones based on NRQCD initial-scale inputs also for the charm fragmentation channel, as proposed in a calculation made during the preparation of the present manuscript (see Ref.~\cite{Bai:2024ezn}).
Color-octet contributions will be considered in future.
The application to investigate the enigmatic $Z_c(3900)$, which has not yet been observed promptly, is also in our plans~\cite{Guo:2013ufa}.

Insights on the inner structure of hadrons will come from a progressive enhancement of our knowledge of the core dynamics underlying quarkonium and exotic matter formation from data to be collected at the FCC~\cite{FCC:2018byv,FCC:2018evy,FCC:2018vvp,FCC:2018bvk} and other future colliders~\cite{Chapon:2020heu,Anchordoqui:2021ghd,Feng:2022inv,AlexanderAryshev:2022pkx,Arbuzov:2020cqg,Accettura:2023ked,InternationalMuonCollider:2024jyv,Black:2022cth,Accardi:2023chb}.
In Ref.~\cite{Flore:2020jau} it was argued that $(\Jpsi + c)$ unresolved photoproduction at the forthcoming EIC~\cite{AbdulKhalek:2021gbh,Khalek:2022bzd,Hentschinski:2022xnd,Amoroso:2022eow,Abir:2023fpo,Allaire:2023fgp} will serve as a promising channel whereby measuring the intrinsic-charm (IC)~\cite{Brodsky:1980pb,Brodsky:2015fna,Jimenez-Delgado:2014zga,Ball:2016neh,Hou:2017khm,Ball:2022qks,Guzzi:2022rca} valence PDF in the proton~\cite{NNPDF:2023tyk}.
In this context, a two-way portal between IC phenomena in the proton (and other hadrons) and the physics of exotics might exist (see, \emph{e.g.}, Ref.~\cite{Vogt:2024fky}).

A link between IC effects and the existence of $| q_1 q_2 q_3 c \bar c \rangle$ pentaquark states was discussed in Ref.~\cite{Mikhasenko:2012km}.
Pentaquarks from intrinsic charms in $\Lambda_b$ baryon decays were considered in Ref.~\cite{Hsiao:2015nna}.
Measurements of multi-$\Jpsi$ rates at NA3 done in the eighties, later supplemented by analyses at the LHC and Tevatron, could provide support for hypotheses involving pion double IC and tetraquark resonances~\cite{NA3:1982qlq,NA3:1985rmd}.
Another engaging challenge is connected with the use of tetraquark and tetraquark-in-jet observables to address the \emph{dead-cone} effect in QCD, theorized in the early nineties as a peculiar feature of heavy-quark fragmentation~\cite{Dokshitzer:1991fd}, and recently observed at the ALICE experiment~\cite{ALICE:2021aqk}.

We believe that future programs aimed at unveiling exotic matter formation and properties \emph{via} high-energy techniques will open a window of opportunities for a joint search for novel physics hidden in the core of the strong force. 

\section*{Data availability}
\label{sec:data_availability}
\addcontentsline{toc}{section}{\nameref{sec:data_availability}}

The current version of the {\tt TQ4Q1.0} release consists of four collinear FF sets in {\tt LHAPDF} format
\begin{itemize}
    \item \; NLO, \,$\TQcZpp$\,: \,{\tt TQ4Q10\_cs\_T4c-0pp\_nlo}\,; \;{\tt TQ4Q10mQ\_cs\_T4c-0pp\_nlo}\,;
    \item \; NLO, \,$\TQcTpp$\,: \,{\tt TQ4Q10\_cs\_T4c-2pp\_nlo}\,; \;{\tt TQ4Q10mQ\_cs\_T4c-2pp\_nlo}\;.
\end{itemize}
Here the ``{\tt cs}'' sublabel is to remark that corresponding initial-scale inputs are based on the color-singlet case, while the ``{\tt mQ}'' one refers to {\tt TQ4Q1.0$^-$} functions evolved from the gluon input only.
They can be publicly accessed from the following url: \url{https://github.com/FGCeliberto/Collinear_FFs/}.

\section*{Acknowledgments}
\label{sec:acknowledgments}
\addcontentsline{toc}{section}{\nameref{sec:acknowledgments}}

We thank Valerio~Bertone, Christian~Biello, Marco~Bonvini, Matteo~Cacciari, Terry~Generet, and Felix~Hekhorn for insightful conversations on heavy-flavor physics.
We are grateful to to Alessandro~Pilloni for useful suggestions on addressing tetraquark phenomenology and to Hongxi~Xing for a discussion on the connection between collinear factorization and the quarkonium theory. 

This work received support from the Atracci\'on de Talento Grant n. 2022-T1/TIC-24176 of the Comunidad Aut\'onoma de Madrid, Spain, and by the INFN/QFT@Colliders Project, Italy.
Feynman diagrams in this work were realized with {\tt JaxoDraw 2.0}~\cite{Binosi:2008ig}.

\bibliographystyle{apsrev}
\bibliography{references}

\end{document}